%% file: sample-acmsmall-conf.tex
\documentclass[acmsmall,nonacm]{acmart}

\setcopyright{none}                
\renewcommand\footnotetextcopyrightpermission[1]{}

\AtBeginDocument{%
  }

\usepackage{titletoc}

\titlecontents{section}
  [1.5em] 
  {\addvspace{4pt}\bfseries} 
  {\contentslabel{1.5em}} 
  {\hspace*{-1.5em}} 
  {\titlerule*[0.4pc]{.}\contentspage} 

\titlecontents{subsection}
  [3.3em] 
  {\addvspace{2pt}} 
  {\contentslabel{2.3em}} 
  {\hspace*{-2.3em}} 
  {\titlerule*[0.4pc]{.}\contentspage} 

\titlecontents{subsubsection}
  [5.5em] 
  {\addvspace{1pt}} 
  {\contentslabel{3.0em}} 
  {\hspace*{-3.0em}} 
  {\titlerule*[0.4pc]{.}\contentspage} 

\usepackage{graphicx}
\usepackage{subcaption}   
\usepackage{caption}
\usepackage{comment}
\usepackage{todonotes}
\usepackage{xcolor}
\usepackage{etoolbox}
\usepackage{siunitx}

\usepackage{amsmath,amssymb}
\usepackage{arydshln}
\usepackage{multirow} 
\usepackage{enumitem} 
\usepackage{marvosym} 
\usepackage{booktabs}  
\usepackage{makecell}  

\usepackage{pifont} 
\usepackage{tikz}
\newcommand{\circlednum}[1]{%
  \tikz[baseline=(char.base)]%
    \node[shape=circle, draw=black, fill=black, text=white, inner sep=1pt] (char) {%
      \sffamily\bfseries\small #1%
    };%
}
\makeatletter
\patchcmd{\@mkauthors@i}
  {\MakeUppercase}
  {}%
  {}{\PackageWarning{acmart}{Failed to patch \\@mkauthors@i, author names may still be uppercase}}
\makeatother

\newcommand{\zuorevise}[1]{\textcolor{black}{#1}}

\newcommand{\system}{CloudMatrix-Infer}

\newcommand{\CMname}{CloudMatrix384}

\begin{document}

\begin{tikzpicture}[remember picture,overlay]
  \node[anchor=north west, xshift=1.35cm, yshift=-0.6cm]
       at (current page.north west)
       {\includegraphics[height=1.2cm]{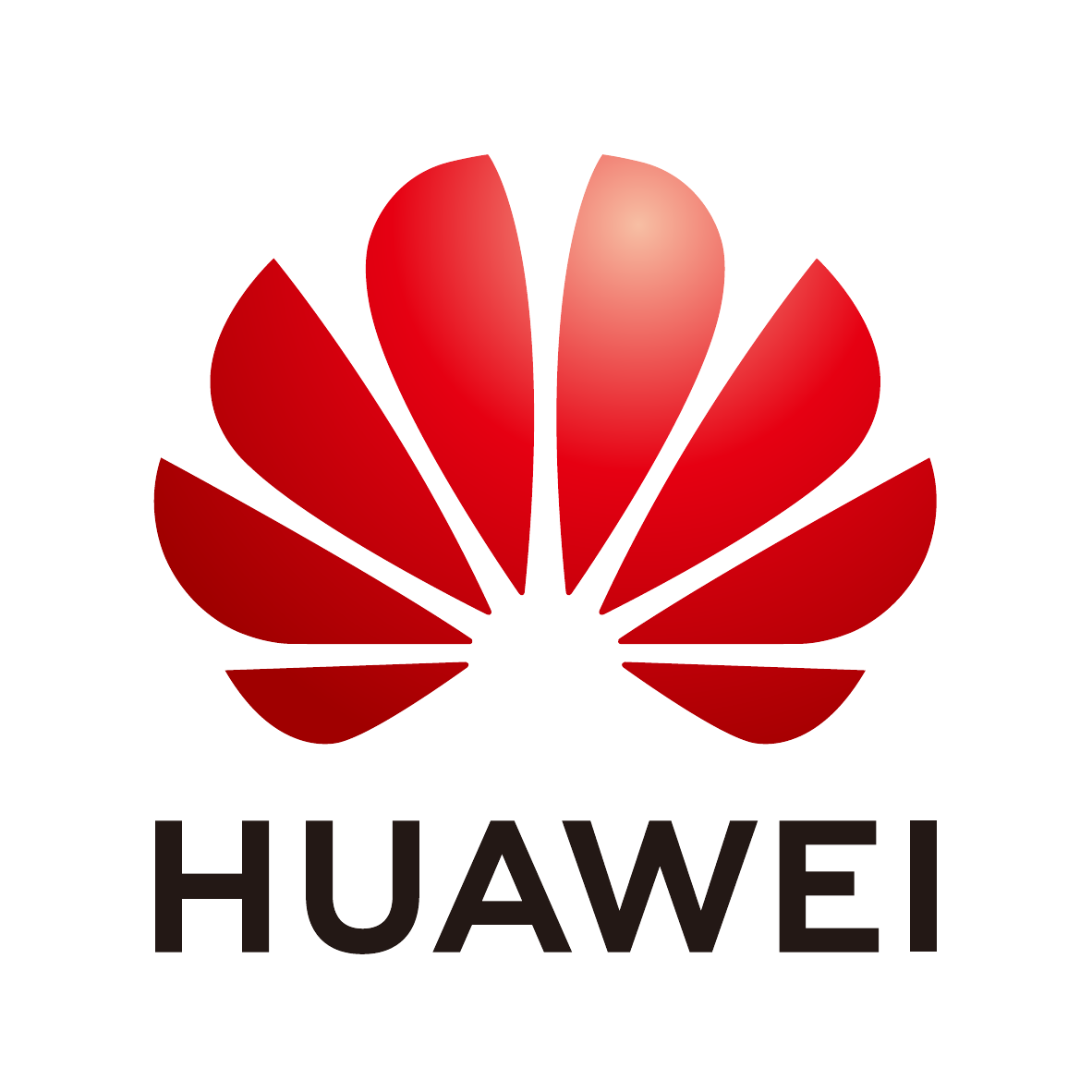}};
\end{tikzpicture}

\vspace{-1.7cm}
\begin{center}
  \rule{1\linewidth}{1pt} 
\end{center}
\vspace{0.4cm}
\title{ Serving Large Language Models on Huawei CloudMatrix384 }

\author{
    \centering
    Pengfei~Zuo, Huimin~Lin, Junbo~Deng, Nan~Zou, Xingkun~Yang, Yingyu~Diao, Weifeng~Gao, Ke~Xu, Zhangyu~Chen, Shirui~Lu, Zhao~Qiu, Peiyang~Li, Xianyu~Chang, Zhengzhong~Yu, Fangzheng~Miao, Jia~Zheng, Ying~Li, Yuan~Feng, Bei~Wang, Zaijian~Zong, Mosong~Zhou, Wenli~Zhou, 
    Houjiang~Chen\textsuperscript{*}, Xingyu~Liao\textsuperscript{*}, Yipeng~Li\textsuperscript{*}, Wenxiao~Zhang\textsuperscript{*}, Ping~Zhu\textsuperscript{*}, Yinggang~Wang\textsuperscript{*}, Chuanjie~Xiao\textsuperscript{*}, Depeng~Liang\textsuperscript{*}, Dong~Cao\textsuperscript{*}, Juncheng~Liu\textsuperscript{*},
    Yongqiang~Yang, Xiaolong~Bai, Yi~Li, Huaguo~Xie, Huatao~Wu, Zhibin~Yu, Lv~Chen, Hu~Liu, Yujun~Ding, Haipei~Zhu, Jing~Xia, Yi~Xiong, Zhou~Yu\textsuperscript{\Letter}, Heng~Liao\textsuperscript{\Letter} \\
    \textit{Huawei}\\
    \textit{\textsuperscript{*}SiliconFlow}\\
}

\thanks{\textsuperscript{\Letter}Corresponding authors: yuzhou@huawei.com, liao.heng@hisilicon.com}

\renewcommand{\shortauthors}{}

\begin{abstract}
\vspace{0.1cm}
{\centering \textnormal{\large \textbf{Abstract}} \\} 
The rapid evolution of large language models (LLMs), driven by increasing parameter scales, adoption of mixture-of-experts (MoE) architectures, and expanding context lengths, imposes unprecedented demands on AI infrastructure.  Conventional AI clusters are increasingly constrained by compute intensity, memory bandwidth limitations, inter-chip communication overhead, and stringent latency requirements. In real-world deployments, these challenges are further compounded by the need to handle diverse, bursty workloads, variable-length inputs, and imbalanced expert activations, while meeting strict service-level objectives. Overcoming these constraints requires a fundamentally re-architected, co-designed hardware and software stack.

To address these challenges, this paper introduces Huawei \textit{CloudMatrix}, a next-generation AI datacenter architecture that embodies Huawei’s vision for reshaping the foundation of AI infrastructure. Huawei CloudMatrix384 represents the first production-grade realization of this vision. It integrates 384 Ascend 910 NPUs, 192 Kunpeng CPUs, and other hardware components into a unified supernode, interconnected via an ultra-high-bandwidth, low-latency Unified Bus (UB) network. Unlike conventional hierarchical designs, this architecture enables direct all-to-all communication via UB, allowing compute, memory, and network resources to be dynamically pooled, uniformly accessed, and independently scaled. These architectural features are particularly beneficial for communication-intensive operations such as large-scale MoE expert parallelism and distributed key-value (KV) cache access, making CloudMatrix384 a scalable and high-performance foundation for next-generation LLM serving. 

To fully harness CloudMatrix384’s capabilities, we propose \textit{\system{}}, a comprehensive LLM serving solution that establishes \textit{a best practice} for deploying large-scale MoE models such as DeepSeek-R1. \system{} incorporates three core innovations. First, we design a \textit{peer-to-peer serving architecture} that disaggregates prefill, decode, and caching into independently scalable resource pools. Unlike existing KV cache-centric architectures, this design enables high-bandwidth, uniform access to cached data via the UB network, thus reducing data locality constraints, simplifying task scheduling, and improving cache efficiency. Second, we design a \textit{large-scale expert parallelism (EP) strategy} that leverages the UB network to achieve efficient token dispatch and expert output combination. This strategy supports a very large EP degree, e.g., EP320, enabling each NPU die to host exactly one expert, thus achieving low decode latency. Finally, we propose a set of \textit{hardware-aware optimizations} tailored to CloudMatrix384, including highly-optimized operators, microbatch-based pipelining, and INT8 quantization, to enhance execution efficiency and resource utilization. 

Our extensive evaluation with the DeepSeek-R1 model shows that \system{} achieves state-of-the-art efficiency without sacrificing accuracy. \system{} delivers a prefill throughput of 6,688 tokens/s per NPU, and a decode throughput of 1,943 tokens/s per NPU (at <50 ms TPOT). These results correspond to compute efficiencies of 4.45 tokens/s/TFLOPS for prefill and 1.29 tokens/s/TFLOPS for decode, both exceeding published results for SGLang on NVIDIA H100 and DeepSeek on NVIDIA H800. \system{} also effectively manages the throughput-latency trade-off, sustaining a decode throughput of 538 tokens/s per NPU even under the stricter sub-15 ms TPOT constraint. Furthermore, the INT8 quantization on Ascend 910 maintains model accuracy comparable to the official DeepSeek-R1 API across 16 distinct benchmarks.

\end{abstract}

\authorsaddresses{}

\maketitle

\clearpage
\tableofcontents           
\clearpage

\input{sections/1-Introduction}
\input{sections/2-LLM}

\input{sections/3-CM384}
\input{sections/4-Serving}

\input{sections/5-Evaluation}

\input{sections/6-Future}

\input{sections/7-Conclusion}

\bibliographystyle{ACM-Reference-Format}
\bibliography{sample-new}

\appendix

\end{document}

%% file: sections/1-Introduction.tex
\section{Introduction}

The landscape of large language models (LLMs) has undergone a dramatic transformation in recent years, driven by several defining trends: the exponential growth of parameter scales, the widespread adoption of mixture-of-experts (MoE) architectures, and the substantial extension of context lengths~\cite{kaplan2020scaling}. Modern LLMs such as DeepSeek-R1~\cite{deepseekai2025deepseekr1}, LLaMA-4~\cite{meta2025llama4}, and Qwen-3~\cite{qwen3_2025} routinely scale to hundreds of billions or even trillions of parameters, placing unprecedented demands on compute power and memory capabilities. MoE models introduce structural sparsity by selectively activating a small subset of experts per token, enabling greater efficiency at scale while introducing new system-level challenges in expert routing and synchronization~\cite{fedus2022switch,jacobs1991adaptive,lepikhin2021gshard}. Simultaneously, context windows have expanded from tens of thousands to over a million tokens~\cite{openai2025gpt45,google2025gemini25}, imposing immense strain on attention computation and key-value (KV) cache storage. \zuorevise{The total KV cache capacity grows linearly with the number of concurrent users, placing significant constraints on how KV cache is distributed, placed, and accessed across the system to support efficient inference.} These trends collectively cause intense pressure on AI infrastructure, requiring massive compute power, high memory capacity and bandwidth, intensive inter-chip communication, and stringent latency constraints, ultimately pushing conventional AI clusters to their scalability limits.

In production environments, serving such models is further complicated by the dynamic and heterogeneous nature of real-world workloads. To be specific, LLM serving systems must accommodate variable-length user inputs, imbalanced expert activations across tokens, and highly bursty user queries, while sustaining stringent latency and throughput targets. Meeting these demands goes beyond simply scaling up hardware resources. It demands a comprehensive hardware and software co-design, including tightly integrated compute, memory, and network hardware resources complemented by intelligent task scheduling, adaptive runtime orchestration, and elastic resource management strategies that dynamically respond to evolving model structures and fluctuating workloads. In summary, as LLMs continue to scale in both size and complexity, it becomes essential to reimagine the design of AI infrastructure from the ground up.

In response to these needs, we present Huawei CloudMatrix, a next-generation AI datacenter architecture built on the principle of fully peer-to-peer high-bandwidth interconnectivity and fine-grained resource disaggregation. We specifically highlight \CMname{}, the first production-grade implementation of this innovative architectural concept. \CMname{} is an AI supernode purpose-built for large-scale AI workloads, featuring a fully peer-to-peer interconnected hardware design. It comprises 384 Ascend 910 NPUs and 192 Kunpeng CPUs, interconnected via an ultra-high-bandwidth and low-latency network named \textit{unified bus (UB)}. In particular, this UB network enables direct all-to-all data exchange across all compute and memory components. Unlike conventional hierarchical architectures with uneven intra-node and inter-node interconnect bandwidth, \CMname{} allows the entire supernode to operate as a logically unified, tightly coupled compute entity, embodying the fully peer-to-peer principle that “everything can be pooled, treated equally, and combined freely”. These architectural features are particularly beneficial for communication-intensive operations such as large-scale MoE expert parallelism and distributed KV cache access, making \CMname{} a scalable and high-performance foundation for next-generation LLM serving.

The initial design of \CMname{} predates the widespread adoption of MoE architectures~\cite{deepseek2024v3,qwen3_2025,meta2025llama4}, as the design and deployment of such a comprehensive supernode system typically spans several years. Nonetheless, \CMname{} was purpose-built to enhance interconnect bandwidth and communication efficiency—core capabilities essential for scaling large training and inference workloads. The emergence of large-scale MoE models such as DeepSeek-R1~\cite{deepseekai2025deepseekr1} validates this architectural foresight, highlighting that communication bandwidth is as crucial as compute and memory bandwidth capabilities in modern LLM deployments.

To fully exploit \CMname{}’s capabilities, we propose \system{}, a comprehensive LLM serving solution that represents \textit{a best practice} for deploying large-scale MoE models such as DeepSeek-R1. \system{} introduces three core innovations.

First, we design a novel \textit{peer-to-peer serving architecture} that disaggregates the LLM inference system into three independent subsystems: prefill, decode, and caching. \zuorevise{Peer-to-peer means that the three subsystems operate as equal and independent resource pools, without being orchestrated around a centralized entity. This contrasts sharply with conventional KV cache-centric architectures~\cite{qin2025mooncake,nvidia_dynamo_2025}, which tightly couple request scheduling to the physical placement of cached KV blocks, adding scheduling complexity and limiting flexibility in resource assignment.} By leveraging the high-bandwidth UB interconnect, we construct a disaggregated memory pool that provides shared caching services across the system. All NPUs in the prefill and decode subsystems can access cached KV data directly from this pool in a peer-to-peer manner, with uniform bandwidth and latency, regardless of where the data was originally computed or stored. This design decouples request scheduling from data locality, greatly simplifying task scheduling logic, improving cache efficiency, and enhancing overall system resource utilization.

Second, we develop a \textit{large-scale expert parallelism (LEP)} strategy specifically optimized for MoE models. \zuorevise{The core principle of LEP is to aggregate compute power and memory bandwidth across a large number of NPUs to accelerate the computation of attention and feed-forward networks. This acceleration comes at the cost of increased communication overhead due to token dispatch and expert output combination. However, \CMname{}’s ultra-high-bandwidth UB interconnect ensures that this communication latency remains bounded and does not become the dominant performance bottleneck. Furthermore, our LEP strategy supports extremely high degrees of expert parallelism, such as EP320, enabling each NPU die to host exactly one expert for DeepSeek-R1. This configuration minimizes serial execution among experts within the same rank, thereby reducing overall MoE execution latency.} Together, these design choices enable low decode latency and deliver substantial end-to-end performance gains for MoE-based inference.

Finally, we introduce a suite of \textit{hardware-aware optimizations} explicitly tailored for \CMname{}, including highly-optimized Ascend operators, microbatch-based pipelining, and INT8 quantization. The optimized operators accelerate end-to-end execution and provide efficient support for LEP. The microbatch-based pipelining design enhances both resource utilization and system throughput by overlapping the processing of two consecutive microbatches. INT8 quantization boosts computational efficiency and substantially reduces memory bandwidth consumption. Collectively, these optimizations are co-designed with the unique architectural features of \CMname{}, including on-chip cube, vector, and communication engines, as well as the high-bandwidth UB interconnect, to maximize overall execution efficiency.

Our evaluation of \system{} on the \CMname{}, using the 671-billion-parameter DeepSeek-R1 model, demonstrates impressive performance and hardware efficiency. In the prefill phase, \system{} achieves a throughput of 6,688 tokens/s per NPU for a 4K prompt length. This translates to a compute efficiency of 4.45 tokens/s per TFLOPS. For the decode phase, the system sustains 1,943 tokens/s per NPU for a 4K KV cache Length while maintaining a time-per-output-token (TPOT) consistently below 50 ms, yielding an efficiency of 1.29 tokens/s per TFLOPS. Notably, the compute efficiency metrics for both phases surpass those of leading frameworks like SGLang on NVIDIA H100 and DeepSeek on NVIDIA H800. \system{} also demonstrates effective management of the fundamental throughput-latency trade-off. To meet a stricter sub-15 ms TPOT requirement, \system{} can dynamically adjust its batch size, achieving a decode throughput of 538 tokens/s per NPU. This highlights its predictable performance and adaptability under varying service-level objectives. Furthermore, the INT8 quantization maintains accuracy comparable to the official DeepSeek-R1 API across 16 representative benchmarks. These results collectively establish \CMname{}, in combination with our peer-to-peer serving solution \system{}, as a scalable, high-throughput, and production-grade platform for large-scale LLM deployment.

The remainder of this paper is organized as follows. Section~\ref{sec:background} begins by reviewing key LLM trends and presenting system-level challenges inherent in conventional datacenter infrastructure. Section~\ref{sec:cm384} describes the vision of Huawei CloudMatrix and details the design of \CMname{}. We then introduce the serving system architecture and optimization techniques employed in \system{} in Section~\ref{sec:deepseek_on_cm384}. A detailed performance evaluation is presented in Section~\ref{sec:evaluation}. Finally, Section~\ref{sec:future_directions} outlines our future research directions before Section~\ref{sec:conclusion} concludes the paper.

%% file: sections/2-LLM.tex
\section{LLM Trends and Their Challenges for Datacenter Infrastructure}
\label{sec:background}

In this section, we first discuss recent trends in large language model (LLM) design that are shaping the landscape of AI computing (\S\ref{sec:llm-trends}). We then present the corresponding system-level challenges these trends impose on conventional datacenter infrastructure (\S\ref{sec:challenge-for-ai-infra}).

\subsection{LLM Trends} 
\label{sec:llm-trends}

The rapid evolution of LLMs has been marked by three prominent trends: the ever-increasing model parameter counts, the adoption of sparsity through Mixture-of-Experts (MoE) architectures, and the extension of context windows. These developments aim to enhance model performance while addressing computational efficiency and scalability.

\textbf{Ever-Larger Parameter Counts.}
Empirical scaling laws suggest that increasing the number of parameters in LLMs leads to improved model performance across various tasks~\cite{kaplan2020scaling}. Recent developments exemplify this trend: Meta's Llama 4 Behemoth boasts nearly 2 trillion parameters, while its counterpart, Llama 4 Maverick, comprises 400 billion parameters~\cite{meta2025llama4}. DeepSeek-V3, developed by DeepSeek-AI, contains 671 billion parameters~\cite{deepseek2024v3}. Google's PaLM model includes 540 billion parameters~\cite{chowdhery2023palm}, and xAI's Grok-1 features 314 billion parameters~\cite{xai2024grok1}. These models underscore the industry's ongoing pursuit of scaling LLMs to enhance capabilities in reasoning, multilingual understanding, and code generation.

\textbf{Sparsity through MoE.}
To manage the escalating costs of training and inference, modern LLMs increasingly adopt sparsely-activated MoE architectures, which decouple total model capacity from per-token computational requirements. Notable implementations include Mixtral 8×7B, which comprises 46.7 billion total parameters but activates only 12.9 billion per token by routing each token to 2 of 8 experts per layer, achieving performance comparable to GPT-3.5 while maintaining computational efficiency~\cite{jiang2024mixtral}. Databricks' DBRX employs a fine-grained MoE architecture with 132 billion total parameters, activating 36 billion per token through the selection of 4 out of 16 smaller experts, enhancing throughput and reducing latency~\cite{dbrx2024}. Meta's Llama 4 series introduces MoE in open-source models, with Llama 4 Maverick utilizing 128 experts and Llama 4 Scout employing 16 experts, both maintaining 17 billion active parameters per token~\cite{meta2025llama4}. DeepSeek-V3 expands upon its predecessor by increasing the number of routed experts per layer from 160 to 256, thereby enhancing model capacity without proportionally increasing computational load~\cite{liu2024deepseekv2,deepseek2024v3}. Alibaba's Qwen3-235B model incorporates 128 experts, activating 22 billion parameters per token, balancing large-scale capacity with computational efficiency~\cite{qwen3_2025}. Huawei’s Pangu Ultra MoE model scales to 718 billion parameters, with 39 billion active parameters per token. It employs an MoE architecture featuring 256 experts per layer, of which 8 are activated per token~\cite{tang2025pangu}. Collectively, these models underscore a paradigm shift in LLM scaling strategies, emphasizing the importance of architectural sparsity over sheer parameter count to achieve enhanced performance and efficiency. 

\textbf{Extension of Context Windows.}
The expansion of context windows in LLMs enables the processing of longer sequences, which is vital for tasks requiring extended reasoning and coherence. Recent advancements reflect this shift: OpenAI's GPT-4.5 supports a context window of 128,000 tokens~\cite{openai2025gpt45}, while Google's Gemini 2.5 Pro offers a context window of up to 1 million tokens~\cite{google2025gemini25}. Benchmarks such as LongBench~\cite{bai-etal-2024-longbench} quantify the benefits of extended context windows for tasks like question answering, summarization, and multi-step reasoning. However, feeding LLMs with long prompts significantly increases computational costs and prolongs inference latency. To mitigate these costs, production systems adopt context caching, wherein key-value (KV) blocks generated from earlier prompt segments are stored and reused across subsequent turns or requests. This approach eliminates redundant attention computations for prompts, thereby reducing latency and improving efficiency~\cite{gao2024cachedattention,qin2025mooncake}.

\subsection{\zuorevise{Challenges for Datacenter Infrastructure}}
\label{sec:challenge-for-ai-infra}

\zuorevise{These LLM trends place stringent new demands on underlying datacenter infrastructure. As model capabilities continue to expand, they drive the emergence of increasingly complex workloads, such as reasoning-intensive inference, reinforcement learning (RL)-based post-training, interactive media generation, and autonomous AI agents. These applications require not only significantly greater compute and memory capacity, but also a fundamental re-architecture of infrastructure to support high-bandwidth communication, low-latency storage access, and sustained throughput, while meeting tight service-level latency objectives under dynamic, heterogeneous real-world conditions. In this context, we identify four key system-level challenges:}

\textbf{Challenge 1: Scaling Communication-Intensive Parallelism.}
As model sizes grow, state-of-the-art AI models often exceed the capacity of a single compute node, necessitating multi-node parallelism strategies. While existing AI clusters support inter-node communication via RDMA networks, their bandwidth and topology are typically optimized for data or pipeline parallelism (DP/PP), which involve modest inter-node traffic. However, tensor parallelism (TP) and expert parallelism (EP) demand frequent, fine-grained, and low-latency communication that is difficult to scale efficiently across node boundaries. This forces many deployments to confine TP/EP groups within a single compute node, limiting scalability.

\textbf{Challenge 2: Maintaining High Utilization under Heterogeneous AI Workloads.}
Modern AI workloads exhibit highly diverse and dynamic resource requirements. Training is typically compute-intensive, inference (particularly the decode phase of LLMs) is often limited by memory bandwidth, and tasks such as autonomous-driving model training involve substantial CPU-side data preprocessing. Fixed node configurations cannot efficiently accommodate this diversity, often leading to over-provisioning or underutilization. To maximize efficiency and adaptability, modern AI infrastructure must enable dynamic, fine-grained composition of heterogeneous resources, e.g., NPUs, CPUs, and memory, adapted to the specific demands of each workload.

\textbf{Challenge 3: Enabling Converged Execution of  AI and Data-Intensive Workloads.}
AI workflows increasingly intersect with traditional data-intensive operations such as data ingestion, preprocessing, retrieval, analytics, and simulation. Meanwhile, general-purpose workloads, e.g., databases, big data, and HPC, are themselves evolving to incorporate AI capabilities. These converged execution patterns demand high-throughput, low-latency communication and flexible resource orchestration. However, legacy datacenter infrastructures primarily optimized for conventional general-purpose workloads struggle to meet these stringent requirements. Enabling efficient convergence of AI and data-intensive tasks requires a fundamentally new infrastructure.

\textbf{Challenge 4: Delivering Memory-class Storage Performance.}
Modern AI pipelines operate at unprecedented data scales that far exceed the capabilities of traditional storage systems. Tasks, such as ingesting petabyte-scale datasets, managing multi-terabyte model checkpoints, and supporting latency-sensitive inference, particularly with large KV caches and retrieval-augmented generation (RAG) modules, require storage subsystems with memory-class bandwidth, latency, and IOPS. Legacy storage hierarchies, designed around disk-based access patterns, frequently become performance bottlenecks, leading to NPU underutilization due to data starvation.

%% file: sections/3-CM384.tex
\section{Huawei CloudMatrix} 
\label{sec:cm384} 

To address these emerging challenges in AI workloads, Huawei proposes CloudMatrix, a next-generation AI datacenter architecture designed to reshape the foundation of AI infrastructure. This architectural vision centers on constructing a unified, tightly-coupled compute fabric that can efficiently support the scale, heterogeneity, and communication demands of modern AI applications. \CMname{} represents the first production-grade realization of this vision, delivering a purpose-built platform optimized for large-scale AI workloads.

This section begins by outlining the CloudMatrix vision (\S\ref{sec:motivations_cm}). We then provide an overview of the fully peer-to-peer hardware architecture of \CMname{} (\S\ref{sec:cm384_hw_overview}), followed by a breakdown of its core hardware components (\S\ref{sec:cm384_hw_components}). Next, we present the software stack that enables CloudMatrix deployment in Huawei Cloud (\S\ref{sec:cm384_software_stack}). Finally, we analyze its suitability for efficiently serving large-scale MoE models like DeepSeek-R1 (\S\ref{sec:applicability_deepseek}).

\subsection{Vision for Huawei CloudMatrix}
\label{sec:motivations_cm}

\begin{figure}[t]
    \centering
    \includegraphics[width=0.95\linewidth]{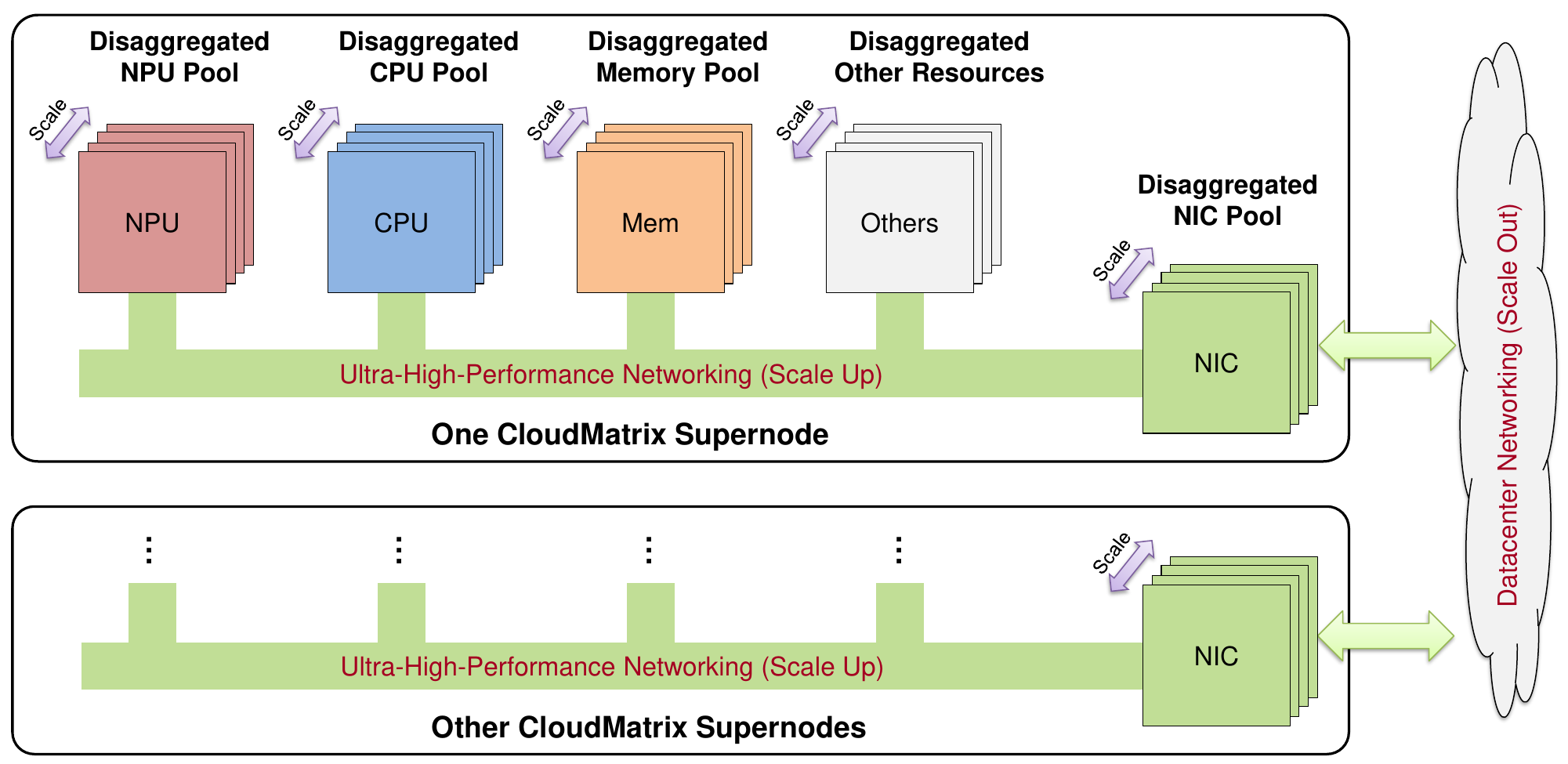}
    \caption{Huawei’s CloudMatrix architecture vision reimagines AI datacenter infrastructure from the ground up. By dismantling traditional siloed designs, it enables full peer-to-peer disaggregation and pooling of CPUs, NPUs, memory, NICs, and other resources over a unified, ultra-high-performance networking, forming the foundation for scalable, AI-native datacenters.}
    \label{fig:cm-conceptual-design}
\end{figure}

In response to the escalating demands of modern large-scale AI workloads, Huawei introduces CloudMatrix, a pioneering next-generation AI datacenter architecture.  This architecture is meticulously designed around the principle of fully peer-to-peer high-bandwidth interconnectivity and fine-grained resource disaggregation.
As conceptually outlined in Figure~\ref{fig:cm-conceptual-design}, CloudMatrix moves beyond traditional CPU-centric hierarchical designs. It facilitates direct, high-performance communication among all heterogeneous system components, including NPUs, CPUs, DRAM, SSDs, NICs, and domain-specific accelerators, notably without requiring CPU mediation.

At the heart of this architecture is the ultra-high-bandwidth, low-latency \textit{Unified Bus (UB)} network, which facilitates efficient, system-wide data movement and coordination. Built upon this interconnect substrate, CloudMatrix delivers four foundational capabilities that collectively define a new paradigm for AI-native infrastructure:

\begin{enumerate} 
    \item \textbf{Scalable Communication for TP/EP.}
    The UB interconnect supports direct, high-throughput peer-to-peer communication across NPUs, enabling TP and EP groups to scale beyond the boundary of a single node. This removes inter-node bottlenecks and allows large models to be efficiently distributed across the supernode.

    \item \textbf{Flexible Resource Composition for Heterogeneous Workloads.}
    CloudMatrix disaggregates CPUs, NPUs, and memory into independently pooled resources, enabling fine-grained, workload-driven composition. This flexibility allows resource allocation at fine granularity based on workload needs, e.g., memory-rich caching nodes, CPU-heavy preprocessing nodes, freeing deployments from fixed node configurations or PCIe-based host-device coupling.

    \item \textbf{Unified Infrastructure for Converged Workloads.} The high-bandwidth UB network supports both AI and data-intensive applications within a single, scale-up infrastructure. This enables converged execution of LLM inference, training, simulation, and analytics workloads, an increasingly common requirement for hybridized AI pipelines.

    \item \textbf{Memory-class Storage via Disaggregated Memory Pool.} CloudMatrix aggregates CPU-attached DRAM across the cluster into a shared, high-performance memory pool accessible via UB. This substrate powers services such as the elastic memory service (EMS)~\cite{ems2025}, which accelerates latency-critical operations like KV cache reuse, parameter loading, and model checkpointing by eliminating conventional I/O bottlenecks.
\end{enumerate}

Huawei \CMname{}, described in the following sections, is the first production-grade realization of this architectural vision. It is specifically engineered to meet the compute, memory, and communication demands of next-generation AI workloads at scale.

\subsection{\CMname{} Overview: A Fully Peer-to-Peer Hardware Architecture} 
\label{sec:cm384_hw_overview}

\begin{figure}[t]
    \centering
    \includegraphics[width=0.95\linewidth]{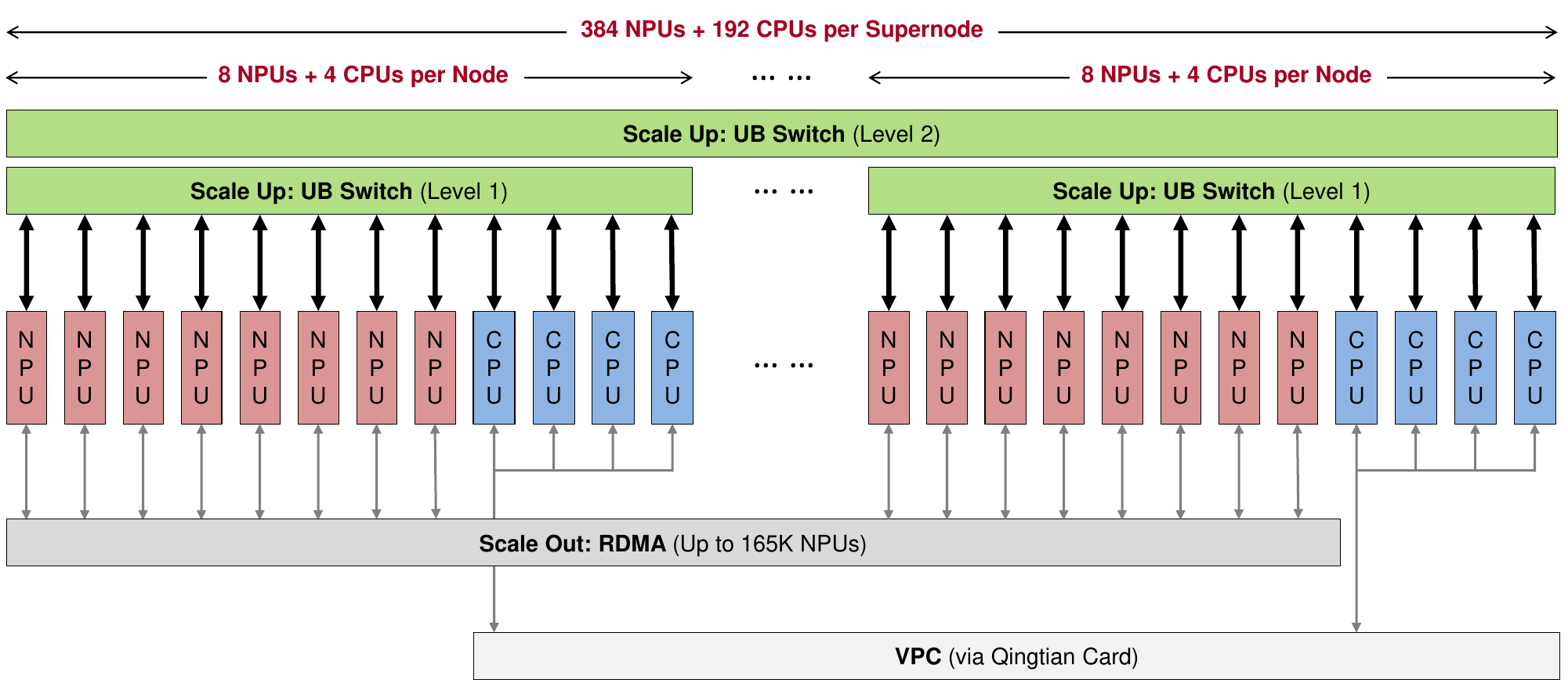}
    \caption{Peer-to-peer hardware architecture of a \CMname{} supernode, featuring an ultra-high-bandwidth Unified Bus (UB) plane for intra-supernode scaling, an RDMA plane for inter-supernode communication, and a Virtual Private Cloud (VPC) plane for integration with the datacenter network. All reported network bandwidth values denote unidirectional bandwidth.}
    \label{fig:cm384-arch}
\end{figure}

\CMname{} is engineered as an AI supernode that integrates 384 Ascend 910 neural-network processing units (NPUs) and 192 Kunpeng central processing units (CPUs), as illustrated in Figure~\ref{fig:cm384-arch}. A defining feature of \CMname{} is its \textit{peer-to-peer, fully interconnected, ultra-high-bandwidth network} that links all NPUs and CPUs via the UB protocol. \CMname{}’s UB design is a precursor to the UB-Mesh proposed in~\cite{liao2025ubmesh_article}. Each of the 384 NPUs and 192 CPUs connects through UB switches, enabling inter-node communication performance that closely approximates intra-node levels. The inter-node bandwidth degradation is under 3\%, and inter-node latency increase is less than 1~\textmu s. Given that modern AI workloads are predominantly bandwidth-intensive rather than latency-sensitive, this marginal latency overhead has a negligible impact on the end-to-end performance of AI tasks. Overall, this design allows \CMname{} to function as a tightly-coupled, large-scale logical node with globally addressable compute and memory, facilitating unified resource pooling and efficient workload orchestration.

To support diverse traffic patterns and maintain compatibility with legacy datacenter networks, \CMname{} incorporates three distinct yet complementary network planes:

\textbf{1) UB Plane.} The UB plane forms the primary ultra-high-bandwidth scale-up fabric within the supernode. It directly interconnects all 384 NPUs and 192 CPUs in a non-blocking all-to-all topology. UB enables: (1) efficient implementation of fine-grained parallelism strategies such as TP and EP, unconstrained by node boundaries; (2) fast peer-to-peer access to pooled memory (spanning both CPU and NPU memory), which is crucial for efficiently caching model weights and KV caches.

\textbf{2) RDMA Plane.}
The RDMA plane enables scale-out communication across \CMname{} supernodes and external RDMA-compatible systems. It currently adopts RDMA over Converged Ethernet (RoCE) to ensure compatibility with standard RDMA stacks.\footnote{An alternative design, \textit{RDMA over UB}, leverages UB’s native support for remote memory access to form a unified UB domain for both intra- and inter-supernode communication. While this approach offers streamlined semantics and avoids protocol translation overhead, the current implementation opts for RoCE to ensure immediate compatibility with existing RDMA libraries and tooling.} NPUs are the sole participants in this plane, isolating RDMA traffic from control and storage operations. Key functions include: (1) high-speed transfer of active KV cache data between prefill and decode NPUs during inference; (2) support for distributed training and inference using RDMA-compliant frameworks; (3) low-latency interconnect across supernodes in multi-cluster deployments.

\textbf{3) VPC Plane.}
The virtual private cloud (VPC) plane connects the \CMname{} supernode to the broader datacenter network via high-speed NICs (Huawei’s \textit{Qingtian} card).  It operates over standard Ethernet and IP protocols, optionally augmented with UB-over-Ethernet (UBoE). The VPC plane handles: (1) management and control-plane operations such as deployment, monitoring, and scheduling; (2) access to persistent storage, including the object storage service (OBS), the elastic volume service (EVS), and the scalable file system service (SFS); (3) external service communication from CPU-resident workloads, e.g., databases and user interfaces.

Although the long-term vision of CloudMatrix aims to converge RDMA and VPC planes into a single unified plane as shown in Figure~\ref{fig:cm-conceptual-design}, the current \CMname{} separates them to ensure backward compatibility with legacy datacenter infrastructure. We discuss the future work of unifying VPC and RDMA planes in \S\ref{sec:future-supernode-unifyed-plane}.

\subsection{Hardware Components}
\label{sec:cm384_hw_components}

\subsubsection{Ascend 910 Chip}
\label{sec:ascend910c_chip}

\begin{figure}[t]
    \centering
    \includegraphics[width=0.55\linewidth]{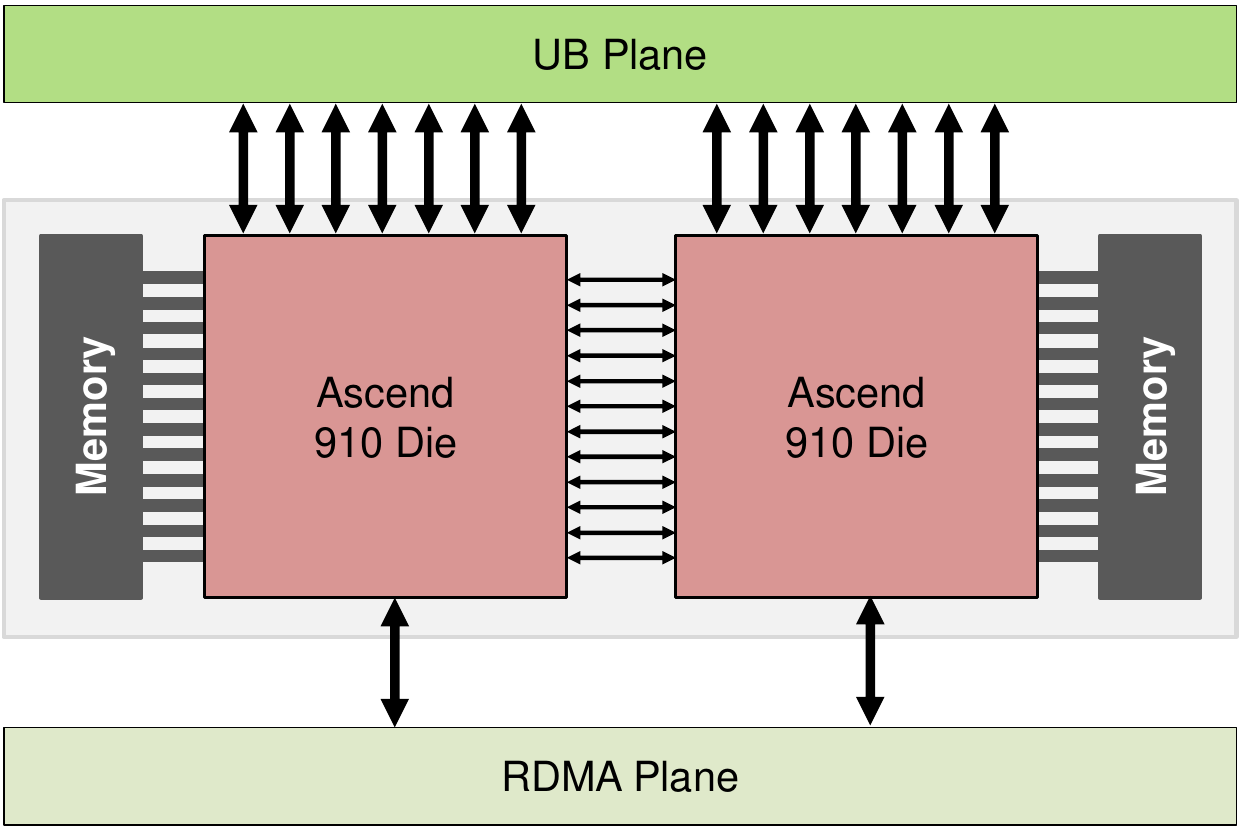}
     \caption{Logical overview of the Huawei Ascend 910 chip, highlighting its dual-die architecture. All reported network bandwidth values denote unidirectional bandwidth.}
    \label{fig:a3}
\end{figure}

At the core of CloudMatrix~384 is the HiSilicon Ascend 910 NPU, Huawei’s 2024‐era flagship AI accelerator. The Ascend 910 is a \textit{dual‐die} package: two identical compute dies are co-packaged, sharing eight on-package memory stacks and connected by a high-bandwidth cross-die fabric, as shown in Figure~\ref{fig:a3}. Each die contains 24 AI cube (AIC) cores, optimized for matrix and convolution workloads, and 48 AI vector (AIV) cores for element-wise operations. All compute engines support FP16/BF16 and INT8 data types. The 8-bit quantization can be implemented with INT8 precision, enabling computational efficiency comparable to native FP8 hardware without requiring dedicated FP8 support. The two dies communicate over an on-package interconnect. Each Ascend 910 die interfaces with two distinct network planes.
\textit{1) UB Plane}: The die integrates seven high-speed transceivers to the scale-up UB plane.
\textit{2) RDMA Plane}: Separately, each die includes a dedicated interface for the scale-out RDMA plane.

\subsubsection{Ascend 910 Node}
\label{sec:ascend910c_server}

\begin{figure}[t]
    \centering
    \includegraphics[width=0.95\linewidth]{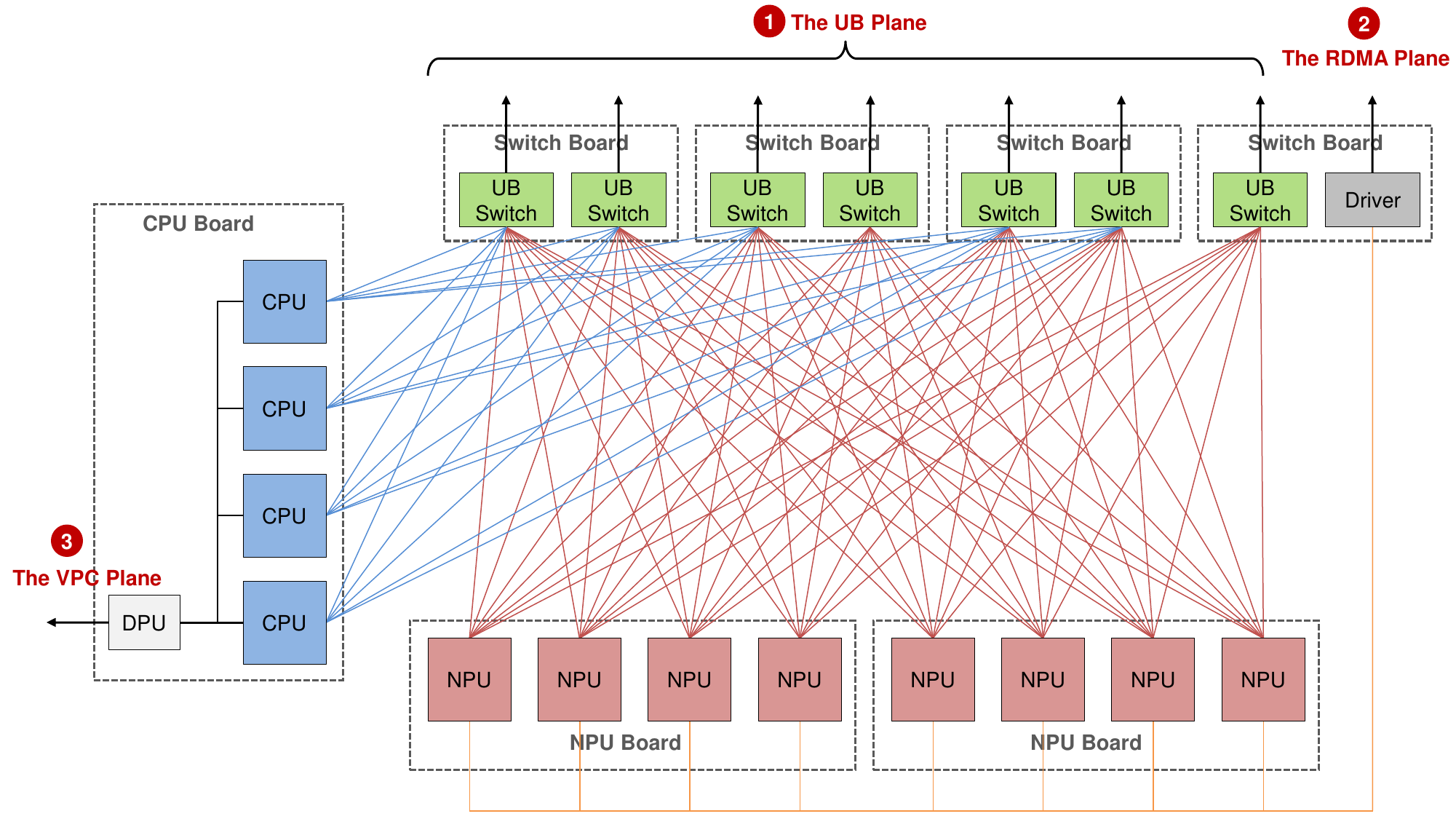}
    \caption{Logical overview of an Ascend 910 node within the \CMname{}. All reported network bandwidth values denote unidirectional bandwidth.}
    \label{fig:a3-server}
\end{figure}

Each compute node in \CMname{} integrates 8 Ascend 910 NPUs, 4 Kunpeng CPUs, and 7 UB switch chips onboard, as illustrated in Figure~\ref{fig:a3-server}. The 12 processors (8 NPUs and 4 CPUs) connect to these on-board switches via UB links, creating a single-tier \textit{UB plane} within the node. Each UB switch chip onboard links to the next switching tier in the supernode fabric. Only NPUs participate in the secondary \textit{RDMA plane}.

Within the CPU complex, the four Kunpeng CPU sockets are interconnected via a full-mesh NUMA topology, enabling uniform memory access across all CPU-attached DRAM. 
One of the CPUs hosts the node’s \textit{Qingtian} card, a dedicated data processing unit (DPU) that not only integrates high-speed network interfaces but also performs essential node-level resource management functions. This \textit{Qingtian} card serves as the primary north–south egress point from the node, interfacing with the third distinct network plane: the datacenter’s \textit{VPC plane}.

\subsubsection{UB Switch System}
\label{sec:ub_switch_system}

The \CMname{} supernode spans 16 racks: 12 \textit{compute racks}, which collectively host the 48 Ascend 910 nodes (384 NPUs in total), and 4 \textit{communication racks}. These communication racks house the second-tier (L2) UB switches that interconnect all the nodes within the supernode.

\begin{figure}[t]
    \centering
    \includegraphics[width=0.95\linewidth]{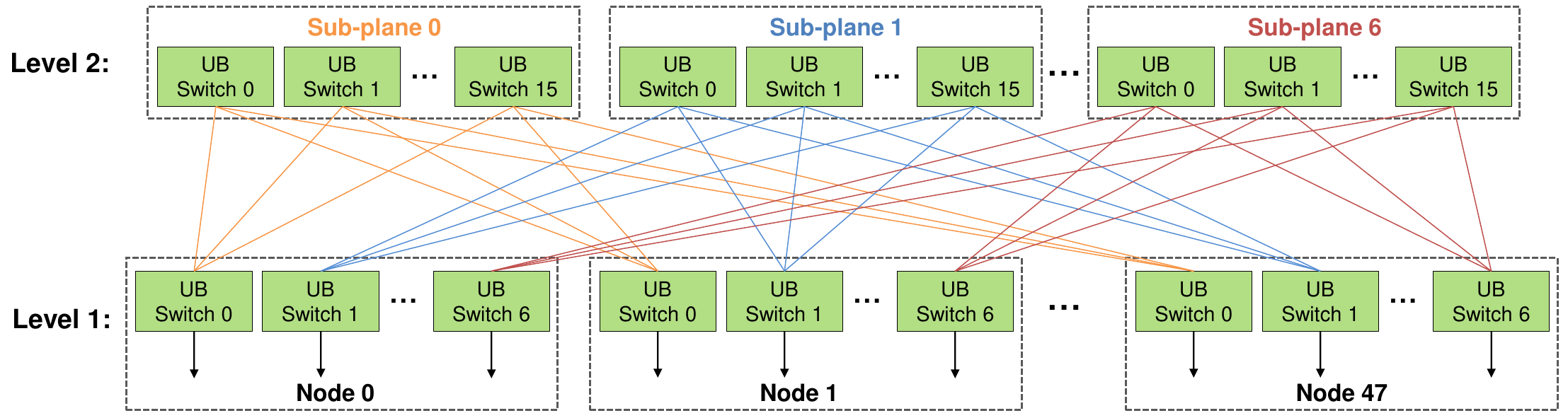}
    \caption{The UB switch system in the \CMname{}. All reported network bandwidth values denote unidirectional bandwidth.}
    \label{fig:a3-switch}
\end{figure}

Figure~\ref{fig:a3-switch} illustrates the topology between the on-board first-tier (L1) UB switches (located inside each Ascend 910 node) and the rack-level L2 UB switches. The network is designed to be non-blocking, meaning there is \textit{no} bandwidth oversubscription at the L2 switching tier. The L2 switches are partitioned into 7 independent \textit{sub-planes}. Each sub-plane contains 16 L2 UB switch chips, and each L2 switch chip provides 48 ports.

Inside each node, the 7 on-board L1 UB switch chips map one-to-one onto these 7 L2 sub-planes. Each L1 switch chip fans out over 16 links (one link to every L2 switch chip in its corresponding sub-plane). This configuration ensures that a node’s aggregate uplink bandwidth to the L2 fabric precisely matches its internal UB capacity, maintaining the non-blocking characteristic across the supernode.

\subsection{Software Stack} 
\label{sec:cm384_software_stack}

\subsubsection{CANN for Ascend NPUs}
\label{sec:cm384_software_stack_cann}

Huawei has developed a comprehensive software ecosystem for Ascend NPUs, known as the compute architecture for neural networks (CANN)~\cite{huawei_cann}. CANN functions as an intermediary software layer, enabling efficient integration between high-level AI frameworks (like PyTorch~\cite{paszke2019pytorch} and TensorFlow~\cite{abadi2016tensorflow}) and the low-level hardware interfaces of Ascend NPUs. By translating abstract computational graphs generated by these frameworks into optimized, hardware-executable instructions, CANN simplifies developer interaction with Ascend hardware, facilitates software-hardware co-design, and aims to maximize application performance on Ascend architectures.

\begin{figure}[t]
    \centering
    \includegraphics[width=0.6\linewidth]{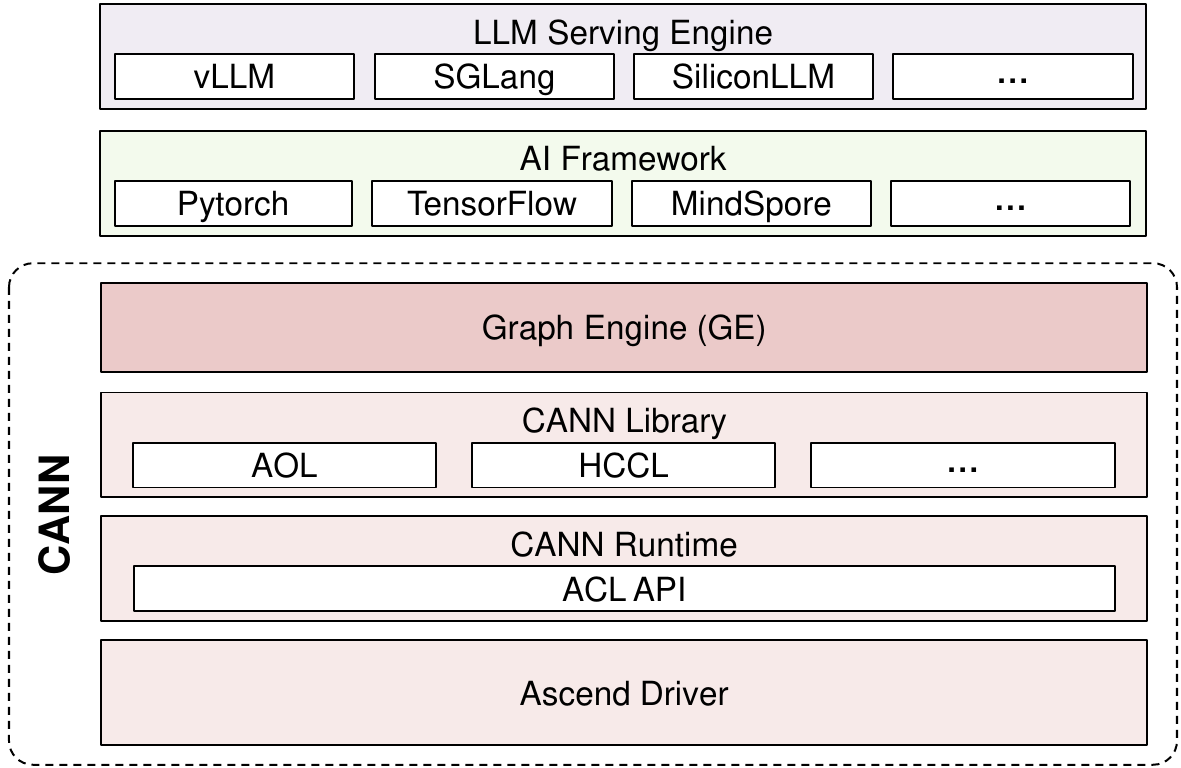}
    \caption{The CANN software stack for Huawei Ascend NPUs.}
    \label{fig:a3-cann}
\end{figure}

\textbf{CANN Architecture.} 
The CANN software stack (Figure~\ref{fig:a3-cann}) is composed of three primary layers: the driver, runtime, and libraries, an architecture analogous to NVIDIA’s CUDA ecosystem~\cite{NVIDIA_CUDAToolkit}.

\textit{1) Driver Layer:} At the foundation, the Ascend NPU driver, comprising kernel modules and firmware, acts as the low-level interface between the operating system and the Ascend NPUs. It manages essential hardware interactions, including device initialization, resource allocation (memory, streams), command scheduling, and inter-NPU communication setup.

\textit{2) Runtime Layer:} The CANN Runtime is the core execution engine for applications on Ascend NPUs. It oversees the application lifecycle, orchestrates model computations, and provides comprehensive device control, memory management, and execution management for models and operators. These functionalities are primarily accessed via the Ascend computing language (ACL) API.

\textit{3) Library Layer:} This layer offers a suite of highly optimized software components to accelerate diverse AI workloads. Key elements include domain-specific acceleration libraries (AOL), the Huawei collective communication library (HCCL) for distributed tasks, an extensive operator package (OPP) with pre-optimized kernels, and engines for neural network acceleration (NNAE) and offline inference (NNRT). Support for custom operator development (e.g., via Ascend C) and integration with third-party libraries to further enhance its capabilities.

Beyond the core layers, the \textit{graph engine (GE)} compiles and optimizes computation graphs from frameworks like PyTorch, TensorFlow, and MindSpore~\cite{mindspore}. It bridges high-level models and low-level execution by applying whole-graph optimizations such as operator fusion, memory planning, dynamic shape handling, and scheduling. These optimizations reduce overhead and improve execution efficiency on Ascend NPUs.

\textbf{Framework Integration.}
CANN offers extensive support for popular AI frameworks, significantly lowering the barrier to entry for adopting Ascend NPUs for existing and new AI projects:

\begin{itemize}
    \item \textit{PyTorch:} Through the PyTorch Ascend adapter (\texttt{torch\_npu})~\cite{Ascend_PyTorch}, developers can seamlessly leverage Ascend NPU acceleration within their existing PyTorch workflows. Huawei provides straightforward installation via pre-built Python wheel packages, comprehensive documentation on API compatibility and best practices, and simplified tools or guidelines for migrating CUDA-based code to CANN.
    \item \textit{TensorFlow:} CANN’s \texttt{TF\_Adapter}~\cite{Ascend_TensorFlow} integrates Ascend acceleration capabilities directly into the TensorFlow framework, enabling high performance and straightforward adoption for TensorFlow-based AI projects with minimal code modification.
    \item \textit{ONNX:} Huawei offers a dedicated CANN execution provider~\cite{ONNXRT_CANN_EP} for the ONNX runtime. This enables efficient execution of models exported in the open neural network exchange (ONNX) format~\cite{onnx}, facilitating broad model compatibility and streamlined deployment across heterogeneous hardware environments that include Ascend NPUs.
    \item \textit{MindSpore:} Developed internally by Huawei, MindSpore provides native and highly optimized integration with Ascend hardware. This framework is designed to deliver potentially superior performance and ease of use within Huawei’s AI ecosystem, offering a tightly coupled software-hardware solution.
\end{itemize}

In summary, CANN delivers a vertically‑integrated software stack including driver, runtime, and libraries comparable to NVIDIA’s CUDA while being tailored to Ascend NPUs. Its GE compiles whole‑graph representations into highly‑optimized execution plans, and rich framework adapters make porting existing workloads almost friction‑free. Together, these components enable developers to harness Ascend hardware with minimal code changes while achieving near‑peak device performance across a broad spectrum of AI applications.

\subsubsection{Infrastructure Software for Cloud Deployment}
\label{sec:infra_software_cloud_deployment}

\begin{figure}[t]
    \centering
    \includegraphics[width=1\linewidth]{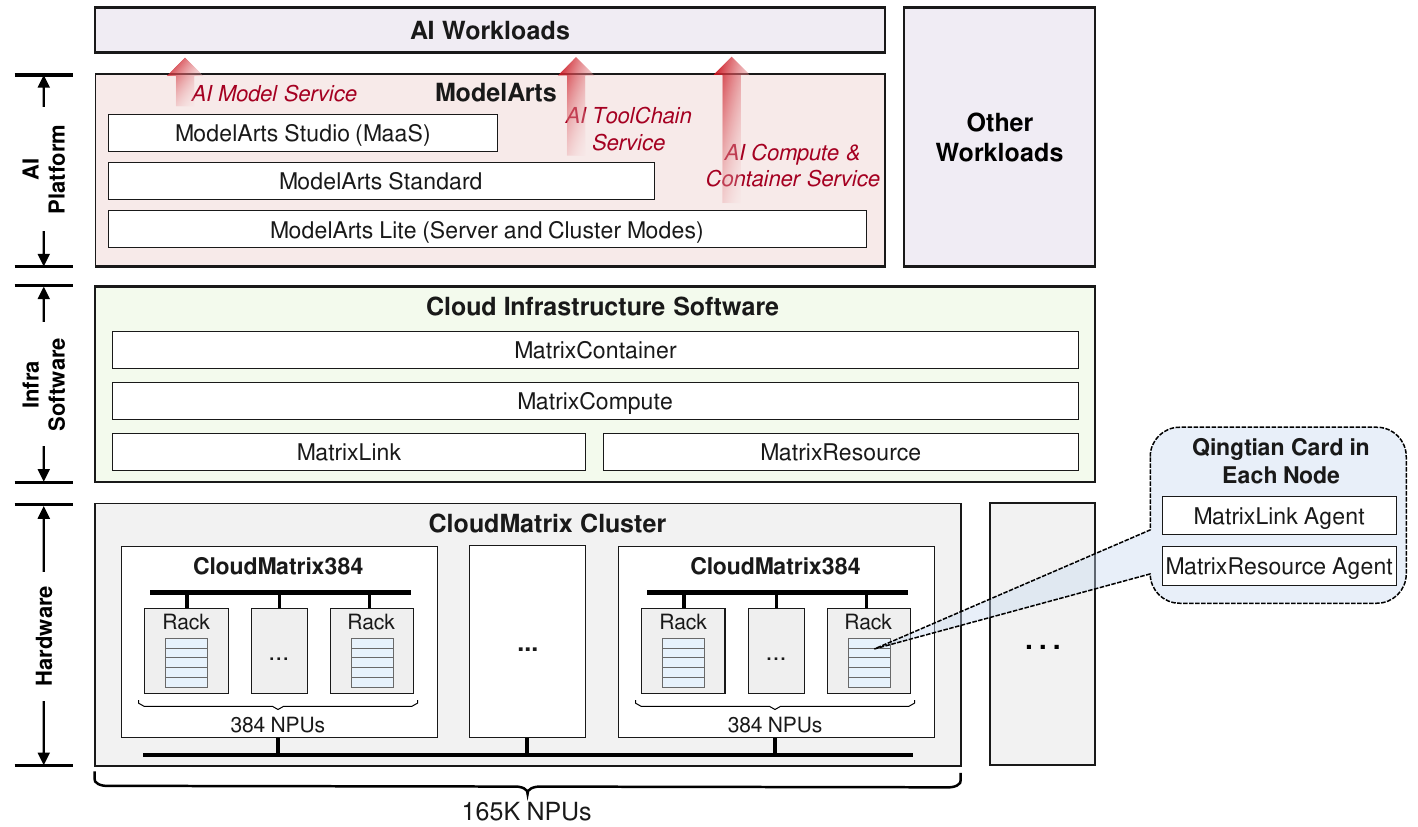}
    \caption{The cloud infrastructure software stack for deploying \CMname{}.}
    \label{fig:cm_infra_sw_stack}
\end{figure}

To enable \CMname{} deployment in cloud environments, Huawei Cloud provides a sophisticated suite of infrastructure software, including MatrixResource, MatrixLink, MatrixCompute, and MatrixContainer, designed to abstract hardware complexity and enable seamless resource orchestration via standard cloud APIs, as illustrated in Figure~\ref{fig:cm_infra_sw_stack}.

\textbf{MatrixResource} manages physical resource provisioning within a supernode, including compute instance allocation based on topology-aware scheduling. The instance provisioning tasks are executed by a MatrixResource agent that runs on the \textit{Qingtian} card in each compute node of the \CMname{}. 

\textbf{MatrixLink} delivers service-oriented networking for the UB and RDMA networks, supporting QoS guarantees and dynamic routing. It manages link-level configurations and enables network-aware workload placement for optimal communication efficiency. These tasks are also executed by a MatrixLink agent on the \textit{Qingtian} card in each compute node.

\textbf{MatrixCompute} coordinates the lifecycle of CloudMatrix instances, from bare-metal provisioning to auto-scaling and fault recovery. It orchestrates resource composition across multiple physical nodes to create tightly-coupled \textit{logical supernode instances}.

\textbf{MatrixContainer} provides container services based on Kubernetes, enhanced with topology-aware scheduling to exploit CloudMatrix’s high-performance interconnect. It enables users to deploy distributed AI workloads using familiar containerized workflows.

\textbf{ModelArts} sits atop the infrastructure stack, offering end-to-end AI platform services~\cite{HuaweiCloud_ModelArts}. It comprises: \textit{ModelArts Lite}, for direct access to Ascend hardware via bare-metal and containerized environments; \textit{ModelArts Standard}, which supports full AI development and MLOps pipelines; \textit{ModelArts Studio}, which delivers Model-as-a-Service (MaaS) capabilities for fast deployment and customization of LLMs and other models.

Together, these components enable users to build and deploy large-scale AI applications efficiently on \CMname{}, abstracting underlying complexity while preserving performance.

\subsection{Suitability Analysis for DeepSeek Models} 
\label{sec:applicability_deepseek}

\subsubsection{\zuorevise{DeepSeek Models and Their Deployment on NVIDIA H800}}
\label{sec:deepseek-model-feature}

DeepSeek-AI has emerged as a significant player in the LLM landscape, particularly with its DeepSeek-V3 and R1 models, which share a common architecture optimized for efficient training and inference~\cite{deepseek2024v3,deepseekai2025deepseekr1}. These models integrate several system-level innovations: a 671B-parameter mixture-of-experts (MoE) architecture that activates only 37B parameters per token using top-8 routing across 256 router experts; multi-head latent attention (MLA) that reduces KV cache size by up to 93.3\%; multi-token prediction (MTP) that enables multi-token generation with decode-time validation; and FP8 quantization to enhance performance while preserving accuracy. Together, DeepSeek's models exemplify a design philosophy centered on training and inference efficiency. These innovations collectively contribute to the models' ability to deliver high-quality outputs with reduced computational and memory requirements.

DeepSeek deploys its V3 and R1 models on clusters of NVIDIA H800 GPUs, each equipped with 80 GB of memory and connected via NVLink within nodes and 400 Gbps InfiniBand across nodes~\cite{deepseek2025inference}. The deployment adopts a disaggregated prefill-decode architecture.  In the prefill phase, DeepSeek organizes four H800 nodes (32 GPUs in total) into a single deployment unit. Within each unit, 256 router experts are strategically distributed across GPUs, with each GPU hosting nine router experts and one shared expert. This configuration, denoted as \textit{DP32+EP32}, employs expert parallelism (EP) across the 32 GPUs, while both the shared expert and the MLA mechanism are replicated via data parallelism (DP) across the same group of GPUs. During the decode phase, DeepSeek expands parallelism further to \textit{DP144+EP144}, grouping 18 nodes for a total of 144 GPUs. Under this larger deployment, each GPU manages two router experts and one shared expert, maintaining a system-wide redundancy of 32 router expert replicas. 

To optimize throughput and latency, DeepSeek employs a dual-microbatch pipeline strategy that overlaps computation and all-to-all communication effectively. Specifically, while one microbatch is involved in MoE-related dispatch and combination, the next microbatch concurrently undergoes local attention or MLP computations.

This carefully orchestrated deployment delivers substantial throughput gains. Each H800 GPU achieves up to 9,213 tokens/s during prefill, aided by a 56.3\% context caching hit rate, resulting in an effective throughput of 4,026 tokens/s when cache hits are excluded. During decoding, each GPU sustains an average throughput of 1,850 tokens/s.

These performance optimization strategies serve as valuable references for the forthcoming deployment of DeepSeek models on Huawei \CMname{}.

\subsubsection{\zuorevise{Architectural Synergy between \CMname{} and DeepSeek Models}}

This subsection uses DeepSeek-R1 as a representative workload to analyze how Huawei \CMname{}’s architectural characteristics align with the demands of large-scale MoE model serving. We focus on four critical dimensions of synergy: MoE communication, memory scalability, cache reuse, and quantization support.

\textbf{MoE Communication Synergy: Efficient Dispatch and Combination.}
DeepSeek-R1 adopts an MoE architecture, which imposes substantial inter-NPU communication demands during token dispatch and expert output combination. \CMname{}’s high-bandwidth, low-latency UB interconnect is particularly well-suited to these requirements. During dispatch, tokens must be routed from routers to selected experts, potentially spanning hundreds of NPUs. The all-to-all UB topology ensures rapid delivery with minimal overhead. Similarly, in the combination phase, multiple experts' outputs must be merged via weighted summation across distributed compute units. The high bandwidth of the UB plane enables efficient collection of expert output, outperforming traditional architectures where network performance can severely hinder MoE inference throughput.

\textbf{Memory Capacity and Management: Accommodating Large Models and KV Caches.}
DeepSeek-R1, with parameter counts approaching 671B, requires vast memory resources for both weights and activations, including attention KV caches. \CMname{} provides a huge amount of total NPU-attached memory, enabling distributed storage of model weights through a combination of tensor, pipeline, and expert parallelism. Beyond model weights, LLMs' attention mechanisms maintain sizable KV caches, especially under long-context or high-batch workloads. \CMname{}’s generous memory footprint supports these scenarios, but efficient partitioning and synchronization of KV caches across NPUs remain essential. 

\textbf{Context Cache Reuse: Accelerating Cache Access.}
LLM workloads, especially in multi-turn dialogue and long-context applications, benefit substantially from prefix cache reuse, with DeepSeek-AI reporting cache hit rates exceeding 56\%. In conventional systems, retrieving historical KV cache from off-chip DRAM or even slower storage layers introduces significant latency, impeding inference performance. \CMname{} mitigates this bottleneck by enabling NPUs to access a disaggregated, CPU-attached DRAM pool directly over the high-bandwidth UB plane (\S\ref{sec:UB-mempool}). This architecture delivers memory-class bandwidth and latency for remote KV cache access. As a result, it minimizes redundant prefill computation, significantly lowers time-to-first-token (TTFT), and scales efficiently to long-context workloads without exhausting limited NPU memory.

\textbf{Quantization for Efficiency: INT8 Support.}
The Ascend 910’s support for INT8 computation (as described in \S\ref{sec:ascend910c_chip}) presents a valuable opportunity for optimizing the inference performance of DeepSeek models. Quantifying model weights and activations from higher precision formats (like FP16 or BF16) to INT8 can significantly decrease the model's memory footprint, reduce computational overhead, and lessen memory bandwidth demands during execution. These benefits can translate into improved throughput and reduced latency.

In summary, \CMname{}’s architecture, including its large-scale NPU compute, extensive memory capacity, high-bandwidth UB interconnect, and DRAM-pool-based caching, is tightly aligned with the needs of large-scale LLM serving. These synergies provide a solid foundation for the optimized inference architecture presented in subsequent sections.

%% file: sections/4-Serving.tex
\section{DeepSeek Serving on Huawei \CMname{}} 
\label{sec:deepseek_on_cm384}

\begin{figure}[t]
\centering
\includegraphics[width=0.85\textwidth]{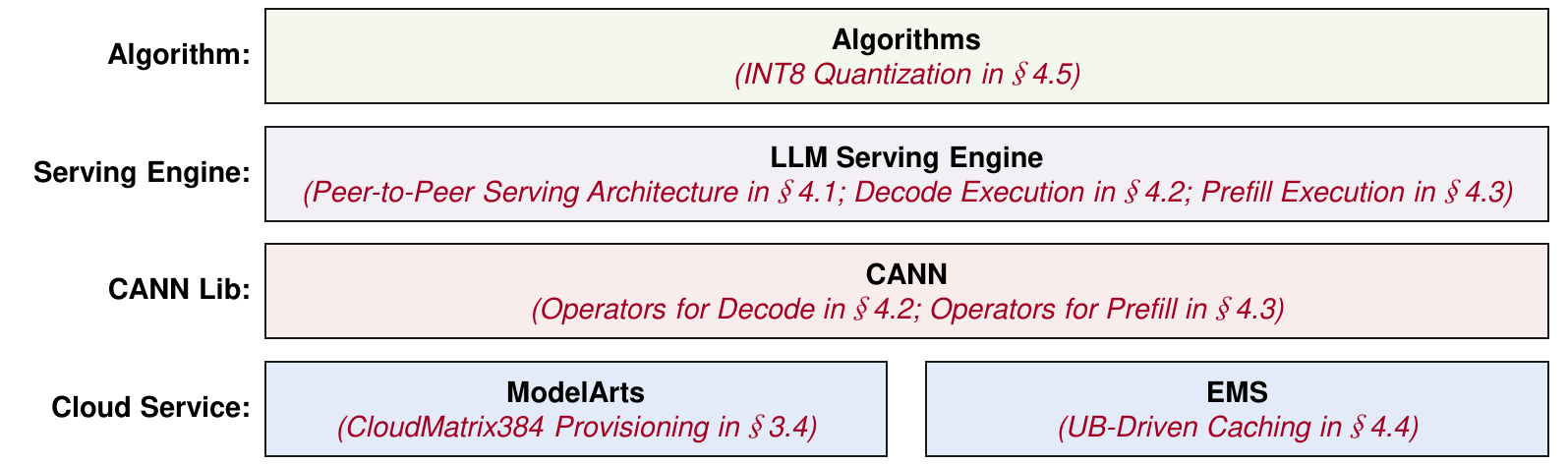}
\caption{An overview of our proposed optimization techniques in different layers of the AI software stack.}
\label{fig:optimization-in-AI-software-stack}
\end{figure}

To fully exploit \CMname{}’s capabilities, we propose \system{}, a comprehensive LLM serving solution that establishes a best practice for deploying large-scale MoE models. We use the DeepSeek-R1 model as a representative example to illustrate our recommended architecture and techniques that exploit cross-layer optimizations for efficient LLM serving on the \CMname{}. Figure~\ref{fig:optimization-in-AI-software-stack} provides an overview of the proposed optimization techniques across multiple layers of the AI software stack. 

In this section, we begin by introducing a novel \textit{peer-to-peer serving architecture} based on prefill-decode-caching (PDC) disaggregation, which decouples prefill, decode, and caching responsibilities and maps them to dedicated NPU and CPU groups connected via high-performance UB interconnects (\S\ref{sec:design-overview}). We then introduce our \textit{tightly-coupled decode optimizations}, which scale large-scale expert parallelism (LEP) across hundreds of NPU dies to accelerate MoE inference (\S\ref{sec:design-decode}). Next, we describe \textit{resource-efficient prefill strategies} that apply hybrid parallelism and pipeline to improve compute efficiency (\S\ref{sec:design-prefill}). We further elaborate on \textit{UB-driven distributed caching mechanisms} that unify memory access across nodes, enabling low-latency access of models and historical KV caches (\S\ref{sec:design-caching}). Finally, we detail the system’s support for \textit{INT8 quantization}, which further boosts end-to-end inference efficiency (\S\ref{sec:design-int8}).

\subsection{Overview: A Peer-to-Peer Serving Architecture with PDC Disaggregation} 
\label{sec:design-overview}

\begin{figure}[t]
\centering
\includegraphics[width=0.9\textwidth]{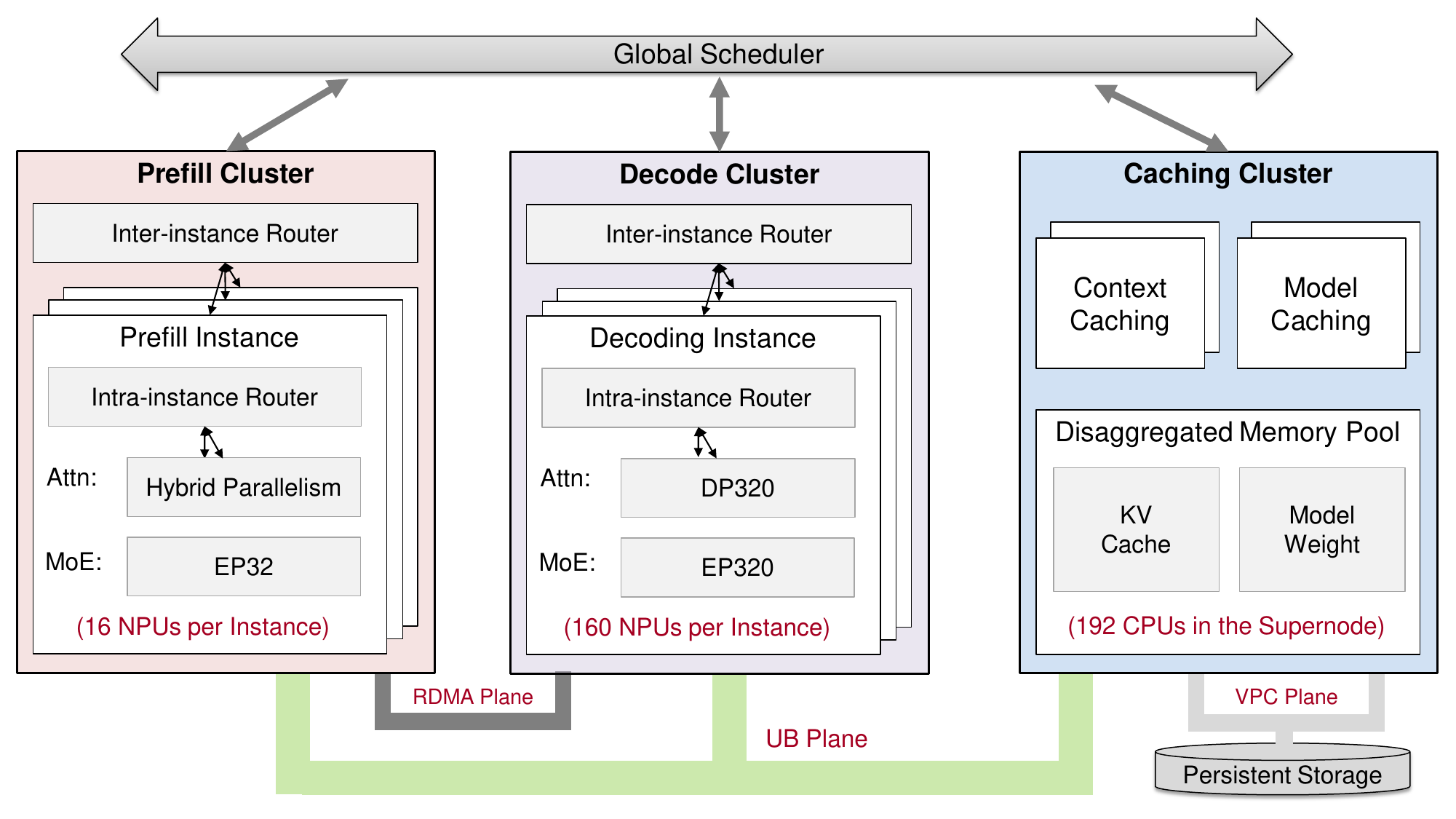}
\caption{Peer-to-peer serving architecture with prefill-decode-caching (PDC) disaggregation on \CMname{}, enabling all NPUs to uniformly access a shared caching cluster backed by a disaggregated memory pool over the ultra-high-bandwidth UB network.}
\label{fig:architecture}
\end{figure}

The architectural design of \system{} is guided by the principles of \textit{disaggregation} and \textit{peer-to-peer communication}, decomposing the LLM inference workflow into independently scalable components while leveraging the high-bandwidth interconnects of \CMname{} for efficient coordination. Building on these principles, we propose a distinctive \textit{peer-to-peer serving architecture} that separates the system into three functional subsystems, i.e., prefill, decode, and caching (PDC), each operating independently and communicating via explicit KV cache transfer interfaces, as shown in Figure~\ref{fig:architecture}. This peer-to-peer design enables each subsystem to scale elastically based on workload demands, maximizing resource utilization and end-to-end performance. These subsystems are interconnected through \CMname{}’s high-bandwidth networking to form a tightly integrated inference 
pipeline:

\begin{itemize}
    \item \textit{Prefill Cluster:} A set of NPUs dedicated to processing the input prompt, consisting of all tokens in the user’s query or context, to generate the first output token and construct the initial KV cache.
    \item \textit{Decode Cluster:} A distinct group of NPUs responsible for autoregressively generating subsequent tokens by consuming and updating the KV cache until an end-of-sequence token is emitted or the output length limit is reached.
    \item \textit{Caching Cluster:} A UB-connected caching layer built on a disaggregated memory pool, providing (i) \textit{context caching} to accelerate prefill through KV cache reuse, and (ii) \textit{model caching} to expedite model block loading and reduce cold-start latency.
\end{itemize}

To better understand the motivation and effectiveness of our proposed design, it is instructive to contrast it with existing \textit{KVCache-centric} architectures~\cite{nvidia_dynamo_2025,qin2025mooncake} that dominate existing LLM serving systems.

\textbf{KVCache-centric vs. Peer-to-Peer Serving Architectures:}
Existing LLM serving systems such as NVIDIA Dynamo~\cite{nvidia_dynamo_2025} and Mooncake~\cite{qin2025mooncake} follow a \textit{KVCache-centric} design, where request scheduling is tightly coupled with KV cache locality. In these systems, requests are typically routed to the specific compute nodes that already hold the corresponding KV cache from previous interactions. This cache-aware scheduling is essential to mitigate the significant performance penalty of remote memory access, as intra-node memory access (e.g., via PCIe at \textasciitilde{}256 GB/s) vastly outpaces inter-node bandwidth (typically at \textasciitilde{}25 GB/s or 200 Gbps). As a result, remote KV cache loading often incurs substantial latency. However, this design introduces non-trivial scheduling complexity and risks degrading load balance, especially under dynamic workloads. Additionally, this design limits global resource efficiency, as DRAM on decode nodes usually remains siloed and underutilized, unable to contribute meaningfully to shared caching capacity.

Our \textit{peer-to-peer serving architecture} in \system{} takes full advantage of the \CMname{}'s ultra-high-bandwidth UB interconnect. This enables uniform access to a distributed caching cluster (Section~\ref{sec:design-caching}) built on a disaggregated memory pool. Crucially, all NPUs, regardless of whether they serve prefill or decode tasks, can directly access this shared disaggregated memory pool, which spans both prefill and decode nodes. This fully peer-to-peer design effectively \textit{flattens the memory hierarchy}, bridging the traditional gap between local and remote access latency.

Decoupling request scheduling from KV cache placement offers several key advantages. First, it enables lightweight, stateless scheduling, allowing inference requests to be dispatched to any available NPU instance without constraints imposed by data locality. This significantly improves system-wide load balancing and NPU utilization. Second, it eliminates the need for complex, affinity-aware scheduling mechanisms, thereby reducing architectural complexity and easing system maintenance. Third, by pooling DRAM resources across prefill and decode nodes, the system forms a unified, elastic caching substrate that enhances memory utilization, increases cache hit rates, and offers greater resilience under skewed or bursty workloads.

\textbf{Prefill and Decode Deployments.} Aligned with prior work~\cite{nvidia_dynamo_2025,qin2025mooncake,zhong2024distserve,patel2024splitwise}, \system{} adopts the strategy of disaggregating the prefill and decode phases across distinct NPU groups. By decoupling these two phases (each characterized by distinct performance bottlenecks), \system{} enables phase-specific hardware allocation, parallelism execution, and independent scalability in response to dynamic workload characteristics.

Each prefill instance is provisioned with 16 Ascend 910 NPUs (32 dies) on \CMname{} and operates with 32-way expert parallelism (EP32). The expert configuration includes 10 experts per rank: one shared expert, eight router experts, and one redundant router expert to support expert parallelism load balancing (EPLB). To further improve efficiency, we employ a hybrid parallelism strategy for MLA computation and apply a microbatch-based pipeline to overlap communication overheads (\S\ref{sec:design-prefill}).

Each decode instance is allocated a significantly larger NPU group, typically 160 Ascend 910 NPUs (320 dies), to meet the high throughput and low latency demands of autoregressive generation. This setup corresponds to 320-way expert parallelism (EP320) for the MoE layers. Each rank hosts one expert, with the overall configuration consisting of 32 shared experts, 256 distinct router experts, and 32 redundant router experts to facilitate EPLB. To further accelerate decoding, we introduce optimized Ascend-native operators, a pipelined decoding strategy, and multiple-token prediction support, as detailed in \S\ref{sec:design-decode}.

\zuorevise{\textbf{Dynamic Adjustment for Asynchronous Real-World Workloads.}
In real-world online serving scenarios, the disaggregated PDC serving architecture enables dynamic, fine-grained adjustment of the numbers of prefill, decode, and caching nodes based on the statistical characteristics of incoming workloads. For example, requests with longer input prompts increase the relative demand for prefill nodes, while workloads generating longer outputs require more decode capacity. These ratios are not fixed but adapt over time to maximize efficiency and maintain latency SLOs.}

\zuorevise{Furthermore, user sessions arrive and depart asynchronously, each with its own start time, prompt length, and generation duration. To cope with this highly dynamic and unpredictable workload pattern, the responsibility of \system{} is to enforce pseudo-synchronous execution through batching and scheduling mechanisms. Specifically, it aligns requests at token boundaries, allowing multiple sessions to be co-scheduled and processed concurrently. This batching strategy amortizes computation, improves throughput, and ensures high resource utilization, even under fully asynchronous request arrival patterns.}

\input{sections/4-2-Decode}

\input{sections/4-3-Prefill}

\input{sections/4-4-Caching}

\input{sections/4-5-INT8}

%% file: sections/4-2-Decode.tex
\subsection{Tightly-Coupled Decode with Large-scale Expert Parallelism}
\label{sec:design-decode}

This section outlines the decode-phase optimizations in \system{} enabled by the tightly-coupled UB plane on the \CMname{}. Minimizing TPOT latency for MoE models requires fine-grained expert parallelism, with each expert placed on a dedicated NPU die. In the DeepSeek-R1 model, 256 router experts are deployed, making large-scale expert parallelism (LEP) a core requirement. However, implementing LEP is non-trivial due to sequential dependencies in token processing and the significant communication overhead incurred when coordinating hundreds of NPU dies.

To address these challenges, we introduce a set of hardware-aware optimization techniques tailored to the \CMname{}. First, we present our fused communication operator design that exploits the UB plane for low-latency, high-throughput MoE execution (\S\ref{sec:decode-comm}). Next, we detail our custom MLA implementation for the Ascend 910 (\S\ref{sec:decode-mla}) and describe a microbatch-based decode pipeline that overlaps two execution streams to hide latency (\S\ref{sec:decode-microbatch}). Finally, we explain how the \system{} supports multiple-token prediction (MTP), a feature leveraged by DeepSeek-R1 to improve decode throughput (\S\ref{sec:decode-mtp}).

\subsubsection{Fused Communication Operators for LEP}
\label{sec:decode-comm}

\begin{figure}[t]
    \centering
    \begin{subfigure}[t]{0.45\linewidth}
        \centering
        \includegraphics[width=\linewidth]{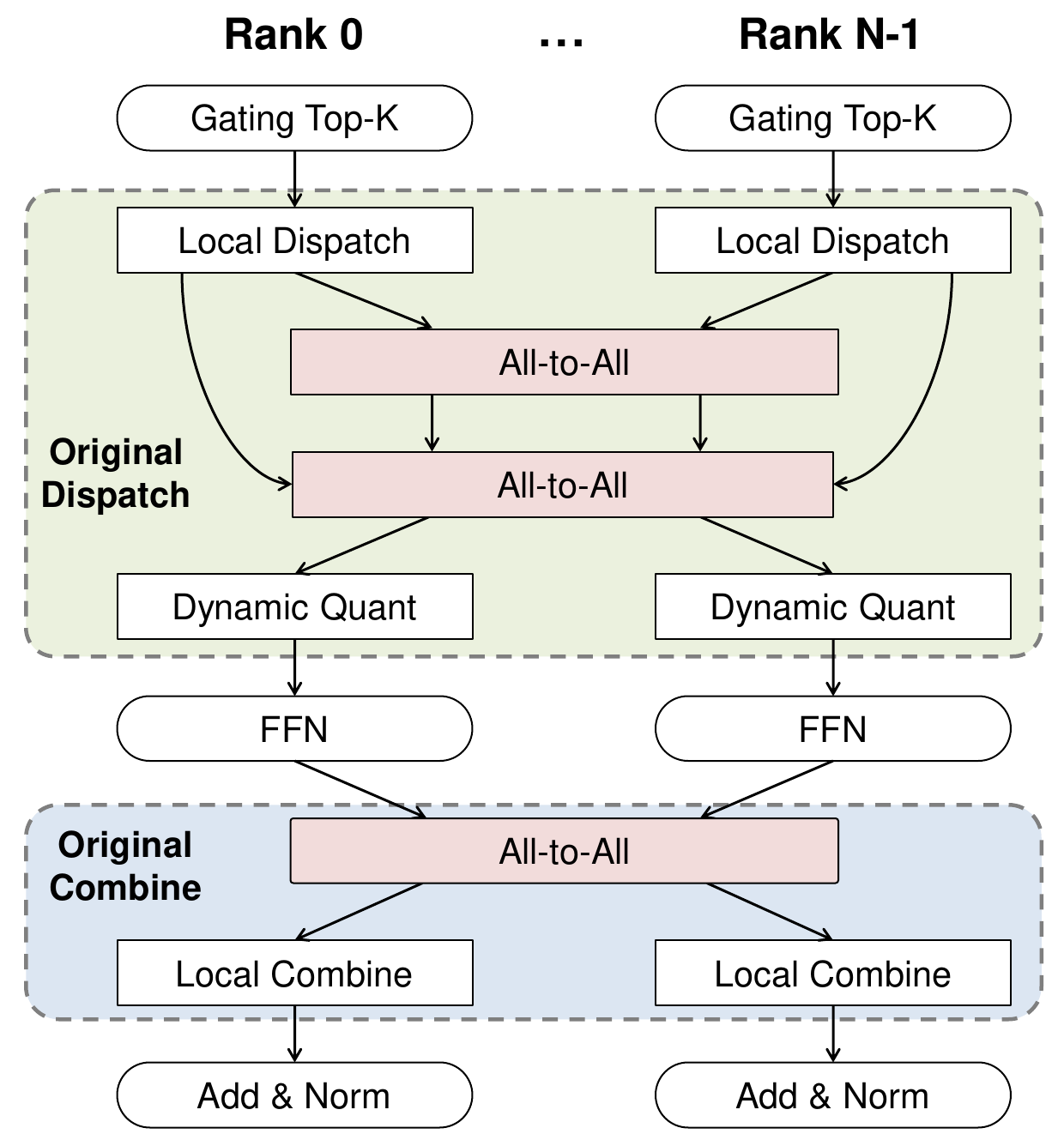}
        \caption{A basic MoE computation flow with all-to-all communications.}
        \label{fig:basic-moe-flow}
    \end{subfigure}
    \hfill
    \begin{subfigure}[t]{0.45\linewidth}
        \centering
        \includegraphics[width=\linewidth]{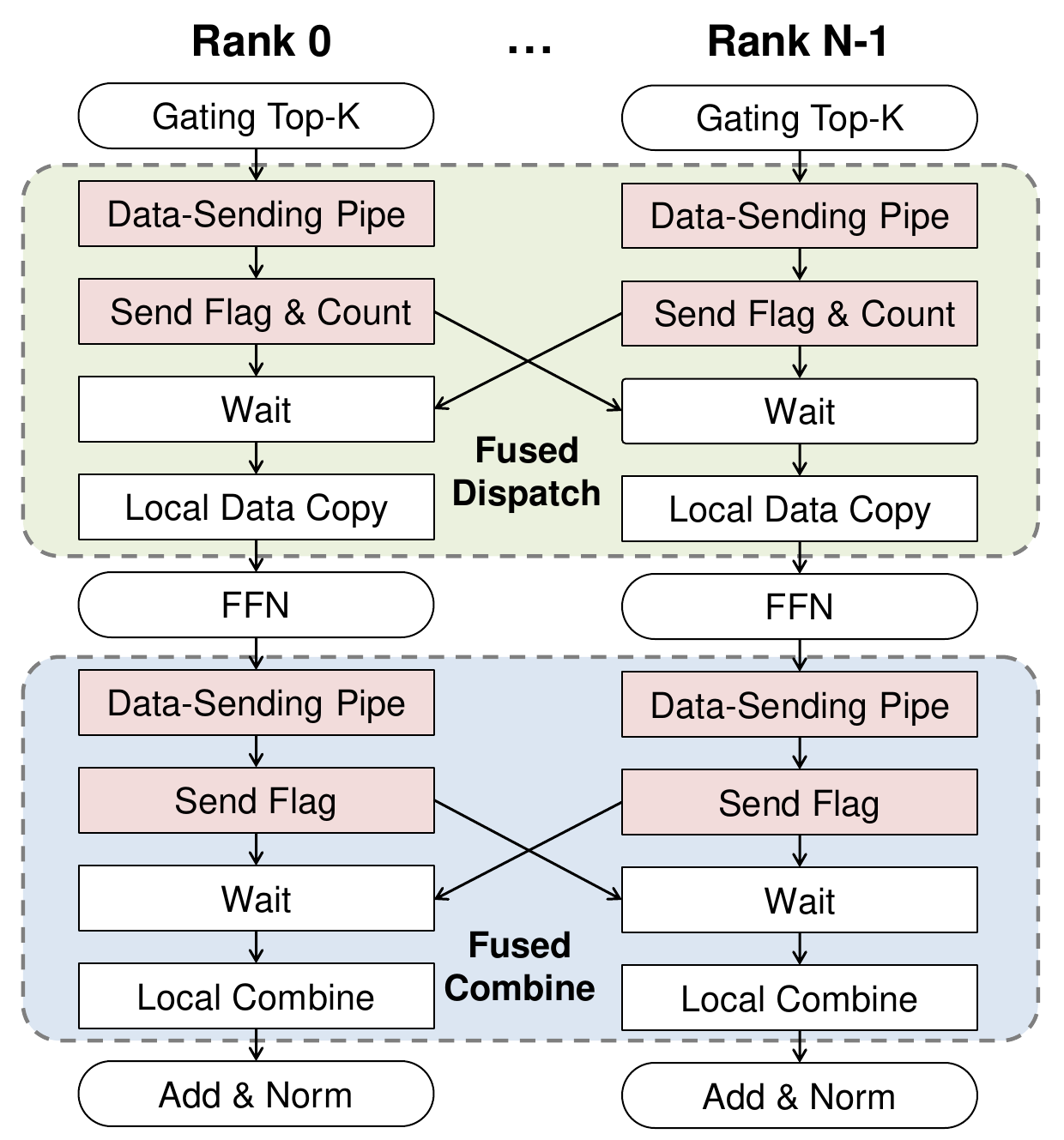}
        \caption{Our proposed MoE computation flow with FusedDsipath and FusedCombine.}
        \label{fig:fused-moe-flow}
    \end{subfigure}
    \caption{Comparison between basic MoE computation flow with all-to-all communications and our proposed MoE computation flow with fused communication operators.}
\end{figure}

Figure~\ref{fig:basic-moe-flow} illustrates a basic MoE computation flow. After the gating mechanism selects the Top-$K$ ($K=8$ in DeepSeek R1) activated experts for each token, two all-to-all communication steps are required before the feed-forward network (FFN) stage. The first all-to-all operation exchanges routing metadata such as token-to-expert assignments across all NPUs. The second all-to-all operation exchanges the actual token data, typically a 7,168-dimensional hidden state vector per token. This data, initially stored in BF16 format, is quantized to INT8 on each NPU to reduce communication and compute costs before being processed by its assigned FFN. After FFN computation, a third all-to-all communication sends the expert outputs back to their source ranks, where each NPU performs the final token combination step to reconstruct the output.
However, this basic MoE implementation suffers from several inefficiencies:

\begin{enumerate}
    \item \textit{Communication Overheads:} The three all-to-all communications introduce significant latency, exacerbated by the large communication domain (hundreds of NPUs).
    \item \textit{Dynamic Shapes:} Data shapes for all-to-all communication are dynamic because the number of tokens assigned to each expert varies per decode iteration. This dynamism reduces execution efficiency due to the need for dynamic memory allocation and frequent CPU-NPU synchronization.
    \item \textit{Sequential Dependencies:} The sequential execution nature of the MoE computation creates dependencies between steps, reducing resource utilization and throughput.
\end{enumerate}

To address these inefficiencies, we developed \texttt{FusedDispatch} and \texttt{FusedCombine}, two fused operators that integrate communication and computation, specifically designed to achieve optimal decode performance on \CMname{}. First, to reduce the overheads of all-to-all communications, the two fused operators replace all all-to-all communications with the send-receive primitive. We further leverage the direct writes among NPUs in the UB plane to reduce the communication latency and move the quantization operation in the dispatch stage before the NPU-to-NPU communication to reduce the message size. Second, to eliminate the overheads related to the dynamic shapes, we pre-allocate all necessary memory space needed for the operators, thus enabling static graph execution. Third, to reduce the overheads of sequential execution, communication and computation steps within the operators are also organized into a pipeline, improving resource utilization and throughput. These optimizations are detailed as follows.

\begin{figure}[t]
    \centering
    \includegraphics[width=0.5\linewidth]{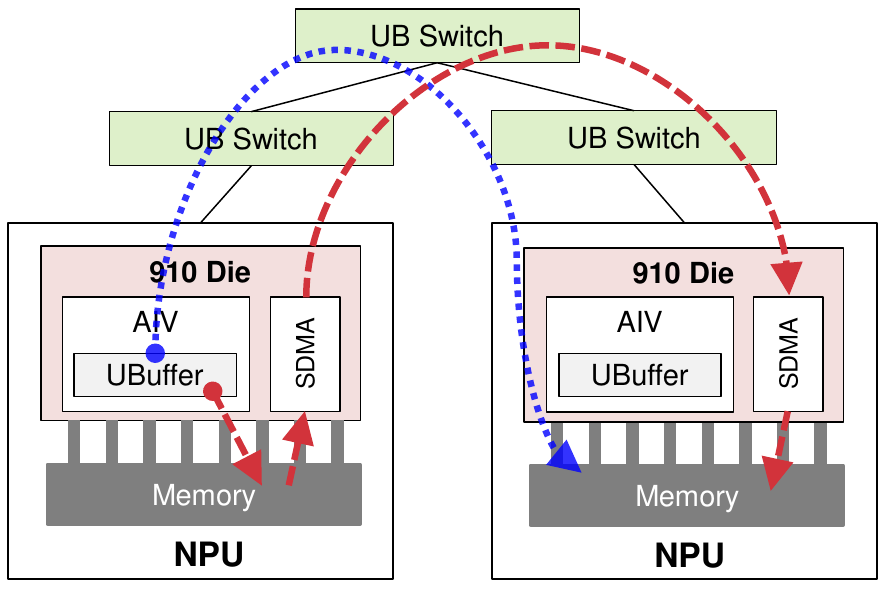}
    \caption{SDMA-based vs. AIV-direct communication across NPUs. The red and blue lines indicate data transmission paths using SDMA and AIV-direct, respectively.}
    \label{fig:fused-operator-direct-write}
\end{figure}

$\circlednum{1}$ \textbf{AIV-Direct Communication across NPUs:}
The conventional all-to-all communication among NPUs typically relies on communication firmware such as a system direct memory access (SDMA) engine to transfer data (red line in Figure~\ref{fig:fused-operator-direct-write}). However, SDMA introduces considerable startup overhead, which becomes a critical performance bottleneck in ultra-low-latency scenarios, particularly during decode. To overcome this bottleneck, we design a new communication mechanism, which we refer to as \textit{AIV-Direct}. AIV-Direct enables AI vector (AIV) cores to directly write data into the memory of remote NPUs via the UB interconnect, completely bypassing the latency-prone SDMA path (blue line in Figure~\ref{fig:fused-operator-direct-write}). By eliminating SDMA’s startup overhead, AIV-Direct provides a fast and lightweight pathway for peer-to-peer communication. This sharply reduces transfer initiation latency and accelerates inter-NPU data exchange, significantly improving performance in latency-sensitive operations such as decode.

$\circlednum{2}$ \textbf{Early Quantization:}
In the original MoE computation flow, as shown in Figure~\ref{fig:basic-moe-flow}, BF16 token data is transmitted during token dispatch, resulting in high communication volume. To mitigate this, we introduce \textit{early quantization} by performing INT8 quantization before sending token data within \texttt{FusedDispatch}. Specifically, instead of sending BF16 data, we transmit INT8-quantized data together with its scaling factor. This reduces the communication payload during the data exchange phase. Given a token data with 7,168 dimensions, the INT8 representation requires 7 KB per token. The scaling factor occupies 4 bytes (INT32), but for alignment, we allocate 512 B. As a result, the transfer message size for each token is 7.5 KB. This optimization substantially reduces communication overhead in the most bandwidth-intensive stage.

$\circlednum{3}$ \textbf{Static Execution via Shared‑Memory Pre‑allocation:}
To avoid dynamic memory allocation and its associated CPU-NPU synchronization overhead, we statically pre‑allocate shared‑memory buffers in each NPU rank for data arriving from every other rank in the MoE layer.  
The required buffer size is:

\begin{equation}
    \mathtt{buffer\_size}= \mathtt{rank\_num} \times \mathtt{max\_tokens} \times \mathtt{msg\_size},
\end{equation}

where 

\begin{equation}
    \mathtt{max\_tokens}= \mathtt{local\_batch}\times\min(\mathtt{topK},\,\mathtt{experts\_per\_die})
\end{equation}
\(\mathtt{max\_tokens}\) is the worst‑case number of tokens an NPU may send to a single peer, and  
\(\mathtt{msg\_size}\) is the per‑token message length (7.5 KB after INT8 quantization for token dispatch and 14 KB for token combine).

With this space pre‑allocated, both \texttt{FusedDispatch} and \texttt{FusedCombine} directly write data into the target NPU memory buffer via AIV-direct communication, avoiding an intermediate local copy and the subsequent remote read, thus reducing memory traffic and synchronization latency.

Because \texttt{FusedDispatch} and \texttt{FusedCombine} execute back‑to‑back, sharing a single buffer would create a race: a faster NPU could launch \texttt{FusedCombine} and overwrite a peer’s buffer before that peer finishes consuming the prior \texttt{FusedDispatch} payload, corrupting data. We eliminate this hazard with \textit{double buffering}: distinct buffers are reserved for \texttt{FusedDispatch} and \texttt{FusedCombine}, ensuring that one buffer is always free for writers while the other is being read.

The pre‑allocation memory overhead is modest. In our experimental setup, each die handles a local batch of at most 96 tokens and hosts up to two experts, yielding
\(\mathtt{max\_tokens}=96\times\min(8,1)=96\).
Across a communication domain of 320 ranks, the dispatch buffer occupies
\(320\times96\times7.5\;\text{KB}\approx225\;\text{MB}\),
and the combine buffer
\(320\times96\times14\;\text{KB}\approx420\;\text{MB}\).
The two buffers together consume only about \(645\;\text{MB}\) memory per die.

\begin{figure}[t]
  \centering
  \begin{subfigure}[b]{0.95\linewidth}
    \centering
    \includegraphics[width=\linewidth]{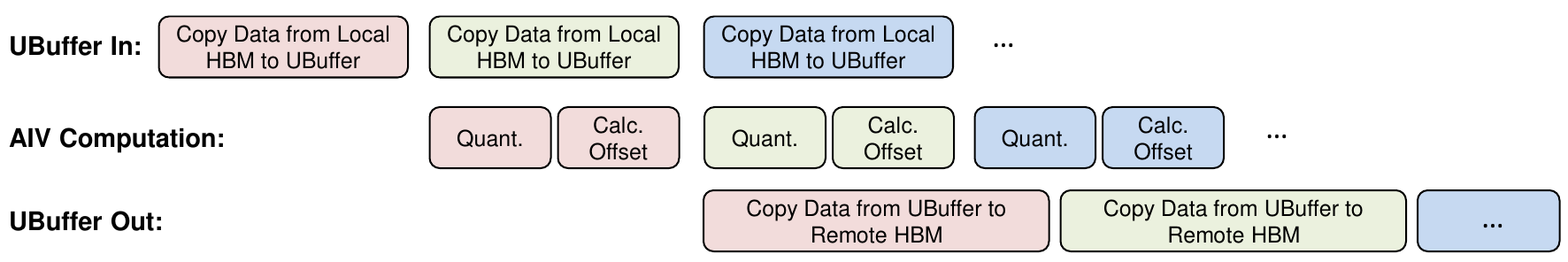}
    \caption{Data‑sending pipeline during dispatch.}
    \label{fig:moe-pipeline-dispatch}
  \end{subfigure}
  \vskip 4pt  
  \begin{subfigure}[b]{0.95\linewidth}
    \centering
    \includegraphics[width=\linewidth]{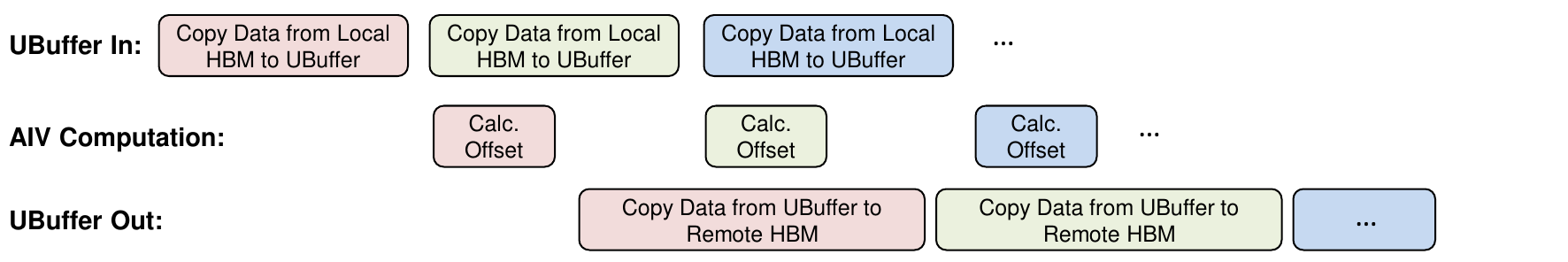}
    \caption{Data‑sending pipeline during combine.}
    \label{fig:moe-pipeline-combline}
  \end{subfigure}
  
  \caption{Data‑sending pipelines for token dispatch, which employs dynamic quantization, and for combine, which transmits unquantized data.}
  \label{fig:moe-pipeline}
\end{figure}

$\circlednum{4}$ \textbf{Data-Sending Pipeline:}
Remote data writes require computing the target offset within a peer NPU’s pre-allocated memory buffer. However, performing this calculation and the transfer sequentially would stall execution. To avoid this, we design a data-sending pipeline inside each fused operator as shown in Figure~\ref{fig:moe-pipeline}, which pipelines the following three stages:
(1) copy the next token into the local UBuffer;  
(2) compute the remote buffer offset and apply INT8 quantization if enabled;  
(3) issue the AIV-Direct write to the peer NPU’s memory.  
Tokens flow through this pipeline as one-token microbatches. While Stage 3 of a microbatch transmits data, Stages 1 and 2 of the following microbatches execute in parallel. This overlap hides both computation and communication latency, enabling continuous and efficient token dispatch.

By combining these techniques, including AIV-direct communication, early quantization, pre-allocated double-buffered memory, and data-sending pipeline, the \texttt{FusedDispatch} and \texttt{FusedCombine} operators significantly reduce the latency of the MoE layer during decode compared to basic implementations.
The workflows of the two fused operators are illustrated in Figure~\ref{fig:fused-moe-flow}.

The \texttt{FusedDispatch} operator proceeds in three main steps. The first step is a pipelined token-sending phase (Opt. $\circlednum{4}$). Each rank iterates over the tokens assigned to remote experts. For each token, the dispatch AIV cores first load the relevant token data from memory into the local UBuffer, then quantize the token data to INT8 format (Opt. $\circlednum{2}$) while appending the associated scale. Routing metadata, including the source rank ID, batch-slot ID, and key offset, is attached to each token data. The system then determines the target rank for each expert ID and writes the data packet into the peer’s pre-allocated shared memory buffer via AIV-direct (Opt. $\circlednum{1}$ and $\circlednum{3}$).
In the second step, once all data packets are issued, a barrier ensures that all token data writes are completed before flags are sent. The dispatch cores compute the token count per expert in parallel, synchronize across cores, and then issue completion flags and token counts to the corresponding peers using AIV-direct (Opt. $\circlednum{1}$ and $\circlednum{3}$).
The final step involves coordination and output assembly. Each rank polls the flags written by remote ranks and waits until all flags are set to ‘1’. It then reads the associated token counts to compute output offsets. Finally, all dispatch cores work in parallel to assemble the received token data, quantization scales, and per-expert token counts from shared memory into contiguous output buffers, ready for the subsequent FFN computation stage.

The \texttt{FusedCombine} workflow similarly consists of three main steps. The first step is a pipelined data-sending phase (Opt. $\circlednum{4}$), in which each combine AIV core loops over its assigned peer ranks. The core reads the corresponding receive count for each peer and copies the associated FFN result data into the local UBuffer. It uses the token’s source metadata—specifically the source rank ID, batch-slot ID, and key offset—to compute the destination address on the peer. The token data is then transmitted back via AIV-direct into the pre-allocated buffer on the originating rank (Opt. $\circlednum{1}$ and $\circlednum{3}$).
In the second step, each token’s metadata is again used to compute the target address for its flag update. The combine AIV core issues an atomic-add operation over AIV-direct to increment the corresponding flag on the peer side, signaling that one contribution has been delivered (Opt. $\circlednum{1}$ and $\circlednum{3}$).
In the final step, each core waits until the flags for its assigned batch are all set to ‘1’, indicating that all expert outputs for that token have been received. The combine core then gathers the expert FFN outputs from shared memory, retrieves the corresponding scale factors from memory, performs element-wise scaling, and sums the results. The combined expert outputs are then added to the shared FFN output to produce the final result for each token.

\subsubsection{MLA Optimization}
\label{sec:decode-mla}

Multi-head latent attention (MLA), introduced by DeepSeek, leverages low-rank compression to reduce the spatial footprint of the KV cache and incorporates weight absorption techniques to lower computational costs. While MLA can be deployed on the \CMname{}, directly migrating DeepSeek’s operators to Ascend 910 NPUs exposes several performance bottlenecks:

\begin{enumerate}[label=(\arabic*)]
    \item \textit{Launch Overhead of Fine-Grained Operators:} 
    MLA introduces numerous fine-grained operations, such as RMSNorm, linear projections, and RoPE encoding, that are typically implemented as separate NPU operators. Each operator invocation incurs non-negligible launch latency, stemming from CPU-side dispatch, parameter loading, instruction scheduling, and tiling configuration. Although capturing these operators into a graph can amortize the CPU dispatch overhead by grouping multiple operations, it does not eliminate the per-operator startup cost on the NPU. As a result, the accumulation of these small kernel launches introduces significant latency in the MLA execution path.
    
    \item \textit{KV Cache Format Conversion Overhead:} To support high-performance matrix computations, the L1 Cache of the Ascend 910 NPU's AI cube cores (AICs) optimally stores data in an NZ format (a specialized hybrid row-major and column-major layout, resulting in a combined N-shaped and Z-shaped traversal path). However, the KV cache is typically stored in the NPU's memory using a standard N-Dimensional (ND) format. Consequently, operator internals often need to explicitly convert KV cache data to the NZ format before AICs can perform matrix calculations. This explicit format conversion consumes memory bandwidth and impacts access efficiency, thereby reducing the effective memory bandwidth available for computation.
    
    \item \textit{Load Imbalance with Multi-Token Prediction (MTP):} When MTP is enabled, the decode phase must validate multiple tokens predicted in the previous step. This results in varying effective sequence lengths for different queries within the same batch (as detailed in \S\ref{sec:decode-mtp}). The original tiling strategies for attention operators, often assuming a BNSD (Batch, Num-heads, Sequence-length, Head-dimension) memory layout, can lead to significant load imbalance. Specifically, without MTP, all queries in a decode step typically have a sequence length of 1, allowing tiling strategies based on $B$ and $N$ axes to create compute tasks of equal size (as $S$ and $D$ are constant per task), thus ensuring load balance. With MTP active, the sequence length $S$ can differ per query. Persisting with $B$-axis and $N$-axis tiling under these conditions leads to substantial load disparities among NPU cores, extending the overall MLA computation time.
\end{enumerate}

\begin{figure}[t]
\centering
\includegraphics[width=0.95\textwidth]{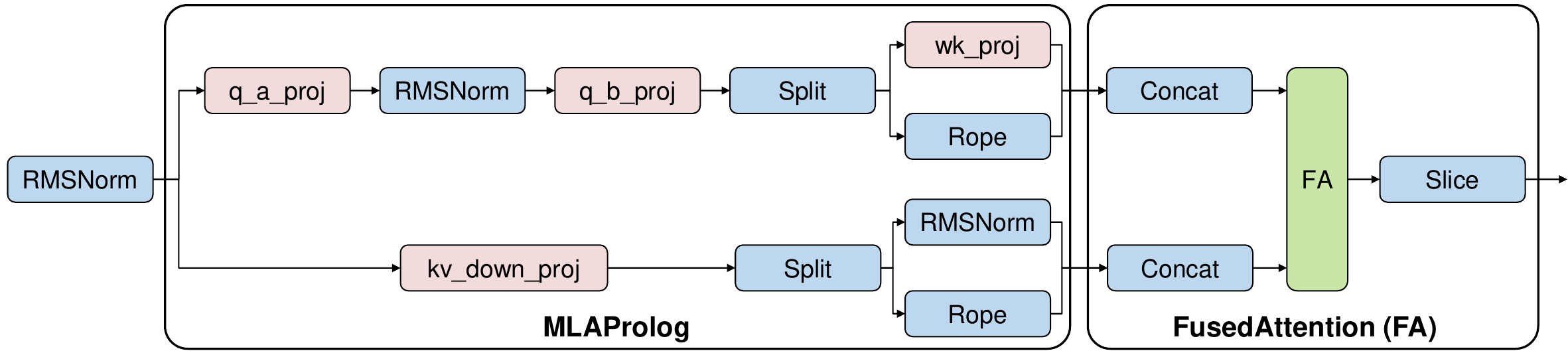}
\caption{The MLAProlog and FA operators, key components of our MLA optimization.} 
\label{fig:design-mla-fused-operator}
\end{figure}

To overcome these limitations and fully exploit the capabilities of Ascend NPUs, we propose the following NPU-friendly optimizations:

\textbf{Fused Operators: MLAProlog and Fused Attention (FA).}
To drastically reduce the launch overhead from numerous small operators in the MLA computation path, we employ aggressive operator fusion, as illustrated conceptually in Figure~\ref{fig:design-mla-fused-operator}.

Firstly, multiple pre-attention operations, including RMSNorm, Q/K/V projections, and RoPE, are consolidated into a single composite operator, termed \texttt{MLAProlog}. This fusion reduces the operator startup costs from those of many individual operators to only one. Furthermore, \texttt{MLAProlog} is designed with internal micro-parallelism, dividing its workload into multiple sub-tasks that are executed in a pipelined fashion across the AIC and AIV units. This fine-grained AIC-AIV parallelism allows the computation times of different sub-tasks on these heterogeneous cores to effectively mask each other, further minimizing the fused operator's execution time.

Secondly, to complement \texttt{MLAProlog}, we developed a fused attention (\texttt{FA}) operator that integrates FlashAttention with adjacent data shaping operations, such as pre-attention \texttt{Concat} (for preparing Q, K, V) and post-attention \texttt{Slice} (for extracting relevant outputs). This further minimizes kernel launches and improves data locality throughout the attention computation path.

\textbf{NZ-Formatted KV Cache.}
To eliminate tensor format conversion overhead, we natively store the KV cache in NZ format within NPU memory. During the MLA computation, the calculated KV tensors are appended to the KV cache directly in this NZ format. In the decode phase, as new KV tensors are generated token by token, they can be efficiently written to NPU memory according to NZ format rules. Ascend NPUs provide data movement interfaces capable of on-the-fly format conversion during memory writes. This \textit{write-with-format-conversion} capability avoids an explicit, separate ND-to-NZ data transformation step for the KV cache, thereby improving effective NPU memory bandwidth utilization.

\textbf{MTP-Aware Tiling with BSND Layout.}
To restore load balance under MTP, we shift from BNSD to BSND memory layout and adopt a dynamic tiling strategy along batch ($B$) and sequence ($S$) axes, which vary across queries. Since the $N$ (number of heads) and $D$ (head dimension) values remain relatively stable during these operations, this ensures better uniformity in task size across AIC cores, reducing tail latency caused by straggling compute tasks.

Together, these three strategies, including operator fusion, native NZ storage, and adaptive tiling, maximize the performance of MLA-based inference on \CMname{}, yielding substantial gains in latency and throughput for DeepSeek models.

\subsubsection{Microbatch-Based Decode Pipeline} 
\label{sec:decode-microbatch}

\begin{figure}[t]
  \centering
  \begin{subfigure}[b]{0.95\linewidth}
    \centering
    \includegraphics[width=\linewidth]{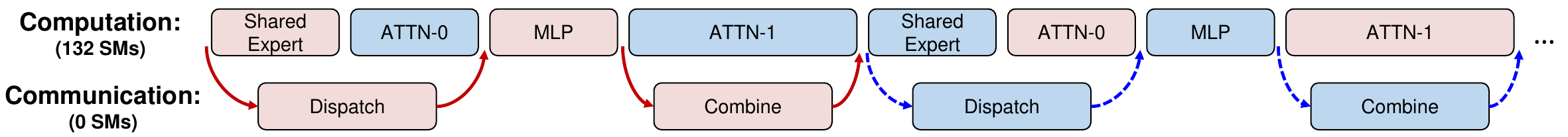} 
    \caption{The DeepSeek's decode pipeline on NVIDIA H800 (\textit{ATTN-0: MLA down/up projection before core attention; ATTN-1: core attention, attention output projection, and MoE routing gate}).}
    \label{fig:design-decode-pipe-base}
  \end{subfigure}
    \vskip 4pt 
  \begin{subfigure}[b]{0.95\linewidth}
    \centering
    \includegraphics[width=\linewidth]{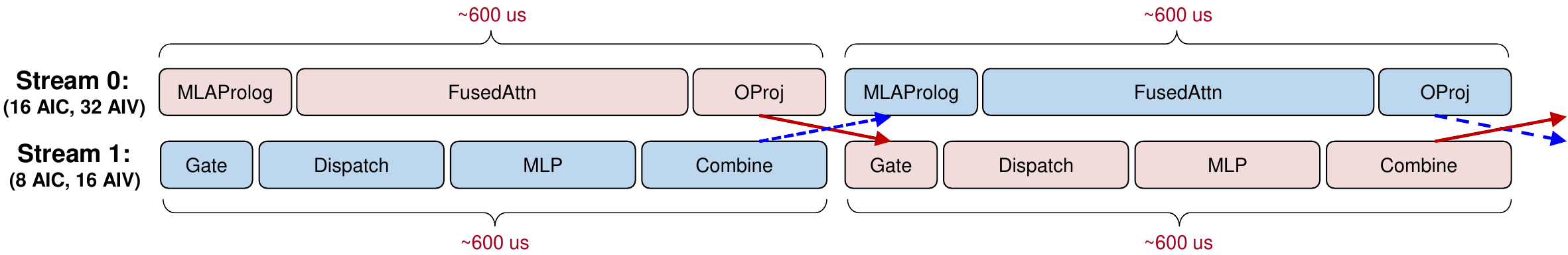} 
    \caption{Our proposed microbatch-based decode pipeline on \CMname{} (\textit{The latency example is for decoding with a 4K sequence length, a batch size of 96 per NPU, and MTP enabled}).}
    \label{fig:design-decode-pipe-ours}
  \end{subfigure}
  \caption{Comparison of decode pipelines: (a) DeepSeek's approach on H800 and (b) our proposed pipeline on \CMname{}. Alternating colors denote two interleaved microbatches.}
  \label{fig:design-decode-pipe} 
\end{figure}

While fused communication operators (\S\ref{sec:decode-comm}) help mitigate some overheads, the latency associated with expert parallelism communication remains a significant factor in the decode phase. To further improve efficiency, inspired by DeepSeek’s microbatch pipelining strategy~\cite{deepep2025}, we design a tailored \textit{microbatch-based decode pipeline} for \CMname{} that maximizes resource utilization and reduces execution latency via fine-grained latency overlap across two streams.

Our proposed resource partitioning and pipelining strategies diverge from DeepSeek's method due to both the unique characteristics of the Ascend NPU and our specific parallelism deployment for MoE models. Unlike DeepSeek's deployment on NVIDIA H800s, which co-locates three experts per GPU (one shared expert and two router experts) as shown in Figure~\ref{fig:design-decode-pipe-base}, our deployment on \CMname{} involves deploying a large expert parallelism degree (EP320) with typically one expert per NPU die for low decode latency. Without the shared expert computation, the compute latency of \texttt{ATTN-0}  alone is insufficient to fully mask the MoE dispatch latency. This necessitates a different load-balanced pipelining strategy. 

To achieve efficient latency overlap under these conditions, we implement a microbatch-based pipeline with \textit{asymmetric AIC and AIV partitioning} for \CMname{}, as illustrated in Figure~\ref{fig:design-decode-pipe-ours}. The pipeline comprises two interleaved execution streams, each responsible for distinct portions of the decode process and provisioned with differing compute capacity:
\begin{itemize}
    \item \textit{Stream 0 (Attention Path):} Executes \texttt{MLAProlog}, \texttt{FusedAttention}, and \texttt{O\_PROJ}. These are compute-heavy or memory-intensive operators and thus assigned more NPU resources—16 AICs and 32 AIVs. Under typical decode conditions (4K sequence, batch size 96, MTP enabled), this stream has a per-microbatch latency of ~600 $\mu$s.

    \item \textit{Stream 1 (MoE Path):} Handles the MoE sequence: \texttt{Gate}, \texttt{Dispatch}, \texttt{MLP}, and \texttt{Combine}. Due to the inclusion of both compute and communication phases, this stream is given 8 AICs and 16 AIVs, half the resources of Stream 0, yet achieves a comparable latency (~600 $\mu$s) owing to lower computational load but higher communication latency.
\end{itemize}

The asymmetric allocation ensures a close per-layer latency when executing Streams 0 and 1, thereby enabling the perfect overlap of two interleaved microbatches. As depicted by alternating colors in Figure~\ref{fig:design-decode-pipe-ours}, Stream 0 processes attention computation for one microbatch while Stream 1 simultaneously performs MoE computation and communication for another.

To accommodate changing runtime conditions, such as variable KV cache lengths, the allocation of compute resources to the two streams can be adjusted adaptively. This elasticity ensures that latency balance is preserved, enabling sustained performance across diverse workloads.

\subsubsection{Multiple-Token Prediction Support} 
\label{sec:decode-mtp}

Multiple-Token Prediction (MTP) is a speculative decoding technique used in DeepSeek-R1, wherein $k$ tokens are predicted during each decode step. These predictions are then validated in subsequent steps. By generating multiple tokens per decode, MTP can significantly improve the throughput. However, enabling MTP in existing inference frameworks often incurs substantial inefficiencies due to tight CPU-NPU synchronization, leading to pipeline interruptions and diminished performance. We refer to this as the \textit{pipeline break problem}.

As shown in Figure~\ref{fig:design-mtp-bubble} (naïve MTP pipeline), MTP typically triggers $k+1$ compute graphs per decode step, $k$ for speculative modules and one for final validation. Each graph dispatch introduces a startup latency of $0.6-0.8$ ms. This overhead, especially under CPU-mediated orchestration, leads to idle bubbles on NPUs, undermining the benefits of MTP. We identify two main sources of these obstacles:

\begin{figure}[t]
  \centering
  \begin{subfigure}[b]{0.85\linewidth}
    \centering
    \includegraphics[width=\linewidth]{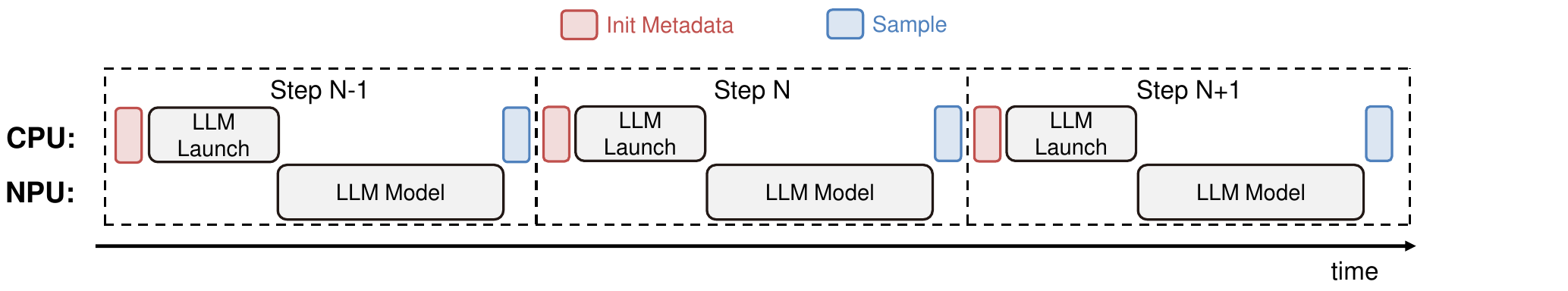}
    \caption{The basic LLM decode workflow without MTP.}
    \label{fig:design-mtp-base}
  \end{subfigure}
  \vskip 4pt  
  \begin{subfigure}[b]{0.85\linewidth}
    \centering
    \includegraphics[width=\linewidth]{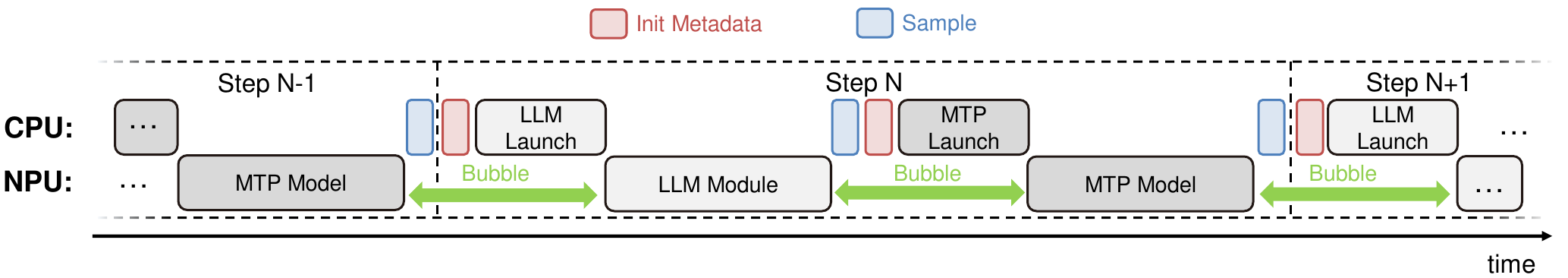}
    \caption{The original LLM decode workflow with MTP.}
    \label{fig:design-mtp-bubble}
  \end{subfigure}
  \vskip 4pt  
  \begin{subfigure}[b]{0.85\linewidth}
    \centering
    \includegraphics[width=\linewidth]{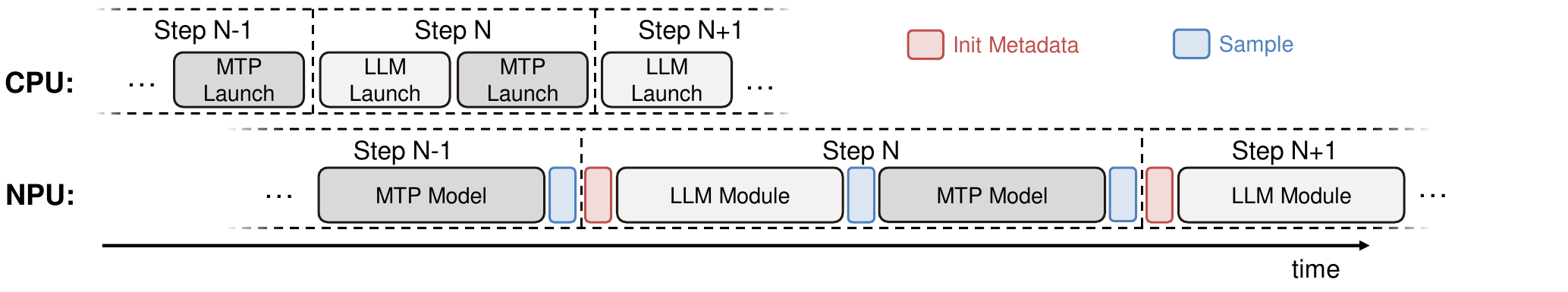}
    \caption{Our proposed LLM decode workflow with pipelined MTP.}
    \label{fig:design-mtp-pipeline}
  \end{subfigure}
  
  \caption{The pipelined MTP optimization on Asend NPUs.}
  \label{fig:design-mtp}
\end{figure}

\begin{itemize}
    \item \textit{CPU Intervention for Dynamic Metadata Initialization:} Both the MTP modules and the main LLM rely on metadata, such as the current sequence length, which changes dynamically during decoding. This metadata can only be finalized after the completion of the preceding module's execution. For example, an MTP module requires the sequence length determined after the previous LLM validation. As shown in Figure~\ref{fig:design-mtp-bubble}, the CPU initializes and transfers this metadata before dispatching each graph, resulting in frequent CPU-NPU synchronization barriers.

    \item \textit{CPU-Intervened Sampling Disrupts NPU Execution:} After MTP modules and the main LLM generate token distributions, sampling is needed to select the actual tokens. This process involves a mix of CPU procedures and discrete NPU operations. These frequent CPU-NPU interactions create overhead from data copying between the host and device. Crucially, because each subsequent computational graph relies on the sampled output from the previous one, this introduces serialization, preventing consecutive NPU execution.
\end{itemize}

To overcome these bottlenecks, we introduce a pipelined MTP execution technique (Figure~\ref{fig:design-mtp-pipeline}) that eliminates these CPU dependencies and enables efficient graph execution:

\textbf{Aggregated Metadata Initialization.} Rather than performing metadata setup separately for each of the $k+1$ graphs, we precompute and batch all metadata tensors at the start of the decode step. These tensors that are stored directly in NPU memory include incremental sequence lengths for each MTP module and a metadata block for the validation graph. This eliminates repeated CPU involvement and enables seamless, metadata-aware execution on the NPU.

\textbf{CPU-Free In-NPU Sampling.}
To eliminate NPU execution stalls frequently caused by CPU-based sampling, we migrate the entire sampling process to the NPU. This strategy involves implementing the necessary sampling operations, such as token probability sorting, cumulative sum calculations, and candidate filtering, as a sequence of NPU operators. Furthermore, to minimize the launch overhead that could arise from dispatching numerous NPU sampling operators, these operators are fused into the MTP and LLM validation graphs. By keeping sampling entirely on-device, we prevent execution stalls between MTP stages and the LLM validation stage, allowing compute graphs to execute back-to-back with no host intervention.

Together, these enhancements eliminate the frequent pipeline breaks caused by CPU-NPU coordination in naïve MTP implementations. As the NPU executes one compute graph, the CPU concurrently schedules the next, enabling sustained parallelism and continuous NPU execution. This achieves a seamless flow of operations on the NPU, maximizing its utilization and fully realizing the potential latency benefits of MTP.

%% file: sections/4-3-Prefill.tex
\subsection{Resource-Efficient Prefill with Hybrid Parallelism and Microbatching} 
\label{sec:design-prefill}

The prefill phase, responsible for processing the input prompt to generate the initial KV cache, significantly impacts time-to-first-token (TTFT) and system throughput. Given its typically compute-intensive nature, achieving high NPU utilization during prefill is paramount. However, this phase often faces challenges such as load imbalances due to heterogeneous input sequence lengths and communication overheads, particularly in complex architectures like MoE models. To address these issues and maximize efficiency on the \CMname{}, we propose three key optimizations in \system{}. First, we introduce a \textit{staged hybrid parallelism} scheme for MLA computation that overcomes the inherent inefficiencies of conventional data parallelism (\S\ref{sec:prefill-mla}). Second, we present a \textit{microbatch-based prefill pipeline} that exploits the heterogeneous compute and communication units of the Ascend 910 NPU to maximize latency overlap and reduce contention (\S\ref{sec:prefill-microbatch}). Finally, we present the transfer optimizations between prefill and decode phases to minimize the interference to decoding (\S\ref{sec:prefill-decode-scheduing}).

\subsubsection{Hybrid Parallelism for MLA Computation}
\label{sec:prefill-mla}

\begin{figure}[t]
  \centering
  \begin{subfigure}[t]{0.42\textwidth}
    \centering
    \includegraphics[width=\linewidth]{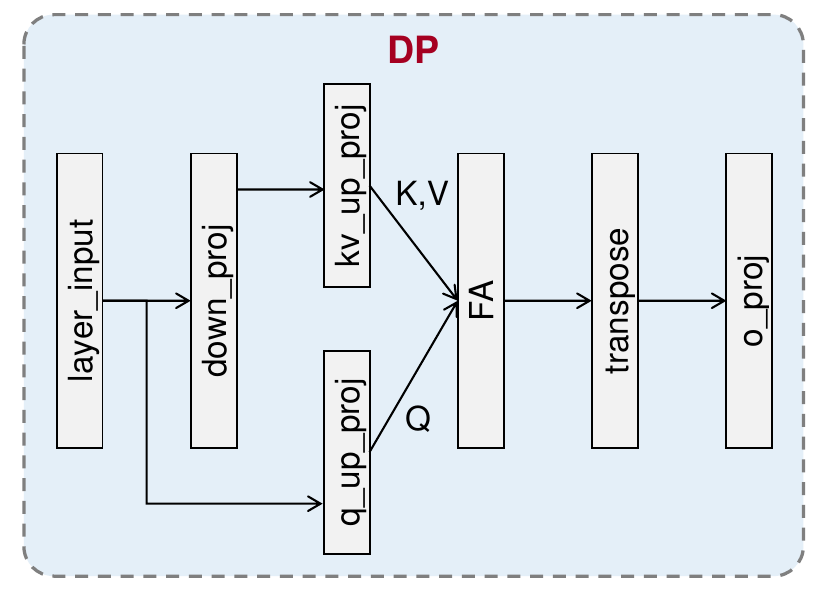}
    \caption{The basic MLA flow with pure DP.}
    \label{fig:prefill-basic-attn}
  \end{subfigure}
  \hfill
  \begin{subfigure}[t]{0.54\textwidth}
    \centering
    \includegraphics[width=\linewidth]{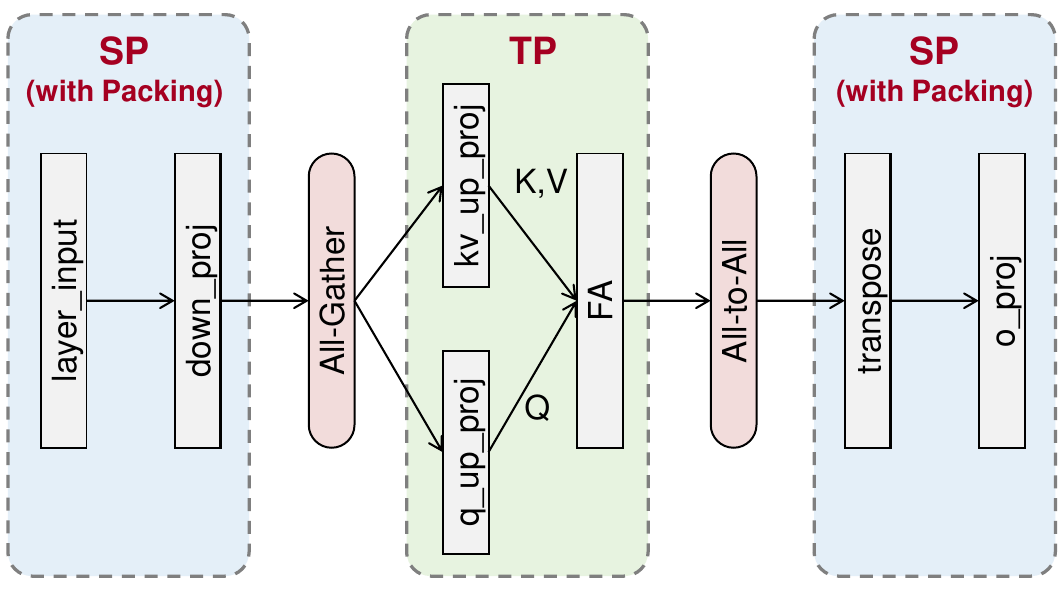}
    \caption{Our proposed MLA flow with hybrid parallelism.}
    \label{fig:prefill-hybrid-parallelism}
  \end{subfigure}
  \caption{Comparison between basic MLA flow using pure DP and our proposed MLA flow leveraging hybrid parallelism during the prefill phase.}
  \label{fig:prefill-attn-comparison}
\end{figure}

Prefill in LLMs presents a significant computational bottleneck. Although \CMname{} offers substantial compute power and high-bandwidth interconnects, we observe that the pure data parallelism (DP) for MLA computation, as originally used in DeepSeek's GPU deployment (\S\ref{sec:deepseek-model-feature}), leads to suboptimal load balancing and resource utilization on Ascend NPUs. This inefficiency stems from two primary reasons:

\begin{enumerate}[label=(\arabic*)]
    \item \textit{Sequence-Length Skew:} In practice, incoming requests often have varying input sequence lengths. With a typical 32-way DP configuration, NPUs assigned shorter sequences complete their work earlier and then idle while waiting for those processing the longest sequence in the batch, leading to wasted compute cycles.
    
    \item \textit{Insufficient Concurrency:} If the number of in-flight requests is less than the DP degree (e.g., fewer than 32 requests for DP32), some DP shards receive no work. Delaying processing to accumulate a full batch of 32 requests increases TTFT, while proceeding with a partial batch underutilizes the NPU resources.
\end{enumerate}

To mitigate these inefficiencies, we introduce a \textit{staged hybrid parallelism} strategy optimized for MLA computation during the prefill phase, visually contrasted with basic DP in Figures~\ref{fig:prefill-basic-attn} and \ref{fig:prefill-hybrid-parallelism}. We decompose MLA into three stages and apply different parallelism schemes to each.

The first stage, which includes processing the layer input and the \texttt{down\_proj} operation, and the third stage, comprising the \texttt{o\_proj} operation, involve computations that are not inherently dependent on token positions within the sequence for their parallelization strategy. For these stages, we leverage \textit{Sequence Parallelism (SP) combined with sequence packing}, replacing pure DP. This method involves concatenating the prompt sequences of multiple requests and then distributing segments of this packed super-sequence across the SP ranks. Consequently, tokens from requests of varying lengths are distributed in an approximately uniform manner among the NPU dies, achieving effective load balancing irrespective of individual request lengths.

The second stage, which includes \texttt{q\_up\_proj}, \texttt{kv\_up\_proj}, and the core FlashAttention mechanism, critically depends on token positions for the attention computation. For this stage, we apply tensor parallelism (TP) to ensure a balanced distribution of the computational load across NPU dies. In our prefill implementation, MLA is typically performed without certain weight matrix absorption to enhance raw computational efficiency, allowing it to be treated effectively as a standard 128-head multi-head attention (MHA) operation. Given that MHA computation is independent for each attention head, we apply TP by distributing these attention heads evenly across the NPU dies.

\begin{figure}[t]
\centering
\includegraphics[width=0.9\textwidth]{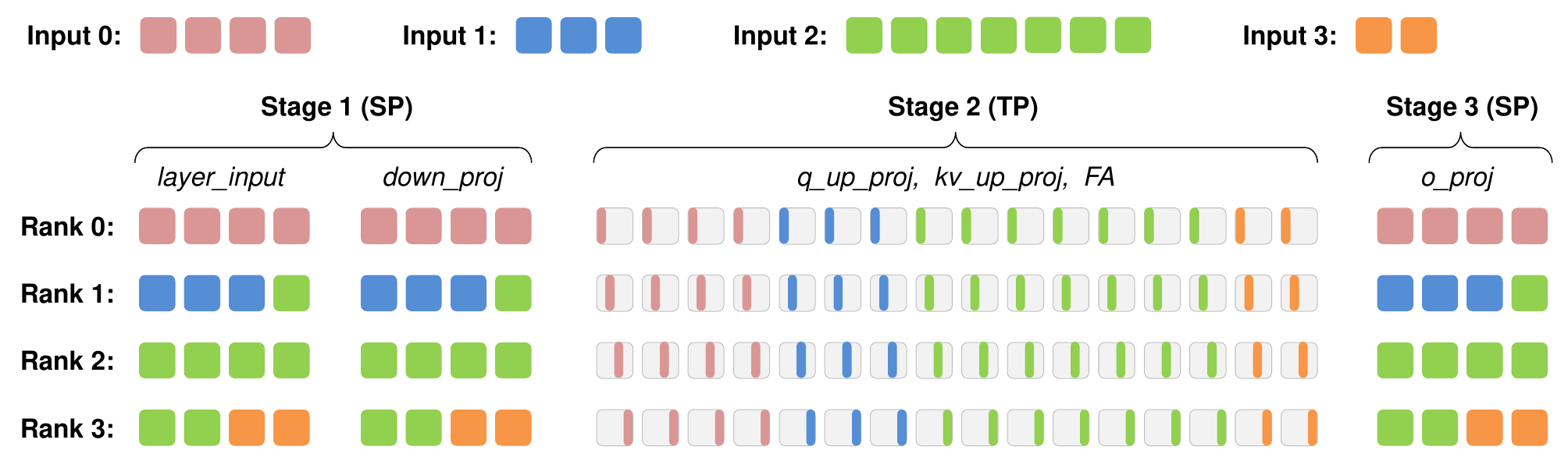}
\caption{Illustrative data flow of the staged hybrid parallelism (SP-TP-SP) for MLA computation in prefill.}
\label{fig:prefill-attn-example}
\end{figure}

Transitioning between these different parallelism strategies across stages necessitates data redistribution. We insert an \texttt{All-Gather} between Stages 1 and 2 and an \texttt{All-to-All} between Stages 2 and 3 to correctly re-shard and distribute the activation data among the ranks.

Figure~\ref{fig:prefill-attn-example} provides an illustrative example of this hybrid parallelism data flow with four inputs of varying lengths (Input 0, Input 1, Input 2, Input 3) processed across four NPU ranks. Initially, in Stage 1, tokens from these inputs are packed and distributed using SP. Each rank processes a contiguous segment of the packed sequence, ensuring that all ranks receive a roughly equal number of tokens, thereby balancing load despite the differing original query lengths. For Stage 2, after an \texttt{All-Gather}, the data is redistributed for TP. Here, each rank processes a shard (e.g., a subset of attention heads) of all tokens from all four inputs. The colored blocks in the figure at this stage indicate how each rank now handles parts of every input. Finally, following an \texttt{All-to-All} operation to gather results from the TP stage, Stage 3 performs its computations with data once again organized according to SP, similar to Stage 1. This example highlights how the hybrid approach maintains load balance throughout the MLA computation.

Compared to a conventional DP strategy (Figure~\ref{fig:prefill-basic-attn}), this hybrid parallelism introduces these two additional collective communication steps. However, their overhead is carefully managed. The \texttt{All-Gather} operation is performed after a dimensionality reduction step (implied by \texttt{down\_proj}), thus operating on potentially smaller tensors. The \texttt{All-to-All} collective primarily redistributes the tensor-parallel shards of the attention mechanism. Since these shards are already reduced in size by the TP degree, the data exchanged per rank during this operation is substantially less than collectives that might handle full, unsharded tensors. On the \CMname{} with its high-bandwidth UB plane, the communication overhead of both operators is relatively small.

\subsubsection{Microbatch-Based Prefill Pipeline}
\label{sec:prefill-microbatch}

\begin{figure}[t]
  \centering
  \begin{subfigure}[b]{0.95\linewidth}
    \centering
    \includegraphics[width=\linewidth]{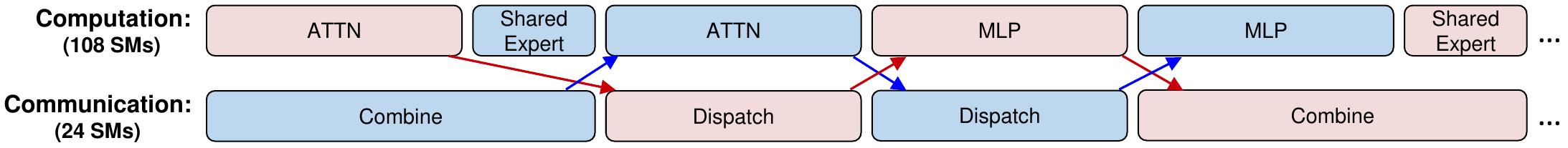} 
    \caption{The DeepSeek's prefill pipeline on NVIDIA H800.}
    \label{fig:design-prefill-pipe-base}
  \end{subfigure}
    \vskip 4pt 
  \begin{subfigure}[b]{0.95\linewidth}
    \centering
    \includegraphics[width=\linewidth]{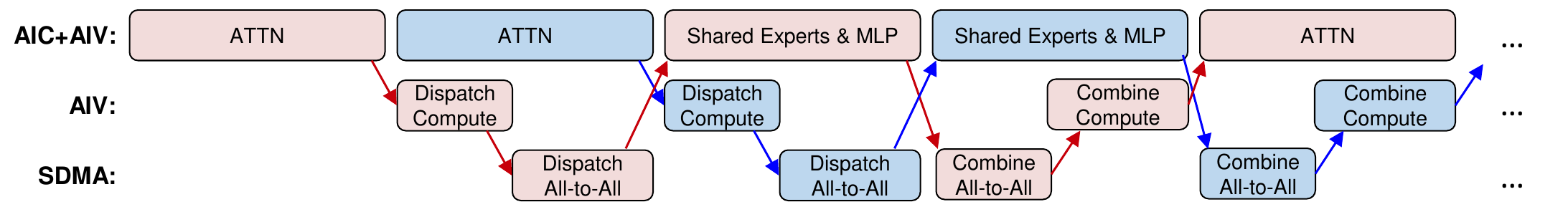} 
    \caption{Our proposed prefill pipeline on \CMname{}.}
    \label{fig:design-prefill-pipe-ours}
  \end{subfigure}
  \caption{Comparison of prefill pipeline strategies: (a) DeepSeek's approach on H800, reserving compute units for communication, versus (b) our proposed pipeline on \CMname{}, leveraging heterogeneous AIC, AIV, and SDMA units for specialized task execution and enhanced computation-communication overlap. In both diagrams, alternating colors are used to distinguish the two interleaved microbatches being processed.}
  \label{fig:design-prefill-pipe} 
\end{figure}

To alleviate the communication overhead introduced by expert parallelism, the original DeepSeek deployment adopts a dual microbatch pipeline. As shown in Figure~\ref{fig:design-prefill-pipe-base}, this approach interleaves computation and communication from two concurrent microbatches on NVIDIA H800 GPUs. By overlapping the computation of one microbatch with the communication overhead (i.e., \texttt{Dispatch} and \texttt{Combine}) of the other, this method improves pipeline efficiency and amortizes latency during the prefill phase.

However, directly porting this strategy to the Ascend 910 NPU on \CMname{} proves inefficient due to architectural mismatches. The pipeline on H800 typically reserves a subset of its streaming multiprocessors (SMs) for communication tasks, enabling concurrency but reducing available compute resources. In contrast, the Ascend 910 offers a heterogeneous compute fabric, which comprises AICs for matrix operations, AIVs for lightweight computation, and SDMA engines for data movement, enabling finer-grained, role-specific task distribution.

To fully exploit this heterogeneity, we introduce an optimized \textit{microbatch-based prefill pipeline} for \CMname{}, illustrated in Figure~\ref{fig:design-prefill-pipe-ours}. Our design orchestrates workload distribution across the AIC, AIV, and SDMA subsystems as follows:

First, we offload low-intensity auxiliary computations to the AIVs, freeing the AICs to focus on compute-intensive operators such as \texttt{ATTN} and \texttt{MLP}. Tasks like token reordering and metadata generation prior to \texttt{Dispatch} (denoted \texttt{DispatchCompute}), and expert output accumulation after \texttt{Combine} (denoted \texttt{CombineCompute}), are assigned to AIVs. These operations are lightweight and vectorizable, making them ideal for AIV execution. As depicted in Figure~\ref{fig:design-prefill-pipe-ours}, AIVs can process \texttt{DispatchCompute} for one microbatch while AICs execute core computations for another microbatch, achieving fine-grained operator-level overlap.

Second, we explicitly route high-volume data transfers, such as All-to-All communication for MoE \texttt{Dispatch} and \texttt{Combine}, to SDMA engines. By isolating these memory operations to a dedicated transfer stream, we prevent contention with AIC and AIV execution. This segregation ensures that compute-heavy operations can proceed uninterrupted, and communication latency is overlapped by concurrently executing AIC/AIV tasks. Given that prefill workloads are dominated by dense matrix operations and communications, this explicit channeling of data flow through SDMA plays a crucial role in preserving peak NPU throughput.

This hardware-aware task assignment, i.e., AIC for primary compute, AIV for auxiliary vector tasks, and SDMA for communications, improves concurrency and minimizes execution stalls. 

Moreover, this design is notably different from our decode-phase pipeline (\S\ref{sec:decode-microbatch}), where communication logic is more tightly coupled with compute streams due to different latency and throughput requirements. In prefill, the need to process longer sequences and larger microbatches makes it more sensitive to compute saturation and bandwidth contention. Thus, separating concerns through dedicated execution units and overlapping tasks at the operator level aligns better with the performance characteristics of \CMname{}.

\subsubsection{Low-interference Transferring between Prefill and Decode}
\label{sec:prefill-decode-scheduing}

In the prefill-decode disaggregated serving architecture, the prefill phase is responsible for generating the first token and producing the corresponding KV cache, which must then be transferred to the decode phase to initiate autoregressive generation. To prevent the performance of latency-sensitive decoding from being disrupted by prefill activities, we introduce three system-level optimizations in \system{}: (1) hardware-level isolation of KV cache transfers via the RDMA plane, (2) asynchronous scheduling to decouple prefill execution from decode scheduling, and (3) model-aware connection grouping to evenly balance prefill-decode communication traffic.

\textbf{RDMA-plane-based KV Cache Transfer.}
Upon completion of the prefill phase, the complete KV cache is transferred to the assigned decode node. To eliminate potential interference with decode-phase communication, this NPU-to-NPU transfer is conducted via the RDMA plane, which is physically and logically decoupled from the UB plane used for bandwidth-intensive decode operations such as token dispatch and expert output combination. Using the dedicated path of the RDMA plane, we isolate the movement of the KV cache from the latency-critical decode traffic. Furthermore, since the KV cache of each request is transferred only once, the RDMA plane offers sufficient bandwidth without becoming a performance bottleneck.

\textbf{Asynchronous Prefill Scheduling.}
To further minimize interference between the two phases, we offload prefill scheduling and KV cache transfer to a dedicated background thread in the decode scheduler. When a new inference request arrives, the inference engine immediately yields control back to the background thread, which asynchronously performs the following steps: (i) allocates a KV cache buffer on the target decode node, (ii) routes the prefill task to a low-load prefill node, and (iii) triggers RDMA-based cache transfer upon completion. This design ensures that decode threads are never blocked by prefill computation or data transfer, thus enabling continuous decode scheduling and improved responsiveness.

\textbf{Load-balanced Prefill-Decode Connection Mapping.}
A common scenario in a PD-disaggregated system is the use of different parallel configurations for the prefill and decode phases. For instance, the decode phase may employ a combination of tensor parallelism (TP) and data parallelism (DP), while the prefill phase typically uses a larger TP degree to accelerate the processing of long input sequences. A key characteristic of the DeepSeek-R1 model, which uses the MLA with a single latent head, is that all ranks within a TP group (\texttt{tp\_rank}) hold an identical, complete copy of the KV Cache.

While this data redundancy provides flexibility, it also introduces a risk of creating network hot spots if not managed correctly. If all ranks of a decode instance are to pull the KV cache from the same source prefill rank, that single network link would become a severe bottleneck. To prevent this, we developed a \textit{deterministic group connection mechanism} that ensures a balanced transfer load. This mapping scheme is calculated as follows:
\begin{itemize}
    \item Let \textit{prefill\_tp\_size} be the TP size of the prefill instance.
    \item Let \textit{decode\_tp\_size} and \textit{decode\_dp\_size} be the TP and DP sizes of the decode instance, respectively.
    \item Let \textit{decode\_tp\_rank\_id} and \textit{decode\_dp\_rank\_id} be the TP and DP rank ids of a specific decode process.
\end{itemize}

First, the grouping parameters are established:
$\textit{ratio} = \frac{\textit{prefill\_tp\_size}}{\textit{decode\_tp\_size}}$
and
$\textit{group\_size} = \frac{\textit{decode\_dp\_size}}{\textit{ratio}}$.

Subsequently, each decode rank determines its source prefill rank using the following mapping:
$\textit{group\_id} = \lfloor \frac{\textit{decode\_dp\_rank\_id}}{\textit{group\_size}} \rfloor $ and 
$ \textit{prefill\_tp\_rank\_id} = (\textit{group\_id} \times \textit{decode\_tp\_size}) + \textit{decode\_tp\_rank\_id} $.
This scheme ensures a balanced connection topology across all prefill-decode links, avoiding communication hotspots and sustaining high throughput.

Together, these three techniques enable a seamless and low-interference handoff from prefill to decode, preserving system efficiency and ensuring high-performance serving of large-scale LLMs under disaggregated architectures.

%% file: sections/4-4-Caching.tex
\subsection{UB-Driven Distributed Caching with Unified Memory Access}
\label{sec:design-caching}

The efficient deployment of LLMs in cloud environments critically depends on high-performance caching strategies. These strategies are essential for accelerating data access and primarily target two key scenarios: historical KV caches to optimize context prefill (Context Caching), and model parameters to facilitate rapid model deployment and switching (Model Caching). Effective implementation of these caching layers significantly reduces redundant computation, curtails model loading latencies, and enhances overall system performance. Supporting such caching functionalities mandates a high-performance, large-capacity, and low-latency intermediate memory tier, strategically positioned to bridge the performance gap between the NPUs' high-speed memory and slower persistent storage services, e.g., object storage services (OBS).

This section details the UB-driven distributed caching for LLM serving on \CMname{}. We first describe the disaggregated memory pooling foundation (\S\ref{sec:UB-mempool}), which leverages the high-bandwidth UB plane to build a disaggregated memory pool with unified memory access. We then introduce two key caching services built atop this pool: Context Caching (\S\ref{sec:kv-cache}) and Model Caching (\S\ref{sec:model-cache}), both delivered via Huawei Cloud’s elastic memory service (EMS)~\cite{ems2025}.

\subsubsection{Disaggregated Memory Pooling}
\label{sec:UB-mempool}

At the heart of EMS caching services is a logically disaggregated memory pool, composed of CPU-attached DRAM aggregated across nodes within a \CMname{}. This pool acts as a unified, high-performance memory substrate for caching historical KV cache and model parameters. A distinguishing characteristic of this memory pool is its deep integration with the UB network plane, enabling efficient, unified memory access to this distributed DRAM and allowing NPUs to rapidly retrieve necessary data regardless of its physical location, facilitating a peer-to-peer serving architecture as presented in \S\ref{sec:design-overview}. The design's efficacy is critically driven by the following UB's hardware capabilities: \textit{1) High-Speed Peer-to-Peer Fabric:} The UB network enables fast inter-node data transfers, allowing any NPU or CPU to access DRAM on other nodes efficiently; \textit{2) DMA over UB:} Zero-copy data transfers are enabled via direct memory access (DMA), bypassing CPU mediation and cutting transfer latencies; \textit{3) Low-Level Memory Primitives:} The UB protocol exposes primitives for remote memory registration and access, allowing the software stack to maintain a global memory view.

\begin{figure}[t]
\centering
\includegraphics[width=0.9\textwidth]{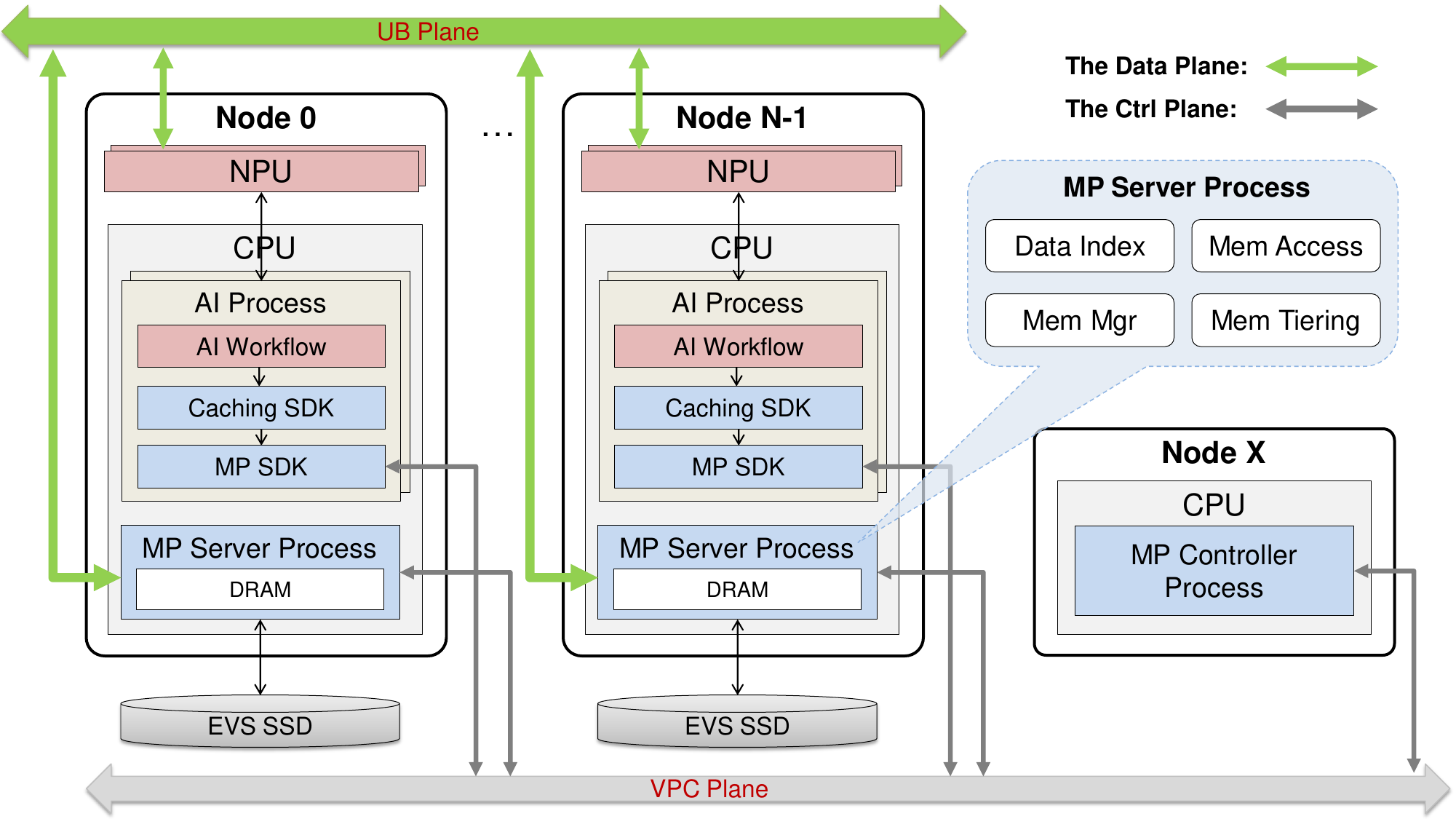}
\caption{The deployment architecture of the UB-driven disaggregated memory pool in EMS.}
\label{fig:mp_architecture}
\end{figure}

As illustrated in Figure~\ref{fig:mp_architecture}, this disaggregated memory pool is managed by a dedicated, three-component software architecture: \textit{1) MP SDK:} Embedded in AI application's processes, it translates upper-layer caching requests into distributed memory operations, exposing key-value store style APIs like \texttt{Put} and \texttt{Get}; \textit{2) MP Controller:} A centralized control plane that maintains metadata (e.g., distributed hash table (DHT) view, namespaces), coordinates operations, and orchestrates resource management; \textit{3) MP Server:} Deployed on DRAM-contributing nodes, it manages local memory, handles tiering and recovery, and participates in load balancing.

The interplay of these software components with the UB plane enables several key operational mechanisms and system features:

\textbf{Distributed Data Indexing and Placement.}
To determine the placement of a key-value pair within the disaggregated memory pool and to efficiently locate it, the memory pool employs a \textit{global consistent hashing} index. This index maps an input key to a responsible MP Server node. A DHT view, whose overall consistency and metadata are managed by the MP Controller, underpins this scheme. Individual MP Servers participate in the DHT by managing their local data portions and responding to routed requests. The MP SDK utilizes this mechanism to distribute keys to specific nodes and DRAM addresses for data access.

\textbf{High-Performance Remote Memory Access.}
A critical function enabled by the UB plane and managed by the software components is direct, high-performance access to remote DRAM by NPUs. This involves a memory mapping and registration process established during the initialization of MP Server instances and MP SDK clients. Control messages are negotiated to exchange physical address ranges of DRAM segments designated for the pool, which are then registered with the UB fabric and the MP Controller. This cross-node mapping capability leverages the \CMname{} supernode's support for global unified memory addressing and routing, allowing UB switches to route NPU SDMA-driven access requests directly to the target MP Server's managed DRAM.

\textbf{Fine-Grained Local Memory Management.}
To effectively manage its allocated DRAM segment and combat fragmentation from variable-sized data objects (such as KV cache blocks or model shards), each MP Server employs a multi-granularity memory allocation system. A key aspect is the use of \textit{huge pages} to reduce the frequency of memory slice allocations and associated management overhead. For data allocation, the system supports variable-length memory partitions, significantly improving memory utilization compared to fixed-size allocators. Furthermore, the MP Server allows dynamic memory flow between different granularities within its managed DRAM, enhancing resource efficiency based on workload-dependent usage patterns.

\textbf{Memory Tiering with Persistence and Recovery.}
To manage storage costs and ensure data persistence, the disaggregated memory pool incorporates an SSD-based tiering layer managed by the MP Server. This layer leverages cloud-provisioned elastic volume service (EVS) SSDs to provide large-capacity, persistent storage. An alternative to EVS-based tiering is using the cloud’s scalable file system service (SFS), which however incurs higher costs. Within this hierarchy, the distributed DRAM pool acts as a fast cache layered above the EVS tier, enabling low-latency access to frequently used data. Persistence is enforced by writing all data to EVS. As EVS volumes have finite capacity, the system employs local eviction policies, e.g., least recently used (LRU), to free space when needed. The MP Server manages DRAM residency independently, using its own LRU eviction logic and capacity thresholds for the DRAM tier. Data evicted from DRAM remains persistently stored in EVS unless it is later removed by EVS’s own space management routines. This tiered structure ensures fault resilience: if in-memory data is lost (e.g., due to node failure), it can be recovered from the EVS tier, assuming it has not been evicted.

Importantly, while the per-node bandwidth to access EVS via the \textit{Qingtian} card is relatively modest, typically under 400 Gbps, the disaggregated memory pool in the \CMname{} aggregates this bandwidth across all 48 nodes, yielding a total EVS access bandwidth of up to $48\times400$ Gbps. Since data is partitioned into fine-grained blocks and distributed across nodes, NPUs can concurrently fetch these blocks from multiple nodes via the high-bandwidth UB plane. This enables high aggregate load bandwidth even when the requested data resides in the EVS tier, effectively amortizing the limitations of per-node EVS access through system-wide parallelism.

\textbf{Namespace Isolation.}
To support multi-tenancy and manage data for different Context Caching and Model Caching instances, the disaggregated memory pool provides KV Namespace isolation. This is primarily orchestrated by the MP Controller, which manages namespace creation, deletion, and metadata. Each MP Server is aware of active namespaces and ensures that data operations are confined to the designated namespace, providing logical data segregation and capacity usage limitation within the shared pool.

In summary, \CMname{}’s UB-driven disaggregated memory pool delivers a high-throughput, scalable memory tier for LLM inference. By combining hardware-level peer-to-peer access with distributed memory management software, the system supports efficient caching for both KV cache and model parameters, forming the backbone of EMS.



\subsubsection{Context Caching}
\label{sec:kv-cache}

The prefill phase of LLM inference, responsible for processing input prompts and generating the initial KV cache, is computationally intensive, particularly for long sequences. Substantial performance gains are possible by reusing historical KV cache from earlier requests. This is especially valuable in scenarios involving recurring prefixes, such as multi-turn conversations, few-shot prompting, and repeated system instructions. Within our architecture, \textit{Context Caching} refers to a dedicated mechanism for storing and efficiently retrieving these historical KV caches. 

Context Caching is implemented by EMS~\cite{ems2025}, a service on Huawei Cloud. EMS leverages the UB-driven disaggregated memory pool (\S\ref{sec:UB-mempool}) to create a shared, distributed repository for historical KV caches. These caches are organized into paged blocks (e.g., 128–512 tokens per block) based on model characteristics and UB transfer efficiency. All NPUs in the serving cluster can access or contribute to this cache via EMS APIs.

\textbf{Indexing, Deduplication, and Retrieval.}
EMS provides a specialized Context Caching SDK (i.e., API layer) to the upper-level LLM serving framework for storing and retrieving historical KV cache blocks. Internally, this EMS SDK utilizes the APIs of the MP SDK (\S\ref{sec:UB-mempool}) to interact with the underlying distributed DRAM and tiered storage.
Each KV cache block is associated with a unique hash key derived from its token sequence and augmented with a prefix hash to enable content-addressable indexing. This allows for fast lookups and deduplication: identical KV blocks are stored once and reused across requests.

The portion of the disaggregated memory pool allocated to Context Caching is subject to capacity constraints. When nearing these limits, the MP Server (\S\ref{sec:UB-mempool}) triggers eviction of colder KV cache blocks from DRAM to the EVS-backed SSD tier. If SSD capacity is also constrained, data is removed entirely based on LRU-style policies. This eviction process ensures fair and efficient resource sharing between context and model caches within the unified pool.

\textbf{Interaction with PDC Disaggregation.}
EMS tightly integrates with the disaggregated prefill and decode pipeline:

\textit{Prefill – Reuse and Store:} Upon receiving a new request, the prefill engine queries EMS with a hash of the input prefix to identify reusable KV cache blocks. If found, these blocks are fetched via the UB plane and loaded directly into NPU memory, bypassing redundant computation. The engine then processes the remaining suffix and generates the corresponding KV cache blocks. These new blocks are asynchronously stored back to EMS, enabling reuse in future requests without stalling ongoing computation.

\textit{Decode – Selective Cache Storage:} KV cache generated during the decode phase can be reused for non-reasoning models, but not for reasoning models like DeepSeek-R1. These reasoning models emit intermediate reasoning tokens followed by final response tokens. Intermediate tokens are typically not re-ingested in subsequent turns, and hence final response tokens shift in position when included in later prompts. Such positional changes disrupt cache validity due to position-sensitive attention. As a result, decode-generated caches are usually excluded from storage. However, if the system adopts approximate KV reuse techniques that tolerate positional shifts, selectively storing final response tokens' cache blocks can offer performance benefits.

\subsubsection{Model Caching} 
\label{sec:model-cache}

Modern LLM serving infrastructures must efficiently support a diverse portfolio of models varying in size, architecture, and task specializations. These infrastructures must also accommodate dynamic model switching in response to fluctuating service demands and continuous model updates. However, loading multi-billion-parameter LLMs from persistent storage, e.g., object storage service (OBS), into NPU memory incurs significant latency. For example, loading a DeepSeek-R1 model with  671B parameters from OBS, assuming a standard 2.5 GB/s access bandwidth per bucket, takes over five minutes. This delay severely limits the practicality of dynamic model switching and impairs service responsiveness, particularly during model updates or A/B testing. Thus, a fast caching mechanism is essential not only to mitigate these overheads but also to ensure responsive, agile model deployment.

To address these challenges, we incorporate Model Caching provided by EMS. At its core, EMS utilizes the UB-driven disaggregated memory pool (\S\ref{sec:UB-mempool}) as a high-performance, distributed caching substrate to support low-latency model access across the system. To integrate with upper-layer serving frameworks, EMS provides a Model Caching SDK that exposes APIs for checking, prefetching, and loading models from the cache. Specifically, the SDK allows users to query whether a model is currently cached in the EMS memory pool, initiate asynchronous prefetching of model blocks from persistent storage into EMS, and trigger model block loading into target NPU memory for inference. When a model is already partially cached, prefetching acts as a hint to promote blocks from slower tiers (e.g., SSD) to faster tiers (e.g., DRAM), further optimizing access latency.

\textbf{Cache Management Policies.}
Internally, EMS decomposes each model into memory blocks and stores them as key-value entries within the disaggregated memory pool. A centralized metadata service tracks the mapping from each model to its corresponding set of blocks, enabling fine-grained, sharded model loading and efficient retrieval during inference.
EMS manages cached model blocks through coordinated policies spanning admission, eviction, and versioning. For admission and prefetching, EMS loads model blocks into DRAM or SSD tiers based on application hints and observed access patterns. Eviction is handled by the native LRU-based policy of the disaggregated memory pool, which, due to the coherent access behavior of model blocks, typically operates at model-level granularity, i.e., entire models or large segments are evicted together, avoiding fragmented state. For versioning, EMS ensures NPUs always execute the correct model version by maintaining version-aware identifiers and associating each model with its corresponding block set. When a new version is deployed, the serving framework requests it explicitly, while stale versions are gradually phased out via natural cache eviction.

\begin{table}[t]
\centering
\footnotesize
\caption{Performance comparison of model loading strategies for loading a 671B INT8 model (approximately 671GB data size) into 8 model instances within a \CMname{} (\textit{The model is originally stored in an OBS bucket with 2.5GB/s bandwidth. We consider two scenarios: 1) Model load: all 8 instances concurrently load the same model using different load strategies for comparing their load latency and DRAM overhead; 2) Model switch: with 8 distinct active models, we compare the model switch latency and cache hit rate when one instance performs a random model switching to one of these 8 models. Latencies are illustrative and representative of defined scenarios.}).}
\label{tab:model_cache_comparison}
\begin{tabular}{@{}llccc@{}}
\toprule
\textbf{Scenario} & \textbf{Metric} & \textbf{No Cache (OBS Load)} & \textbf{Local DRAM Cache} & \textbf{EMS} \\
\midrule
\multirow{3}{*}{\begin{tabular}{@{}c@{}}Model \\ Load\end{tabular}} & Cold Start Latency {\footnotesize (Initial OBS to NPU, s)} & \textasciitilde{}2,560\textsuperscript{1} & \textasciitilde{}2,560\textsuperscript{1} & \textasciitilde{}320 \\
& Warm Start Latency {\footnotesize (DRAM to NPU, s)} & N/A & \textasciitilde{}5 & \textasciitilde{}5 \\
& DRAM Capacity Overhead {\footnotesize ($\times$ Model Size)} & 0 & 8 $\times$ & 1 $\times$ \\
\midrule
\multirow{2}{*}{\begin{tabular}{@{}c@{}}Model \\ Switch\end{tabular}} & Cache Hit Rate {\footnotesize (\%)} & 0 & 12.5\% & 100\%\textsuperscript{2} \\
& Average Latency to Switch {\footnotesize (s)} & \textasciitilde{}320 & \textasciitilde{}281 & \textasciitilde{}5 \\
\bottomrule
\end{tabular}
\vspace{1ex} 
\parbox{\linewidth}{\footnotesize%
\textsuperscript{1} When 8 instances concurrently load the same model from the shared OBS bucket, reflecting significant contention.
\par\vspace{0.25ex}%
\textsuperscript{2} Assumes the capacity of EMS exactly holds all 8 distinct 671B active model versions.
}
\end{table}

\textbf{Benefits of Model Caching with the UB-driven Disaggregated Memory Pool.}
EMS leverages the UB-driven disaggregated memory pool to achieve two key advantages for model caching. First, the high-bandwidth, low-latency UB plane facilitates fast transfer of model blocks from EMS memory tiers (e.g., DRAM or SSD) to NPU memory, substantially reducing model loading latency. Second, EMS uses a unified, cluster-wide memory pool that eliminates data redundancy, allowing a single cached model version to be shared by all NPU instances. This design reduces both the pressure on persistent storage bandwidth and the cumulative DRAM and SSD footprint required for caching, resulting in improved scalability and resource efficiency.

Table~\ref{tab:model_cache_comparison} quantifies these benefits through a performance comparison across different model loading strategies for a 671B-parameter model with INT8 quantization. When no caching is used, all 8 model instances concurrently loading the model from OBS experience a cold start latency of approximately 2,560 seconds each, due to severe contention on the shared 2.5 GB/s OBS bandwidth. Local DRAM caching offers no improvement in this cold start latency, as each node still independently fetches the full model from OBS. In contrast, EMS reduces cold start latency to only \textasciitilde{}320 seconds by enabling shared loading through the memory pool and reusing model blocks across instances.

Beyond latency, EMS also improves memory efficiency. Local DRAM caching results in an 8$\times$ DRAM overhead where each of the 8 instances stores a full model replica. EMS, in comparison, requires only 1$\times$ DRAM footprint to serve all instances, while maintaining an identical warm start latency of \textasciitilde{}5 seconds. In model switching scenarios, EMS achieves a 100\% cache hit rate with an average switch latency of \textasciitilde{}5 seconds, significantly outperforming local DRAM caching, which yields only a 12.5\% hit rate and a latency of \textasciitilde{}281 seconds. These results highlight EMS as a highly effective solution for minimizing both model access latency and memory resource overhead in large-scale inference environments.

%% file: sections/4-5-INT8.tex
\subsection{INT8 Quantization}
\label{sec:design-int8}

To achieve high-throughput, low-latency inference for large-scale MoE models such as DeepSeek-V3/R1 on the Ascend 910 platform, we have designed and implemented a training-free, hierarchical INT8 quantization scheme for model weights and activations. This scheme is engineered to maximize computational efficiency and reduce memory footprint while carefully managing potential accuracy degradation. The core components of our approach are detailed below:

\textbf{Mixed-Precision Strategy.}
Our quantization scheme employs a \textit{mixed-precision strategy} that classifies different operators within the model based on a trade-off between their impact on overall performance (e.g., computational load, memory access) and their sensitivity to numerical precision. The most computationally intensive operations in the critical execution path, such as large matrix multiplications in feed-forward networks (FFNs) and attention mechanisms, are quantized to INT8 to leverage the highest throughput. Conversely, sub-modules or specific operations that are more sensitive to quantization errors but constitute a smaller fraction of the overall memory access or computational burden (e.g., certain normalization layers or critical gating mechanisms) retain higher precision using BF16 or FP32. This flexible partitioning of bit-widths ensures that the entire model can execute efficiently within a unified hardware pipeline, while precision bottlenecks in critical, numerically sensitive pathways are avoided.

\textbf{Adaptive Scale Search.}
Effective quantization requires careful alignment of the dynamic range of floating-point values to the limited range of INT8 integers. For each weight tensor and activation tensor destined for INT8 quantization, we introduce a lightweight, \textit{adaptive scale search process}. This process automatically determines the optimal scaling factor \(s^*\) that minimizes the quantization error, effectively aligning the value distributions before and after quantization. The scale search is formulated as an optimization problem:
\begin{equation}
s^{*} = \arg\min_{s} \mathcal{L}(s), \quad \text{where } \mathcal{L}(s) = ||Q(W \cdot s)(s^{-1} \cdot X) - WX||
\label{eq:scale_search}
\end{equation}
Here, \(W\) represents the weights, \(X\) the activations, and \(Q(\cdot)\) denotes the quantization function. This formulation seeks to find scales \(s\) for weights (and \(s^{-1}\) for activations) such that the output of the quantized operation $Q(W \cdot s)(s^{-1} \cdot X)$ is closest to the original floating-point output $WX$. This entire scale determination process is performed offline during a post-quantization calibration step and therefore incurs no additional runtime overhead during inference. This concept involves transforming $X$ and $W$ with appropriate scales before quantized multiplication.

\textbf{Outlier Suppression and Structural Transformation.}
Certain components within large models, particularly specific expert subnetworks or gating structures in MoE architectures, can exhibit activation or weight distributions with long tails or significant outliers. These outliers can disproportionately affect the quantization range, leading to a loss of precision for the majority of values. To mitigate this, we employ an \textit{outlier suppression technique involving structural transformations}. Prior to quantization, simple linear transformations (conceptually similar to applying learned orthogonal basis rotations or absorbing scaling factors into preceding/succeeding layers) are introduced. These transformations aim to redistribute the extreme values into a more balanced and quantization-friendly range without altering the underlying mathematical function of the layer. By reducing the impact of outliers, this method minimizes the risk of large quantization errors and curtails subsequent error amplification through the network.

\textbf{Efficient INT8 Matrix Multiplication Kernels.}
The performance benefits of INT8 quantization are critically dependent on highly optimized execution kernels. We leverage a \textit{mixed-granularity quantization scheme for matrix multiplications}: activations (\(X\)) are quantized on a per-token basis (dynamic range determined per token), while weights (\(W\)) are quantized on a per-channel basis (typically per-output-channel, static range per channel). This approach balances the need to adapt to rapidly changing activation statistics with the desire to maintain stable weight representations. This mixed granularity, combined with carefully aligned memory layouts for both weights and activations, allows for full utilization of the specialized integer matrix multiplication instructions available on Ascend NPUs. Compared to equivalent BF16 or FP16 implementations, these optimized INT8 kernels can deliver severalfold increases in inference throughput on the same hardware, while ensuring that any accuracy degradation remains within application-acceptable tolerances.

\textbf{Block-Level Clipping and Error Compensation.}
To further refine accuracy and handle local variations within large weight tensors, we implement \textit{block-level clipping and error compensation}. Weights are statistically analyzed and partitioned into smaller blocks. For each block, a distinct, tolerable clipping range is established. This range can be determined by optimizing a scaling factor \(\alpha\) for clipping (e.g., $W_{\text{clip\_max}} = \alpha \cdot \max(W_{\text{block}})$ and $W_{\text{clip\_min}} = \alpha \cdot \min(W_{\text{block}})$) that minimizes the quantization error for that specific block, for instance, by solving:
\begin{equation}
\arg\min_{\alpha} ||\text{Block}(X;W) - \text{Block}(X;Q(W;\alpha))||
\label{eq:block_clipping}
\end{equation}
Here, $Q(W;\alpha)$ represents quantizing the weights $W$ within the block using the clipping factor $\alpha$. Concurrently, lightweight error compensation terms are strategically inserted into the inference computation graph. These terms aim to counteract or partially correct the systematic errors introduced by quantization at different points in the model, thereby mitigating the cumulative impact of quantization noise on the final model output. A significant advantage of this method is that it requires no modifications to the original model training process and does not depend on additional fine-tuning stages, facilitating rapid deployment and iteration.

In concert, these five strategies form a robust and hierarchical INT8 quantization framework that enables high-performance inference for massive models like DeepSeek-V3/R1 on Ascend hardware, carefully balancing computational efficiency with the preservation of model accuracy.

%% file: sections/5-Evaluation.tex
\section{Evaluations} 
\label{sec:evaluation}

In this section, we present a comprehensive performance evaluation for our proposed serving system \system{}, previously detailed in \S\ref{sec:deepseek_on_cm384}, when deployed on the \CMname{}. We begin by outlining the common experimental setup used for our evaluation (\S\ref{sec:evaluation-steup}). Subsequently, we analyze several key aspects of performance and efficacy: this includes the overall system performance (\S\ref{sec:overall_performance}); the inference accuracy achieved with our INT8 quantization scheme (\S\ref{sec:evaluation_accuracy}); an ablation study that investigates the specific contributions of different optimization techniques employed (\S\ref{sec:evaluation-ablation-study}); and finally, a look at the performance of critical underlying operators (\S\ref{sec:operator_performance}).

\subsection{Experimental Setup}
\label{sec:evaluation-steup}

Our evaluation is conducted on a Huawei \CMname{} supernode, provided by the ModelArts Lite (Cluster Mode) service in Huawei Cloud. Specifically, we utilize a configuration comprising 256 Ascend 910 NPUs and their associated host Kunpeng CPUs from a single \CMname{}. The serving system consists of an LLM inference engine optimized by Huawei and SiliconFlow\footnote{\url{https://www.siliconflow.com/}} together, deployed with the requisite Huawei CANN software packages. The elastic memory service (EMS) in Huawei Cloud, providing distributed caching capabilities as detailed in \S\ref{sec:design-caching}, is pre-deployed across the allocated compute nodes. The entire deployment adheres to our proposed peer-to-peer serving architecture with PDC disaggregation (\S\ref{sec:design-overview}), with the following specific configurations for each logical cluster:

\textbf{Decode Cluster:} We deploy a single decode instance utilizing 160 Ascend 910 NPUs (across 20 compute nodes, yielding 320 NPU dies). This instance employs an expert parallelism degree of 320 (EP320) for the sparse MoE layers. For other components like MLA and dense FFN layers, a data parallelism degree of 320 (DP320) is used across the NPU dies. Within these 320 EP ranks, we deploy one expert instance per rank. The expert configuration comprises 32 copies of the shared expert, 256 distinct router experts, and an additional 32 redundant router experts to facilitate expert parallelism load balancing (EPLB).

\textbf{Prefill Cluster:} The prefill cluster consists of 6 prefill instances, each allocated 16 Ascend 910 NPUs (two compute nodes per instance, yielding 32 NPU dies). In total, the prefill cluster uses 96 NPUs. Each prefill instance employs an expert parallelism degree of 32 (EP32) for sparse MoE layers. MLA components within prefill instances utilize a hybrid parallelism strategy detailed in \S\ref{sec:prefill-mla}. 
For expert deployment within each 32-rank prefill instance, we configure 10 experts per rank, consisting of 1 shared expert, 8 router-selected experts, and 1 redundant router expert for effective EPLB.

\textbf{Caching Cluster (EMS):} The distributed caching provided by EMS is realized by leveraging the host CPU DRAM of all physical compute nodes that constitute the prefill and decode clusters. The Kunpeng CPUs and their associated DRAM on these 32 compute nodes (20 for the decode cluster + 12 for the prefill cluster) collectively form the UB-driven disaggregated memory pool. EMS utilizes this pool for both Model Caching (\S\ref{sec:model-cache}) and Context Caching (\S\ref{sec:kv-cache}). Access to this shared memory pool from all NPUs is facilitated by the \CMname{}'s high-speed UB plane.

This experimental configuration serves as the basis for the accuracy and performance evaluation in subsequent sections. The DeepSeek-R1 model evaluated is the 671B parameter version, which has been quantized to INT8 (\S\ref{sec:design-int8}) for execution on the Ascend 910 NPUs.

\subsection{Overall Performance}
\label{sec:overall_performance}

In this section, we evaluate \system{}'s overall performance against leading baselines, measuring both raw throughput and hardware efficiency (tokens/s/TFLOPS), for both prefill and decode phases. We compare these metrics with publicly available performance data for DeepSeek serving on NVIDIA H800 GPUs~\cite{DeepSeekAI_ProfileData_2025} and SGLang on NVIDIA H100 GPUs~\cite{LMSYS_SGLang_Team_2025_DeepSeek_EP}. Our evaluation independently assesses the performance of the prefill and decode phases, mirroring the experimental setups reported in the comparative sources to facilitate a clear analysis.

\textbf{Prefill Throughput.}
We begin by examining prefill throughput, a critical factor for efficiently processing input prompts. Table~\ref{tab:prefill_throughput_comparison} details these per-accelerator comparisons. Effective MoE model serving during prefill also significantly depends on robust EPLB, as highlighted by SGLang's analysis~\cite{LMSYS_SGLang_Team_2025_DeepSeek_EP}. The DeepSeek (Profile) data, with its high throughput (7,839 tokens/s per GPU), may reflect performance under near-ideal expert load balancing. To provide a comparable analytical baseline against such optimized scenarios, Table~\ref{tab:prefill_throughput_comparison} includes \textit{Perfect EPLB} configurations for both SGLang and \system{}. These results represent projected performance under an idealized assumption of perfect load distribution across experts.

\begin{table}[htbp]
  \centering
  \footnotesize
  \caption{Overall prefill throughput (per accelerator) for DeepSeek-R1.}
  \label{tab:prefill_throughput_comparison}
  \renewcommand{\theadfont}{\bfseries}
  \begin{tabular}{llccc}
    \toprule
    \thead{Method} & \thead{Batch\\Size} & \thead{Input\\Length} & \thead{Throughput \\ (tokens/s)} & \thead{Throughput \\ per TFLOPS} \\
    \midrule
    DeepSeek on H800 (Blog) & N/A & N/A & 4,026 & 2.03 \\
    \addlinespace
    SGLang on H100 (Default) & 16,384 & 4,096 & 6,288 & 3.18 \\
    \addlinespace
    \system{} (Default) & 16,384 & 4,096 & 5,655 & 3.76 \\
    \midrule
    DeepSeek on H800 (Profile) & 16,384 & 4,096 & \textbf{7,839} & 3.96 \\
    \addlinespace
    SGLang on H100 (Perfect EPLB) & 16,384 & 4,096 & 7,417 & 3.75 \\
    \addlinespace
    \system{} (Perfect EPLB) & 16,384 & 4,096 & 6,688 & \textbf{4.45} \\
    \bottomrule
  \end{tabular}
\end{table}

In its default configuration, \system{} processes 5,655 tokens/s per NPU, yielding a compute efficiency of 3.76 tokens/s per TFLOPS. This is significantly more efficient than SGLang's default configuration on NVIDIA H100 (3.18 tokens/s per TFLOPS), despite the latter having slightly higher raw throughput.
When tested under an idealized "Perfect EPLB" condition, \system{} achieves 6,688 tokens/s per NPU, translating to an efficiency of 4.45 tokens/s per TFLOPS, surpassing both SGLang's ideal efficiency on H100 (3.75 tokens/s per TFLOPS) and the DeepSeek profile on H800 (3.96 tokens/s per TFLOPS). These comparisons underscore the strong potential of \system{}, while the gap between our default and ideal results highlights the opportunity for further improvement in our load-balancing algorithms.

\textbf{Decode Throughput.}
Next, we analyze performance during the auto-regressive decode phase, as detailed in Table~\ref{tab:decode_throughput_comparison}. We assess absolute decode throughput (tokens/s) targeting a time-per-output-token (TPOT) SLO of below 50 ms, and also evaluate throughput normalized by the accelerator's computer power (tokens/s per TFLOPS) as an indicator of compute efficiency. Notably, both the SGLang (Simulated MTP) and \system{} configurations utilize multi-token prediction (MTP) with an assumed effective acceptance rate of 70\% for a single speculative token.

\begin{table}[htbp]
  \centering
  \footnotesize
  \caption{Overall decode throughput (per accelerator) for DeepSeek-R1.}
  \label{tab:decode_throughput_comparison}
  \renewcommand{\theadfont}{\bfseries}
  \begin{tabular}{lccccc}
    \toprule
    \thead{Method} & \thead{Batch\\Size} & \thead{KV Cache\\Length} & \thead{TPOT\\(ms)} & \thead{Throughput \\ (tokens/s)} & \thead{Throughput \\ per TFLOPS} \\
    \midrule
    DeepSeek (Blog) on H800        & N/A   & 4,989 & \textasciitilde{}50.0 & 1,850 & 0.93 \\
    \addlinespace
    DeepSeek (Profile) on H800     & 128   & 4,096 & \textasciitilde{}50.2 & \textbf{2,325} & 1.17 \\
    \addlinespace
    SGLang (Simu. MTP) on H100 & 128   & 4,000 & \textasciitilde{}55.6 & 2,172 & 1.10 \\
    \addlinespace
    \system{}               & 96    & 4,096 & 49.4 & 1,943 & \textbf{1.29} \\
    \bottomrule
  \end{tabular}
\end{table}

\system{}, configured with a batch size of 96 per NPU and a KV cache length of 4,096 tokens, achieves an excellent TPOT of 49.4 ms. In terms of absolute system throughput, \system{} yields 1,943 tokens/s per NPU with its batch size of 96. This is higher than the DeepSeek (Blog) H800 baseline (1,850 tokens/s per GPU). While numerically lower than DeepSeek (Profile) on H800 (2,325 tokens/s per GPU) and SGLang on H100 (2,172 tokens/s per GPU), these latter systems were benchmarked with a larger batch size of 128. 
The throughput per TFLOPS metric offers further insight into system compute efficiency. \system{} achieves the highest compute efficiency (1.29 tokens/s per TFLOPS), which is higher than SGLang on H100 (1.10 tokens/s per TFLOPS),  DeepSeek (Blog) on H800 (0.93 tokens/s per TFLOPS), and DeepSeek (Profile) on H800 (1.17 tokens/s per TFLOPS). This indicates that our serving solution effectively utilizes the available compute power of the \CMname{} during decoding.

\begin{table}[htbp]
  \centering
  \footnotesize
  \caption{The decode throughput of \system{} under different TPOT SLOs and prompt/output lengths.}
  \begin{tabular}{cccccc}
    \toprule
    \makecell{\textbf{TPOT SLO}\\\textbf{(ms)}} & 
    \makecell{\textbf{Prompt}\\\textbf{Length}} & 
    \makecell{\textbf{Output}\\\textbf{Length}} &
    \makecell{\textbf{Batch}\\\textbf{Size}} & 
    \makecell{\textbf{Achieved TPOT}\\\textbf{(ms)}} & 
    \makecell{\textbf{Throughput per NPU}\\\textbf{(tokens/s)}} \\
    \midrule
    50 & 1,024 & 1,024 & 128 & 46.8 & 2,733 \\
    \addlinespace
    50 & 2,048 & 256 & 112 & 47.4 & 2,360 \\
    \addlinespace
    50 & 4,096 & 256 & 96 & 49.4 & 1,943 \\
    \midrule
    30 & 4,096 & 256 & 24 & 24.6 & 974 \\
    \addlinespace
    15 & 4,096 & 256 & 8  & 14.9 & 538 \\
    \bottomrule
  \end{tabular}
  \label{tab:tpot_slo}
\end{table}

\zuorevise{We also evaluate the decode throughput of \system{} under varying TPOT service-level objectives (SLOs) and different prompt and output lengths, as shown in Table~\ref{tab:tpot_slo}. The results show a clear trend: decode throughput significantly increases with shorter combined prompt and output lengths. For instance, with prompt and output lengths of 1,024 tokens each, the decode throughput reached 2,733 tokens/s per NPU. This dropped to 2,360 tokens/s per NPU when the prompt length increased to 2,048 tokens and the output to 256 tokens. This improvement is attributed to shorter total lengths reducing the KV cache space required per request, which in turn allows for larger batch sizes.
Moreover, as the TPOT SLO becomes more stringent, from 50 ms to 15 ms, \system{} adjusts the batch size accordingly to meet latency targets. Under a relaxed SLO of 50 ms, \system{} supports a batch size of 96 and achieves a throughput of 1,943 tokens/s per NPU while satisfying the latency constraint. As the SLO tightens to 30 ms and 15 ms, the batch sizes reduce to 24 and 8 respectively, resulting in lower throughputs of 974 and 538 tokens/s per NPU. These findings demonstrate \system{}'s ability to meet diverse latency constraints by dynamically scaling batch sizes, all while maintaining high decoding throughput even under stringent real-time demands.}

\subsection{Accuracy} 
\label{sec:evaluation_accuracy}

To comprehensively assess the inference accuracy of DeepSeek-R1 when quantized to INT8 and deployed on \CMname{}, hereafter referred to as DeepSeek-R1 (INT8) for brevity, we conduct extensive tests based on widely used benchmarks. Our evaluation focuses on comparing the accuracy of the INT8 quantization implemented by SilliconFlow (\S\ref{sec:design-int8}) against results from the official DeepSeek-R1 API~\cite{deepseekapi2025} and results published in its technical report~\cite{deepseekai2025deepseekr1}. Given that the original DeepSeek-R1 technical report does not fully disclose all testing parameters for each benchmark, which can lead to variations in direct replication, we adopt a side-by-side evaluation methodology against the live DeepSeek-R1 API to ensure a fair and direct comparison of practical performance.

Our evaluation suite is derived from the extensive list in the DeepSeek-R1 technical report and other widely utilized benchmarks, comprising 16 distinct benchmarks for a multifaceted assessment. Exclusions include AlpacaEval 2.0~\cite{Dubois2024LengthControlledAA} and Arena-Hard~\cite{Li2024Crowdsourced}, due to their reliance on GPT-4 for evaluation (which is outside our current setup), and CodeForces\footnote{https://codeforces.com} because of the lack of readily available automated evaluation scripts. The selected benchmarks cover a broad range of capabilities:
English (MMLU~\cite{hendrycks2020mmlu}, MMLU-Pro~\cite{cui2024mmlupro}, DROP~\cite{dua2019drop}, IFEval~\cite{zhou2023instruction}, GPQA Diamond~\cite{rein2023gpqa}, SimpleQA~\cite{OpenAISimpleQA}),
Code Generation (LiveCodeBench~\cite{liu2024livecodebench}, HumanEval~\cite{chen2021humaneval}),
Mathematics (AIME 2024~\cite{aime2024problems}, MATH-500~\cite{hendrycks2021measuringmath}, CNMO 2024~\cite{cnmo2024problems}, MGSM~\cite{shi2022mgsm}),
and Chinese (CLUEWSC~\cite{xu2020clue}, C-Eval~\cite{huang2023ceval}, C-SimpleQA~\cite{he2024chinesesimpleqa}, C-MMLU~\cite{li2023cmmlu}).
We believe this curated set provides a robust basis for evaluating accuracy.

\begin{table}[t]
    \centering
    \caption{Accuracy comparison of DeepSeek-R1 with INT8 quantization on Ascend 910, the official DeepSeek-R1 API~\cite{deepseekapi2025}, and results reported in the DeepSeek-R1 technical report~\cite{deepseekai2025deepseekr1} across multiple benchmarks (\textit{Results from benchmarks with testing configurations deemed inconsistent have been excluded.}).}
    \footnotesize
    \begin{tabular}{l l c c c}
    \toprule
    \textbf{Category} & \textbf{Benchmark {\tiny (Metric)}} & \textbf{DeepSeek-R1 (INT8)} & \textbf{DeepSeek-R1 API} & \textbf{DeepSeek-R1 Report} \\
    \midrule
    \multirow{6}{*}{English} 
    & MMLU {\tiny (Pass@1)} & 90.82 & 91.05 & 90.8 \\
    & MMLU-Pro {\tiny (EM)} & 83.91 & 83.82 & 84.0 \\
    & DROP {\tiny (3-shot F1)} & 90.42 & 91.02 & 92.2 \\
    & IF-Eval {\tiny (Prompt Strict)} & 83.55 & 83.92 & 83.3 \\
    & GPQA Diamond {\tiny (Pass@1)} & 71.66 & 71.77 & 71.5 \\
    & SimpleQA {\tiny (Correct)} & 30.60 & 30.69 & -- \\
    \midrule
    \multirow{2}{*}{Code}
    & LiveCodeBench {\tiny (Pass@1-COT)} & 63.80 & 63.44 & 65.9 \\
    & HumanEval {\tiny (Pass@1-COT)} & 91.83 & 91.85 & -- \\
    \midrule
    \multirow{4}{*}{Math}
    & AIME 2024 {\tiny (Pass@1)} & 78.96 & 78.12 & 79.8 \\
    & MATH-500 {\tiny (Pass@1)} & 94.46 & 94.62 & -- \\
    & CNMO 2024 {\tiny (Pass@1)} & 77.95 & 76.70 & 78.8 \\
    & MGSM                      & 92.40 & 92.65 & -- \\
    \midrule
    \multirow{4}{*}{Chinese}
    & CLUEWSC {\tiny (Test)} & 94.67 & 94.98 & -- \\
    & C-Eval {\tiny (EM)} & 82.05 & 79.92 & -- \\
    & C-SimpleQA {\tiny (Correct)} & 74.70 & 75.43 & -- \\
    & C-MMLU                   & 90.76 & 90.84 & -- \\
    \bottomrule
    \end{tabular}
    \label{tab:deepseek_comparison}
\end{table}

For consistency in evaluation, prompts for benchmarks such as MMLU, DROP, MGSM, GPQA Diamond, HumanEval, MATH-500, SimpleQA, and C-SimpleQA are sourced from the simple-evals framework. Others, including CMMLU, C-Eval, LiveCodeBench, IFEval, and CLUEWSC, utilize the OpenCompass framework\footnote{https://github.com/open-compass/opencompass}. Adhering to the methodology in the DeepSeek-R1 technical report, MMLU-Pro, C-Eval, and CLUEWSC are tested in a zero-shot setting, while other test sets follow their original protocols. Mathematics competition benchmarks (AIME 2024 and CNMO 2024) undergo 32 repeated test runs each to accurately estimate their pass@1 metrics. For \textit{MATH-500}, \textit{SimpleQA}, and \textit{C-SimpleQA} benchmarks where official evaluations reportedly utilize various GPT-4 versions, we employ Qwen2.5-72B-Instruct\footnote{https://huggingface.co/Qwen/Qwen2.5-72B-Instruct} as the grading model for assessing the outputs of both DeepSeek-R1 (INT8) and the DeepSeek-R1 API. While this choice ensures internal consistency for our study, it may contribute to variations when comparing our scores to those in the DeepSeek-R1 technical report, which relies on GPT-4-based grading. Key generation parameters include a temperature of 0.6 and top-p of 0.95, aligning with settings specified in the DeepSeek-R1 technical report~\cite{deepseekai2025deepseekr1}.

The comparative accuracy results are presented in Table~\ref{tab:deepseek_comparison}. Overall, our DeepSeek-R1 (INT8) implementation on Ascend 910 demonstrates performance largely comparable to both the official DeepSeek-R1 API and the metrics reported in the original technical paper. This indicates that the INT8 quantization applied for deployment on Ascend 910 effectively preserves the model's capabilities across a diverse range of tasks.

\subsection{Ablation Study}
\label{sec:evaluation-ablation-study}

To understand the individual contributions and effectiveness of key optimization techniques employed in \system{}, we conduct a series of ablation studies. These studies isolate the impact of our microbatch-based pipeline strategies for both prefill and decode phases, the Multi-Token Prediction (MTP) mechanism, and the EMS-based Context Caching.

\subsubsection{Microbatch-based Pipeline}
\label{sec:ablation_microbatch_pipeline}

This ablation study quantifies the performance impact of the microbatch-based pipeline strategies by comparing system performance with and without these microbatch optimizations.

\begin{figure}[t]
  \centering
  \begin{subfigure}[t]{0.37\linewidth}
    \centering
    \includegraphics[width=\linewidth]{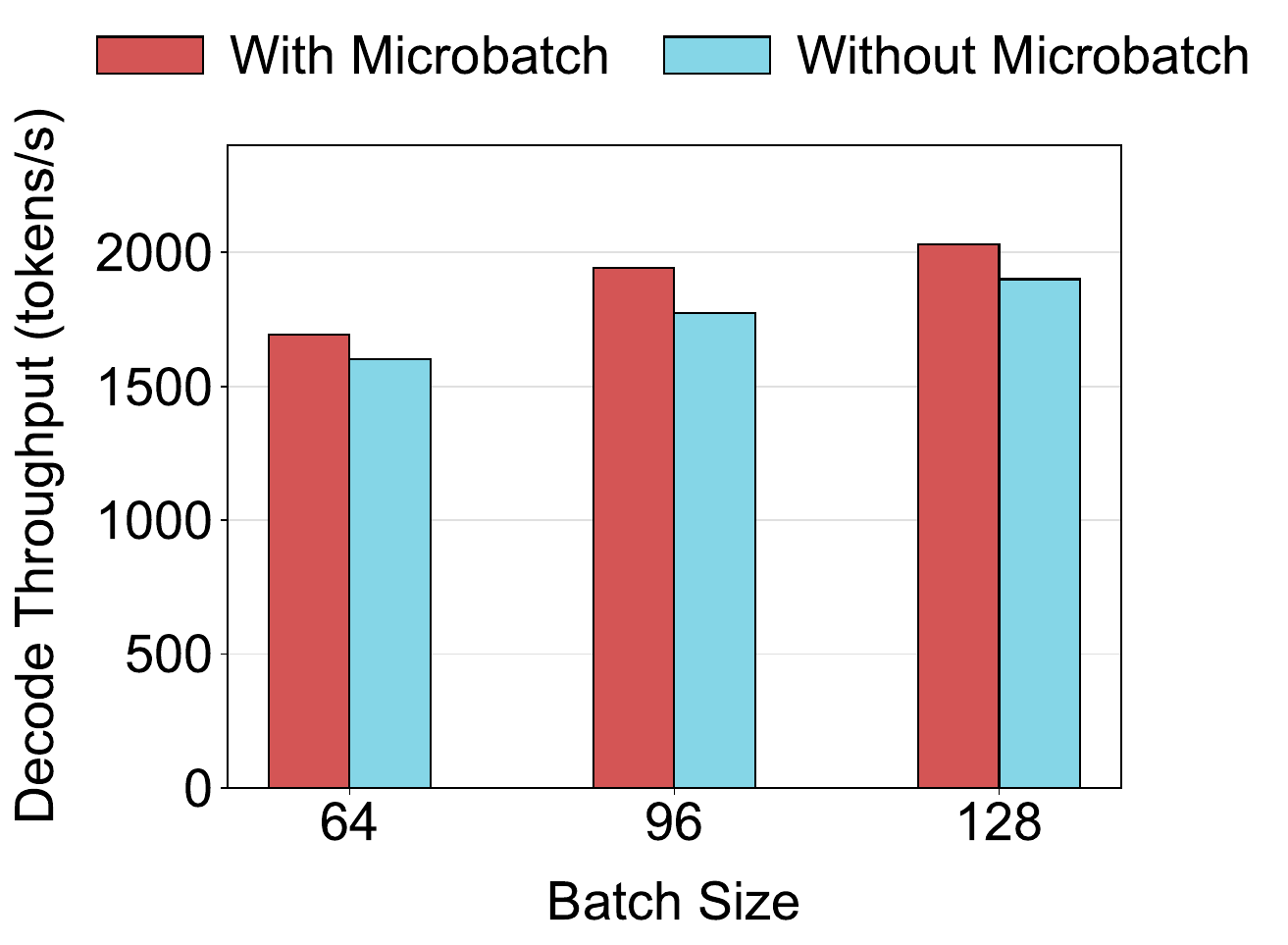} 
    \caption{Decode throughput.}
    \label{fig:eval-ablation-decode-pipeline}
  \end{subfigure}
  \begin{subfigure}[t]{0.61\linewidth}
    \centering
    \includegraphics[width=\linewidth]{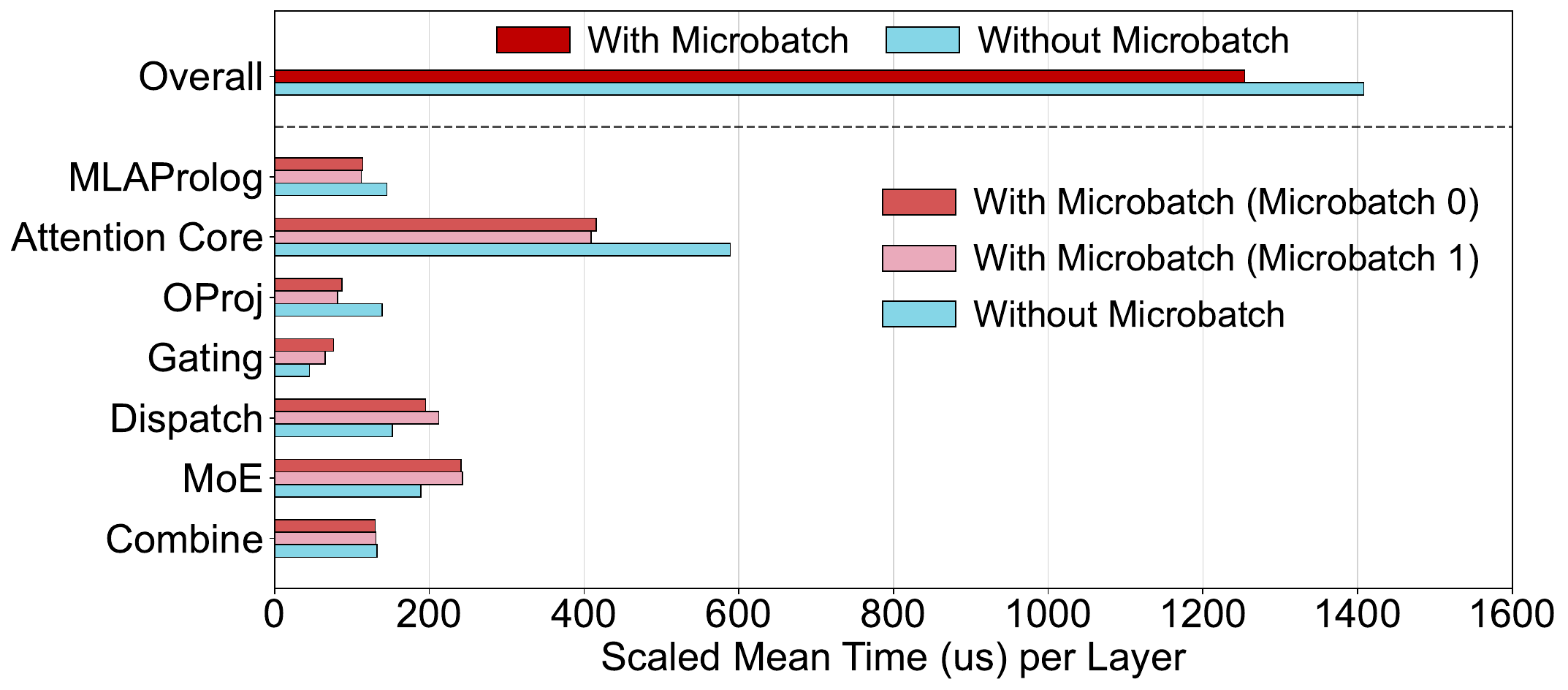} 
    \caption{Decode latency breakdown (batch size 96).}
    \label{fig:eval-ablation-decode-pipeline-latency}
  \end{subfigure}
  \caption{Decode throughput and per-layer latency breakdown with and without the microbatch-based pipeline. All requests have a 4,096-token KV cache length. In (b), the ``Overall'' with Microbatch indicates the per-layer latency after overlapping two microbatches (Microbatch 0 and Microbatch 1).}
  \label{fig:design-decode-pipe}
\end{figure}

\textbf{Decode Pipeline.}
We first evaluate our microbatch-based pipeline for the decode phase, previously detailed in \S\ref{sec:decode-microbatch}. The ablation compares system performance with and without this microbatch optimization. Figure~\ref{fig:eval-ablation-decode-pipeline} illustrates the decode throughput across various batch sizes. We observe that enabling the microbatch-based pipeline improves decode throughput by 5.8\%, 9.4\%, and 6.9\% for batch sizes of 64, 96, and 128, respectively. This gain, while beneficial, is relatively more modest when compared to potential improvements reported for other platforms (e.g., SGLang~\cite{LMSYS_SGLang_Team_2025_DeepSeek_EP} cited ~35\% on NVIDIA H100 clusters). This difference is primarily attributed to the inherently lower MoE dispatch and combine communication overheads on the \CMname{} with its high-performance UB plane (as detailed in Section~\ref{sec:op_perf_communication}), compared to NVIDIA GPU clusters typically utilizing RDMA. With smaller MoE communication stalls on the UB plane, the improvement ceiling from communication hiding via microbatching is naturally more constrained for the \CMname{}.

Figure~\ref{fig:eval-ablation-decode-pipeline-latency} provides a per-layer latency breakdown for decode execution with a batch size of 96. It reveals that although individual microbatch execution latency for stages like \texttt{Gating}, \texttt{Dispatch}, and \texttt{MoE} is marginally increased due to decreased per-stream compute resources (e.g., AICs from 24 to 16), the microbatch-based pipeline significantly benefits overall performance. This is achieved by effectively overlapping the attention path (Stream 0) and MoE path (Stream 1) for different microbatches, leading to an approximate 10\% reduction in overall per-layer latency and a corresponding considerable enhancement in end-to-end decode throughput.

\begin{figure}[t]
  \centering
  \begin{subfigure}[t]{0.375\linewidth}
    \centering
    \includegraphics[width=\linewidth]{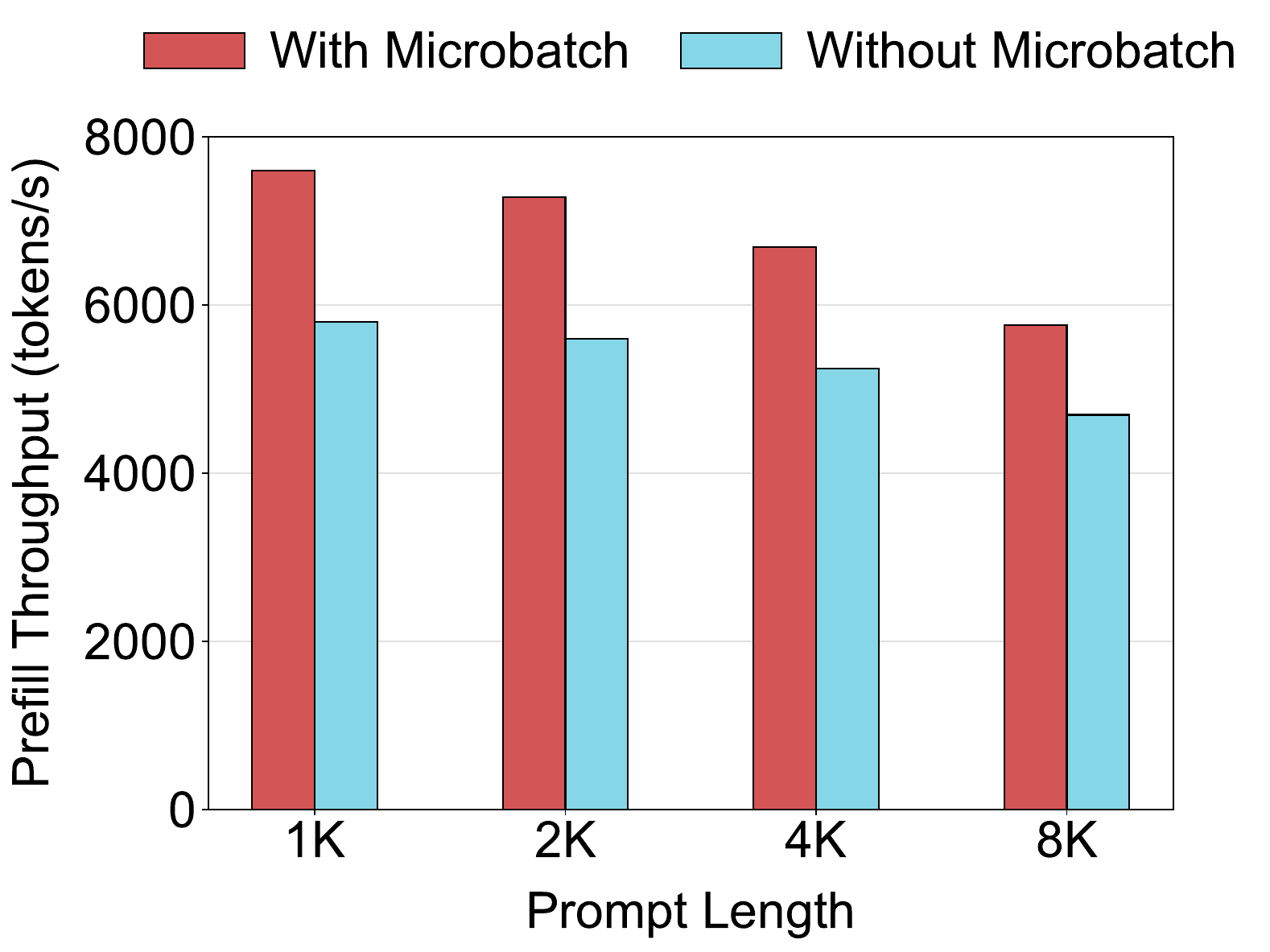} 
    \caption{Prefill throughput.} 
    \label{fig:eval-ablation-prefill-pipeline}
  \end{subfigure}
  \hspace{0.01\linewidth}
  \begin{subfigure}[t]{0.6\linewidth}
    \centering
    \includegraphics[width=\linewidth]{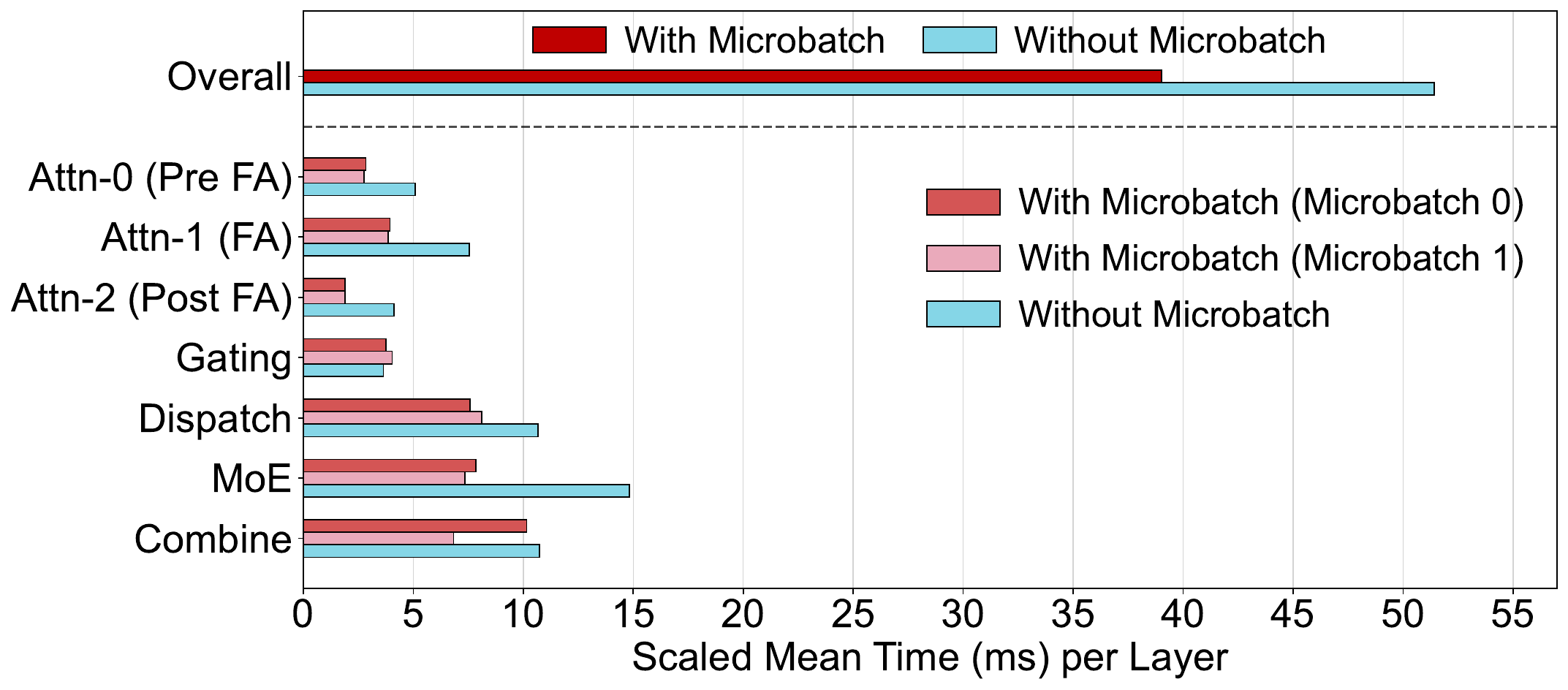} 
    \caption{Prefill latency breakdown (4K prompt length).}
    \label{fig:eval-ablation-prefill-pipeline-latency}
  \end{subfigure}
  \caption{Prefill throughput and per-layer latency breakdown with and without the microbatch-based pipeline. All experiments are executed with a batch containing 16K total tokens per NPU. In (b), the ``Overall'' with Microbatch indicates the per-layer latency after overlapping Microbatch 0 and Microbatch 1.}
  \label{fig:design-prefill-pipe-ablation}
\end{figure}

\textbf{Prefill Pipeline.}
Next, we examine the impact of our proposed microbatch-based prefill pipeline, detailed in Section~\ref{sec:prefill-microbatch}. Figure~\ref{fig:eval-ablation-prefill-pipeline} shows the prefill throughput under various prompt lengths, comparing performance with and without this pipeline. We observe that enabling the microbatch-based pipeline significantly improves prefill throughput by 23\% to 31\% across the tested configurations. Moreover, prefill throughput decreases as prompt lengths increase. This trend occurs because the per-token execution latency of attention computation increases with prompt length.

Figure~\ref{fig:eval-ablation-prefill-pipeline-latency} provides a corresponding per-layer latency breakdown for request execution with a 4K prompt length. The data reveals that the overall execution latency per layer is reduced by approximately 24\% when the microbatch pipeline is active. This substantial gain is primarily achieved by offloading lightweight computational tasks associated with communication (e.g., \texttt{DispatchCompute}, \texttt{CombineCompute}) to AIVs, and dedicating SDMA engines for bulk data transfers (e.g., All-to-All for MoE). This strategy allows their execution latency to be effectively overlapped with the core computations (like \texttt{ATTN} and \texttt{FFN}) performed on the AICs, leading to higher NPU utilization and reduced end-to-end prefill time.

\subsubsection{MTP}
\label{sec:ablation_mtp}

To specifically quantify the performance contribution of the MTP mechanism under typical conditions, we conduct a targeted ablation study. This evaluation focuses on the scenario where MTP generates a single speculative token per decoding step, using a consistent input sequence length of 4K tokens on the \CMname{}. We compare performance with MTP enabled against a baseline of standard single-token autoregressive decoding (i.e., MTP disabled) under identical workload parameters. 

\begin{figure}[t]
    \centering
    \begin{subfigure}[t]{0.35\linewidth}
        \centering
        \includegraphics[width=\linewidth]{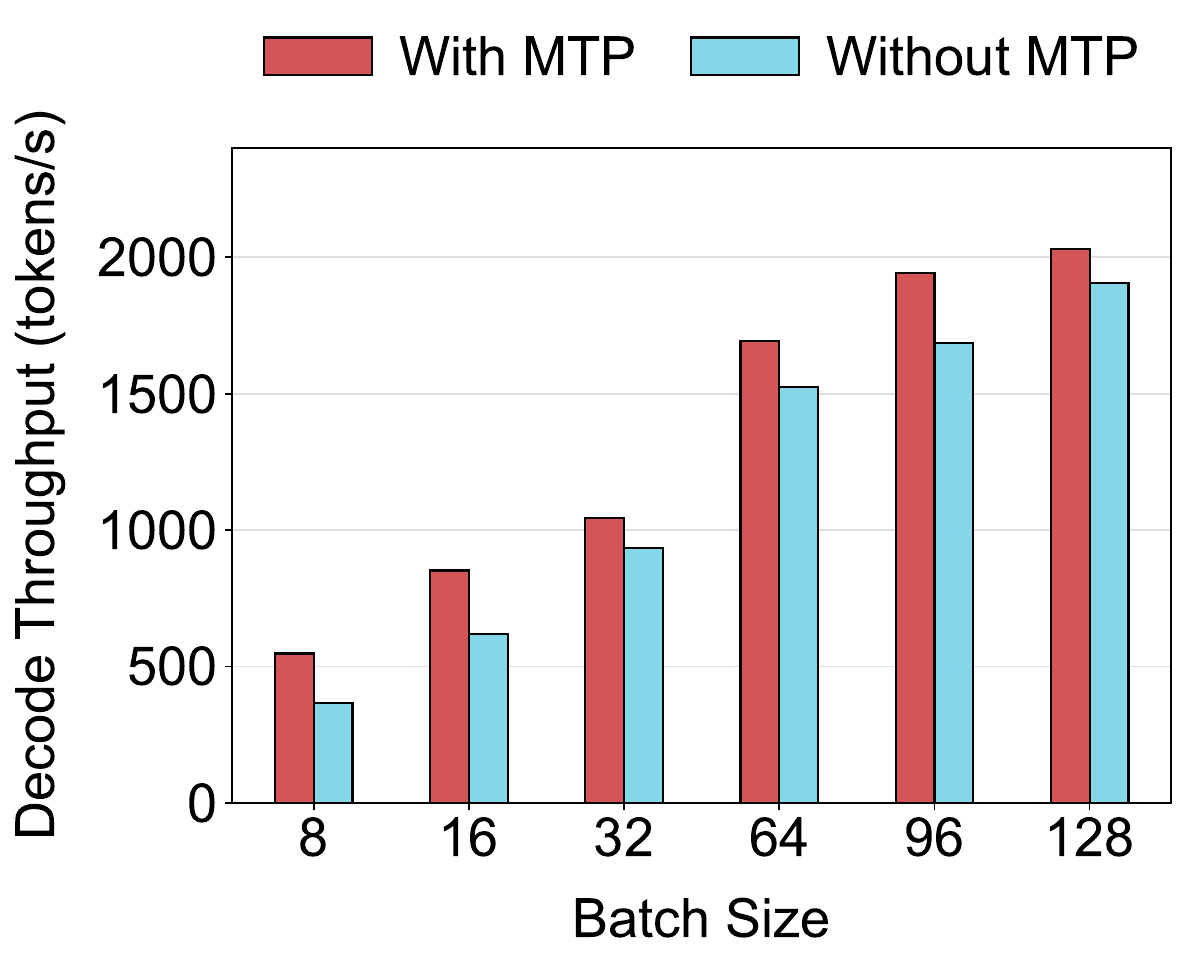} 
        \caption{Decode throughput.}
        \label{fig:eval-ablation-mtp-throughput}
    \end{subfigure}
    \begin{subfigure}[t]{0.63\linewidth}
        \centering
        \includegraphics[width=\linewidth]{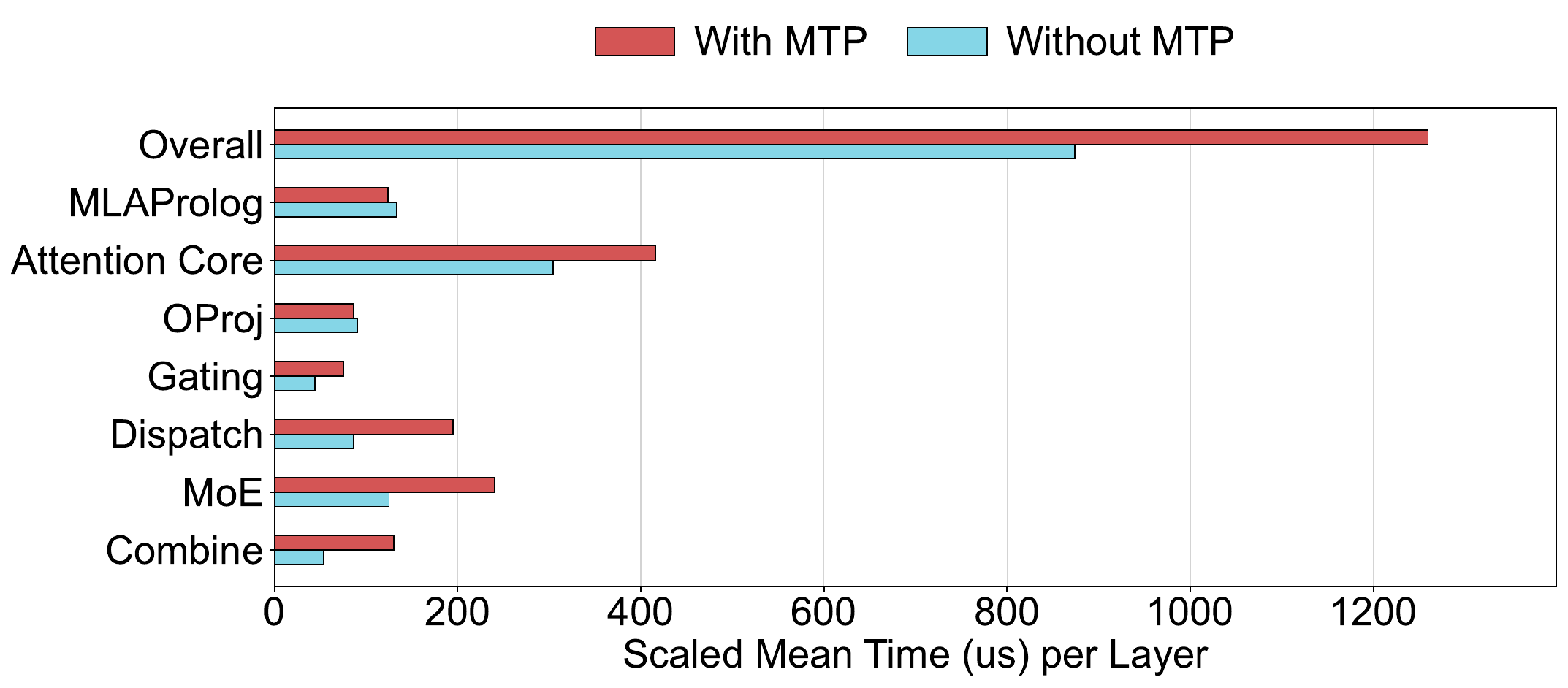} 
        \caption{Decode latency breakdown (batch size 96).}
        \label{fig:eval-ablation-mtp-latency}
    \end{subfigure}
    \caption{Decode throughput and per-layer latency breakdown with and without MTP. All experiments use an input sequence length of 4,096 tokens. In (b), ``Overall'' refers to the per-layer latency after overlapping two microbatches, and the operator latency represents the latency of a single microbatch.}
    \label{fig:mtp_ablation_main}
\end{figure}

As shown in Figure~\ref{fig:eval-ablation-mtp-throughput}, we observe that enabling MTP with a single speculative token improves decode throughput by 6\% to 49\% compared to the non-MTP baseline across different batch sizes. This speedup ratio is observed to be more pronounced for smaller batch sizes. This phenomenon may occur because at smaller batch sizes, the baseline non-MTP system is further from its peak efficiency (e.g., due to fixed per-iteration overheads being less amortized). The additional token accepted via MTP (at the 70\% rate) then provides a larger relative throughput gain. As batch sizes increase, while MTP can still offer an absolute benefit, its relative speedup may diminish as the baseline system itself becomes more saturated or as MTP's own overheads become more prominent.

However, this throughput enhancement is accompanied by an increase in the execution latency per decode layer iteration when MTP is active. As depicted in Figure~\ref{fig:eval-ablation-mtp-latency} for a batch size of 96, using MTP increases the per-layer execution latency by approximately 44\% (e.g., from a baseline of 874\,µs to 1,260\,µs with MTP). This is primarily because each MTP-enabled LLM decode step processes two input tokens per request from the last iteration: a base token and a speculative token. This larger effective batch size per iteration naturally leads to longer execution times for core operations such as Attention Core, Gating, Dispatch, MoE, and Combine.

Despite this increase in per-iteration latency, the overall throughput improves. The successful validation of speculative tokens at a 70\% acceptance rate means that, on average, 1.7 tokens (1 base token + 0.7 speculative token) are produced per MTP-enabled iteration. This gain in tokens per iteration outweighs the approximate 44\% longer iteration time, confirming the net positive impact of our MTP implementation for 4K sequence length workloads on \CMname{}.

\subsubsection{Context Caching}
\label{sec:ablation_context_caching}

The \textit{EMS-Context Caching} mechanism, introduced in \S\ref{sec:design-caching}, accelerates the prefill phase by storing and reusing KV cache blocks from previous requests. This ablation study quantitatively evaluates the effectiveness of EMS-Context Caching on \CMname{}, with a particular focus on how the underlying network fabric impacts cache access performance. Specifically, we measure prefill throughput and time-to-first-token (TTFT) using inputs with a 4K token length and a batch size containing 16K total tokens per NPU. To evaluate the performance under varying cache hit rates, we adjust the token reuse rate, which controls the proportion of historical KV prefixes reused. A central goal of this study is to compare EMS performance under two network configurations: one utilizing the high-bandwidth UB interconnect, and the other falling back to the slower VPC network plane for cache access.

\begin{figure}[t]
    \centering
    \begin{subfigure}[t]{0.48\linewidth}
        \centering
        \includegraphics[width=\linewidth]{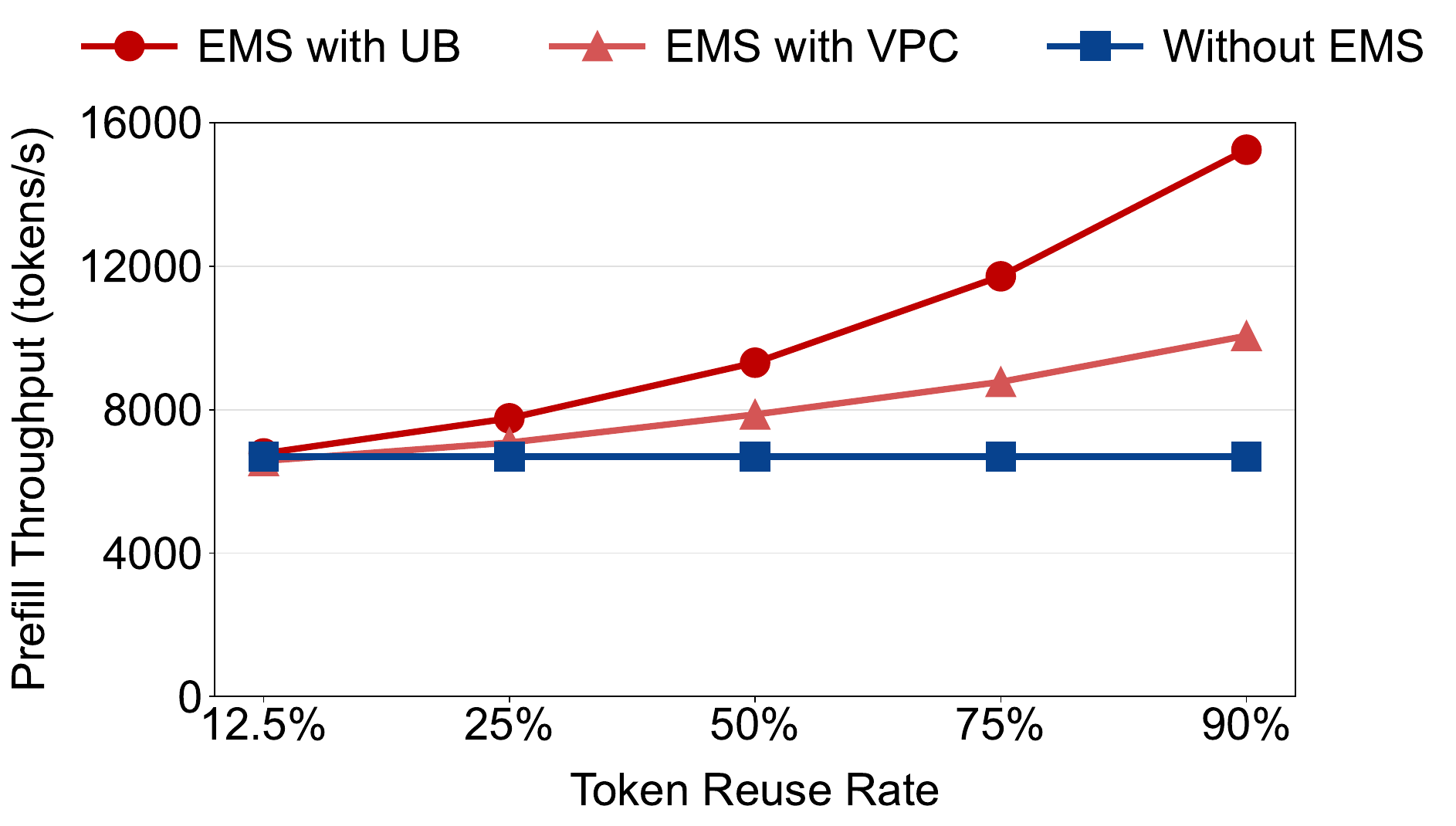}
        \caption{Prefill throughput.}
        \label{fig:eval-abla-ems-throughput}
    \end{subfigure}
    \hfill
    \begin{subfigure}[t]{0.48\linewidth}
        \centering
        \includegraphics[width=\linewidth]{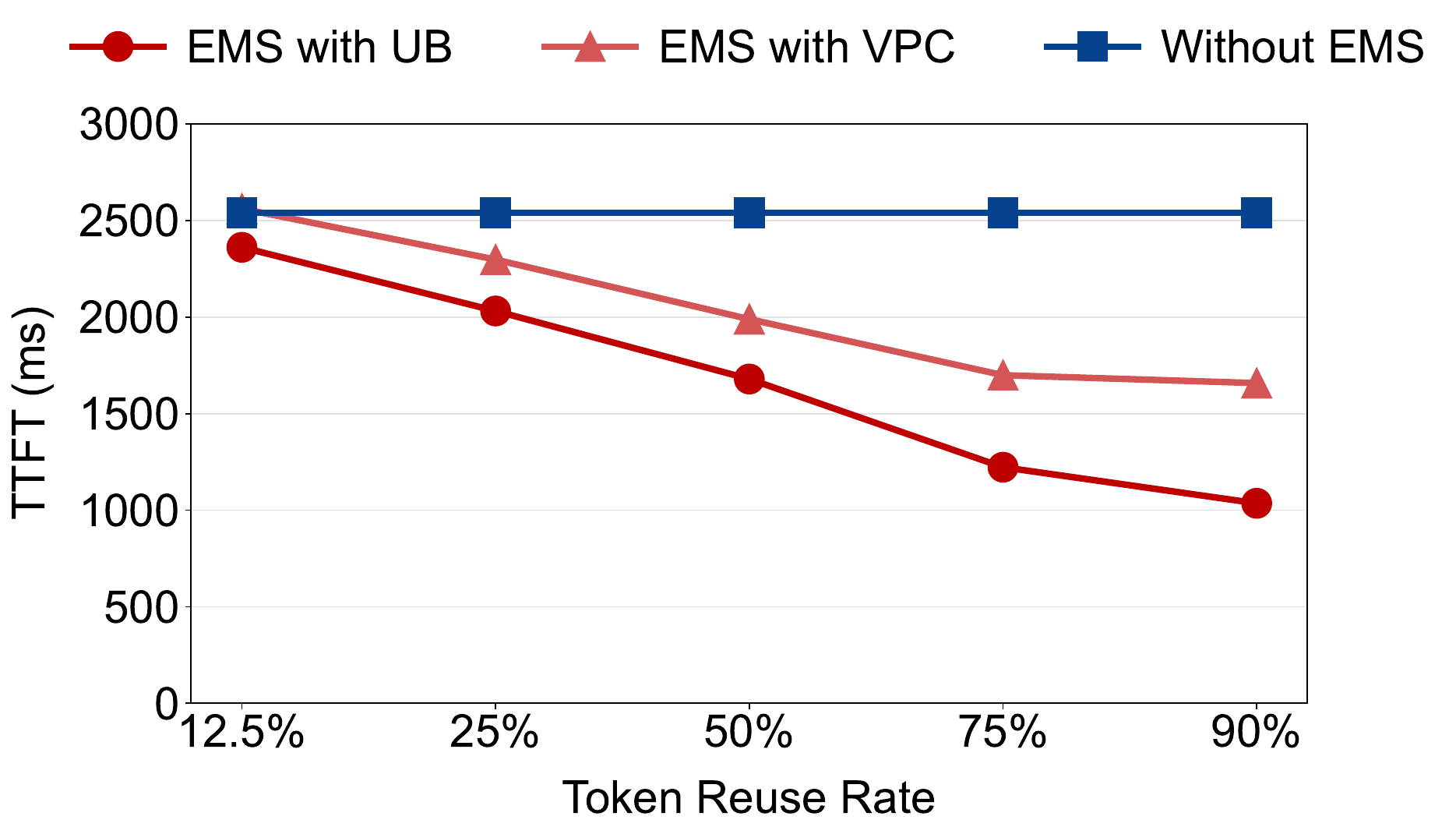}
        \caption{TTFT.}
        \label{fig:eval-abla-ems-latency}
    \end{subfigure}
    \caption{The overall prefill throughput and TTFT using EMS-Context Caching with different configurations.}
    \label{fig:context_caching_ablation_main}
\end{figure}

Figure~\ref{fig:context_caching_ablation_main} illustrates the performance trends as a function of the token reuse rate for these different EMS configurations. As shown in Figure~\ref{fig:eval-abla-ems-throughput}, there is a strong positive correlation between throughput and the reuse rate for both network configurations. For EMS with UB, increasing the reuse rate from 12.5\% to 50\% resulted in a 1.42$\times$ increase in prefill throughput. At a 90\% reuse rate, the throughput improved by 2.28$\times$ over the baseline without EMS. This substantial improvement occurs because a higher reuse rate translates to a larger portion of the input sequence's KV cache being loaded directly from the EMS cache rather than being recomputed, significantly reducing the computational load on prefill NPUs. Furthermore, when comparing the two network configurations, EMS with UB consistently outperforms EMS with VPC. Using the UB plane improves prefill throughput by up to 1.52$\times$. This gain is directly attributable to the significantly higher bandwidth and lower latency of the UB plane, which accelerates the loading of KV cache blocks from the distributed EMS cache to the NPUs.

Concurrently, TTFT significantly decreases as the token reuse rate increases, as depicted in Figure~\ref{fig:eval-abla-ems-latency}. For instance, with EMS on UB, a 50\% token reuse rate reduced TTFT by 861~ms (34\%) compared to no context caching, while a 90\% reuse rate led to a 1,505~ms (59\%) decrease. This marked reduction in TTFT is a direct consequence of bypassing substantial prefill computation when a cache hit occurs. The ability to quickly load historical KV cache from EMS, particularly when accessed via the high-bandwidth UB plane and potentially served from the DRAM tier of the disaggregated memory pool, translates directly into faster initial token generation. Similarly, accessing the EMS cache via the UB plane yields consistently lower TTFT compared to the VPC plane across all reuse rates, underscoring the importance of a high-performance interconnect for latency-sensitive cache retrieval.

\subsection{Performance of Operators} 
\label{sec:operator_performance}

Understanding the performance of fundamental computation and communication operators is key to diagnosing system bottlenecks and guiding software optimization efforts. In this subsection, we present a micro-benchmark analysis of critical operators relevant to LLM serving, specifically MoE communication primitives, MLA computations, and general matrix multiplication (GEMM) kernels. We evaluate their performance on the \CMname{} (per Ascend 910 die) and compare them against representative performance on NVIDIA H800 GPUs.

\subsubsection{Communication Operators}
\label{sec:op_perf_communication}

\begin{table}[t]
  \centering
  \footnotesize
  \caption{Communication operator performance (latency and bandwidth per rank) on NVIDIA H800 (RDMA) and \CMname{} (UB plane) for \texttt{Dispatch} and \texttt{Combine} operations across different EP degrees.}
  \label{tab:operator_breakdown}
  \renewcommand{\theadfont}{\bfseries}
  \begin{tabular}{@{}lccccc@{}}
    \toprule
    \multirow{2}{*}{\textbf{Operator}} & \multirow{2}{*}{\textbf{\#EP}} & \multicolumn{2}{c}{\textbf{DeepSeek DeepEP on H800~\cite{deepep2025}}} & \multicolumn{2}{c}{\textbf{CANN EP on CM384}} \\
    \cmidrule(lr){3-4} \cmidrule(lr){5-6}
    & & \textbf{Latency (\textmu s)} & \textbf{Bandwidth (GB/s)} & \textbf{Latency (\textmu s)} & \textbf{Bandwidth (GB/s)} \\
    \midrule
    \multirow{6}{*}{\textbf{Dispatch}} 
    & 8   & 163 & 46 & 116 & 71 \\
    & 16  & 173 & 43 & 131 & 63 \\
    & 32  & 182 & 41 & 133 & 62 \\
    & 64  & 186 & 40 & 141 & 58 \\
    & 128 & 192 & 39 & 152 & 54 \\
    & 256 & 194 & 39 & 152 & 54 \\
    \midrule
    \multirow{6}{*}{\textbf{Combine}}  
    & 8   & 318 & 46 & 118 & 131 \\
    & 16  & 329 & 44 & 132 & 117 \\
    & 32  & 350 & 41 & 146 & 105 \\
    & 64  & 353 & 41 & 150 & 103 \\
    & 128 & 369 & 39 & 150 & 103 \\
    & 256 & 360 & 40 & 149 & 103 \\
    \bottomrule
  \end{tabular}
\end{table}

We benchmark key MoE communication operators, specifically \texttt{Dispatch} and \texttt{Combine}, on our \CMname{} using the CANN implementation. This implementation is detailed in our design of fused communication operators (\S\ref{sec:decode-comm}). The performance is compared against DeepSeek's DeepEP implementation on NVIDIA H800 GPUs~\cite{deepep2025}, as shown in Table~\ref{tab:operator_breakdown}. The table presents latency and per-rank achieved bandwidth across various EP degrees (\#EP), from 8 to 256 ranks, with a batch of 128 per rank.

For the \texttt{Dispatch} operation, the CANN EP implementation on \CMname{} (CM384) consistently demonstrates lower latencies compared to DeepEP on H800 across all tested EP degrees. For example, at an EP degree of 8, CM384 achieves a latency of 116\,\textmu s, while the H800 records 163\,\textmu s. This latency advantage for CM384 persists as the EP degree increases, with CM384 showing 152\,\textmu s at EP256 versus H800's 194\,\textmu s. In terms of per-rank bandwidth for \texttt{Dispatch}, CM384 exhibits superior performance, at smaller EP degrees (e.g., 71~GB/s vs. 46~GB/s at EP8). However, under large EP degrees, we observe a significant decline in the effective bandwidth of CANN EP on \CMname{}. This degradation highlights a scalability bottleneck in the current EP implementation, which we leave as an avenue for future optimization.

The \texttt{Combine} operation reveals an even more pronounced performance advantage for CANN on CM384. Latencies are significantly lower on CM384 across all EP scales. For instance, at EP8, CM384's latency is 118\,\textmu s compared to H800's 318\,\textmu s. This substantial latency reduction is maintained up to EP256 (149\,\textmu s on CM384 vs. 360\,\textmu s on H800). Furthermore, the achieved per-rank bandwidth for \texttt{Combine} on CM384 is markedly higher than on H800. At EP8, CM384 delivers 131~GB/s per rank, nearly three times the 46~GB/s achieved on H800. This bandwidth superiority continues across all tested EP degrees, with CM384 providing a strong 103~GB/s per rank at EP256, while the H800 offers 40~GB/s.

These results underscore the efficiency of the CANN collective communication library and the high-performance capabilities of the UB plane in \CMname{} for MoE-specific communication patterns. The consistently lower latencies and higher per-rank bandwidth achieved on  \CMname{} are crucial for mitigating communication bottlenecks inherent in large-scale expert parallelism.

\subsubsection{MLA Operator}

\label{sec:op_perf_mla}

We evaluate the TFLOPS utilization and memory bandwidth utilization of our CANN MLA implementation on the CM384 against DeepSeek's FlashMLA on an NVIDIA H800, under both compute-intensive and memory-intensive settings.

\begin{table}[t]
  \centering
  \footnotesize
  \caption{Utilization of MLA operators on NVIDIA H800 and an Ascend 910 die (\CMname{}) in compute-intensive settings (BF16/FP16 precision).}
  \label{tab:tflops_performance}
  \begin{tabular}{@{}ccc@{}}
    \toprule
    \textbf{Operator Implementation} & \textbf{Precision} & \textbf{Compute Utilization (\%)} \\
    \midrule
    DeepSeek FlashMLA on H800 & BF16/FP16 & 66.7 \\
    \addlinespace
    CANN MLA on Ascend 910 die & BF16/FP16 & 65.4 \\
    \bottomrule
  \end{tabular}
\end{table}

\begin{table}[t]
  \centering
  \footnotesize
  \caption{Memory bandwidth utilization of MLA operators on NVIDIA H800 and an Ascend 910 die (\CMname{}) in memory-intensive settings.}
  \label{tab:bandwidth_performance}
  \begin{tabular}{@{}cc@{}}
    \toprule
    \textbf{Operator Implementation} & \textbf{Utilization} \\
    \midrule
    DeepSeek FlashMLA on H800  & 89.6\% \\
    \addlinespace
    CANN MLA on Ascend 910 die  & 84.1\% \\
    \bottomrule
  \end{tabular}
\end{table}

Table~\ref{tab:tflops_performance} presents the TFLOPS utilization for MLA operators when the workload is primarily compute-bound. The DeepSeek FlashMLA on H800 achieves a compute utilization of 66.7\%. Our CANN MLA on \CMname{}, also operating in BF16/FP16, achieves a comparable utilization of 65.4\%. This indicates that the efficiency in utilizing the available compute power for MLA is similar between the two platforms in compute-intensive scenarios.

In memory-intensive settings, the efficiency of utilizing available memory bandwidth is paramount. Table~\ref{tab:bandwidth_performance} shows this comparison. The DeepSeek FlashMLA on H800 achieves an 89.6\% utilization of its hardware memory bandwidth. Our CANN MLA implementation on \CMname{} achieves a similarly high utilization of 84.1\%.

\subsubsection{GEMM Operator}
\label{sec:op_perf_gemm}

General Matrix Multiplication (GEMM) is a fundamental compute kernel in virtually all deep learning models. The efficiency of GEMM operations, particularly at lower precisions like INT8, is critical for achieving high inference throughput. We benchmark the performance of INT8 GEMM kernels provided by CANN on a single Ascend 910 die (within the \CMname{} system) across a range of matrix dimensions. The results, detailed in Table~\ref{tab:perf_metrics}, showcase achieved compute utilization and the sustained memory bandwidth during these operations. These tests are conducted using common GEMM tiling dimensions (BM $\times$ BN = 128 $\times$ 152), with the operations involving INT8 inputs and BF16 outputs.

As indicated in Table~\ref{tab:perf_metrics}, the CANN INT8 GEMM kernels on the Ascend 910 die demonstrate consistently high compute utilization, ranging from 77.4\% to 82.7\% across various matrix shapes (M, N, K) and group counts. For example, with 4 groups and dimensions M=7168, N=4096, K=8192, the kernel achieves an 82.7\% utilization. This high efficiency is maintained across different configurations, indicating robust performance of the INT8 compute units on the Ascend 910 die.

The table also reports the sustained memory bandwidth achieved during these GEMM operations, which ranges from 195~GB/s to 327~GB/s. These values are substantially below the Ascend 910 die's peak memory bandwidth. This observation, when coupled with the high compute utilization figures, strongly suggests that these INT8 GEMM operations are predominantly compute-bound rather than memory-bandwidth-bound for the tested matrix dimensions. Such a characteristic indicates efficient data reuse within the NPU's internal cache hierarchy and registers, allowing the compute units to operate at a high fraction of their peak capability without being consistently starved for data transfers from memory.

\begin{table}[t]
  \centering
  \footnotesize
  \caption{INT8 GEMM achieved memory bandwidth and compute utilization on an Ascend 910 die (\CMname{}) across different configurations, using INT8 inputs and BF16 outputs. Tiling: BM $\times$ BN = 128 $\times$ 152.}
  \label{tab:perf_metrics}
  \renewcommand{\theadfont}{\bfseries\footnotesize}
  \begin{tabular}{@{}cccccc@{}}
    \toprule
    \thead{Groups} & \thead{M} & \thead{N} & \thead{K} & \thead{Memory Bandwidth (GB/s)} & \thead{Compute Utilization (\%)} \\
    \midrule
    4 & 7168 & 4096 & 4096 & 260 & 79.4 \\
    4 & 2048 & 7168 & 4096 & 325 & 77.4 \\
    4 & 7168 & 4096 & 8192 & 195 & 82.7 \\
    4 & 2048 & 7168 & 8192 & 266 & 81.1 \\
    8 & 7168 & 4096 & 4096 & 261 & 79.6 \\
    8 & 2048 & 7168 & 4096 & 327 & 77.9 \\
    \bottomrule
  \end{tabular}
\end{table}

%% file: sections/6-Future.tex
\section{Discussions on Future Directions} 
\label{sec:future_directions}

The rapid evolution of AI models and their pervasive application continue to impose increasingly stringent demands on AI infrastructure.  While \CMname{} represents a major architectural milestone in scaling tightly-coupled AI computation, further evolution is necessary to meet the needs of emerging workloads. In this section, we discuss potential future directions for both the CloudMatrix architecture and the LLM serving systems built upon it, aiming to further enhance scalability, flexibility, efficiency, and performance.

\subsection{Future CloudMatrix Evolutions}
\label{sec:future_supernode}

The supernode concept embodied by \CMname{}  can be extended along multiple dimensions to accommodate future AI workloads. 

\subsubsection{Unifying VPC and RDMA Planes} 
\label{sec:future-supernode-unifyed-plane}

As described in \S~\ref{sec:cm384_hw_overview}, \CMname{}  currently employs separate network planes for scale-out (RDMA) and VPC traffic. However, CloudMatrix enables the potential integration of scale-out communication into the VPC network. In typical AI training and inference workloads, bandwidth-intensive communication phases such as tensor, expert, and sequence parallelism (TP/EP/SP) are predominantly contained within the supernode. In contrast, cross-supernode communication, primarily arising from data and pipeline parallelism (DP/PP), typically exhibits much lower bandwidth demands. With hierarchical DP communication and communication-hiding techniques, the VPC network can adequately meet the inter-supernode communication demands of most AI workloads.

Building on this observation, a unified network architecture based on the VPC plane can enable the construction of large-scale AI clusters at the availability zone (AZ) scale. It accommodates heterogeneous generations of AI hardware, enables flexible and modular expansion using supernodes as the basic unit, and supports seamless interconnection across regions through data center network (DCN) technologies.

\subsubsection{Larger-scale Supernodes}
\label{sec:larger_supernodes}

Although \CMname{} provides a substantial scale with 384 NPUs, next-generation AI models and application scenarios are anticipated to necessitate even larger-scale supernodes. Several key factors drive this trajectory towards increased scale:

\textbf{1) Scaling to Match Model Evolution:}
As LLMs continue to scale in parameter size and architectural sophistication, the infrastructure required to serve them must evolve accordingly. Future models are expected to feature significantly larger parameter counts, longer input sequences, and an increasing number of sparsely activated experts, particularly in MoE designs. These trends place growing demands on compute, memory, and interconnect bandwidth within each inference session. Moreover, emerging architectural patterns, such as modular sub-networks for specialized reasoning, retrieval-augmented generation, or hybrid dense–sparse computation, require tighter coupling between model components, leading to increased intra-model communication and synchronization. Efficiently supporting these workloads necessitates co-locating compute and memory within a single, tightly integrated supernode to minimize communication latency and maintain high throughput. As a result, scaling up supernode capacity is critical not only to meet raw resource requirements but also to sustain the fine-grained locality and performance characteristics demanded by next-generation LLMs.

\begin{figure}[t]
\centering
\includegraphics[width=0.75\textwidth]{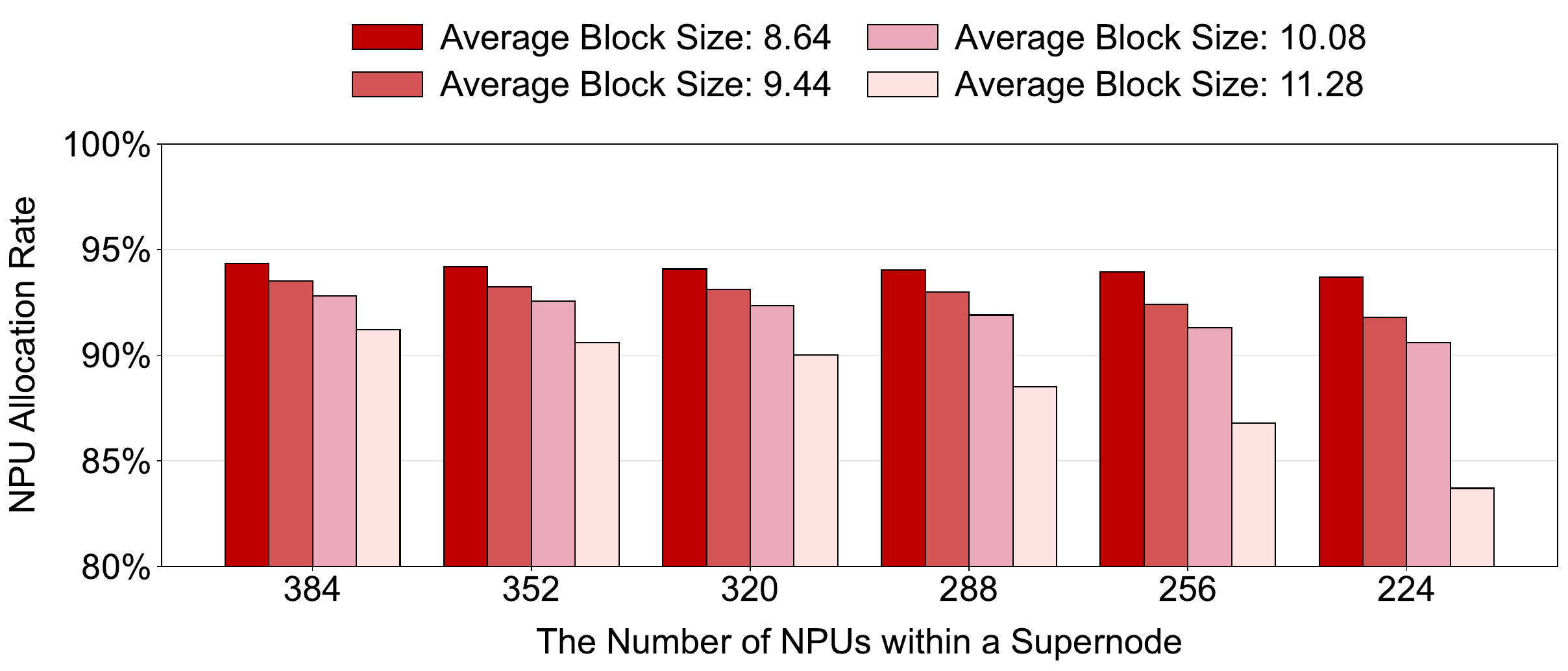}
\caption{NPU allocation rates under different supernode scales and tightly-coupled block sizes.}
\label{fig:future-allocaiton-util}
\end{figure}

\textbf{2) Improved Resource Allocation Efficiency:}
Scaling up supernode size also enhances system-wide resource utilization in real-world heterogeneous workload conditions. Based on real production traces, we simulate future NPU request patterns by modeling each AI task as a set of \textit{tightly-coupled blocks}, where each block is a contiguous group of NPUs that must be provisioned within a single supernode to meet intra-job bandwidth and latency constraints. As shown in Figure~\ref{fig:future-allocaiton-util}, larger supernodes consistently achieve higher NPU allocation rates across a broad range of average block sizes. For instance, at an average block size of 10.08, a 384-NPU supernode achieves over 94\% allocation, while a 224-NPU supernode drops below 91\%. This improvement stems from reduced fragmentation and better statistical multiplexing—larger resource pools offer greater placement flexibility for non-uniform job sizes. Conversely, for a fixed supernode size, increasing block size leads to lower allocation efficiency due to packing difficulty. When the average block size reaches 11.28, the allocation rate of the 224-NPU supernode drops below 85\%. These results highlight that scaling supernode size significantly improves system throughput and efficiency under realistic workload distributions.

\textbf{3) Nearly Constant Amortized Network Cost:}
Scaling up the size of a supernode does not inherently lead to higher per-NPU network costs. Given the same network architecture, e.g., a 2-tier Clos-like switching topology, the amortized cost of network infrastructure per NPU remains nearly constant across different supernode sizes as long as the configuration achieves full switch port utilization. As shown in Table~\ref{tab:cost-supernode-scale}, configurations with 192, 288, or 384 NPUs all achieve 100\% switch utilization with the same per-NPU amortized switch cost. Intermediate configurations, such as 256 or 352 NPUs, suffer from underutilized switches, slightly increasing per-node costs. These results suggest that scaling supernode size to the upper end of a given switching tier does not introduce additional cost overhead, making it a cost-effective strategy from a networking perspective.

\begin{table}[h]
  \centering
  \small
  \caption{Switch utilization across different supernode scales. Note that each logical switch consists of two physical switch chips presented in \S\ref{sec:ub_switch_system}.}
  \label{tab:cost-supernode-scale}
  \begin{tabular}{@{}cccc@{}}
    \toprule
    \textbf{\makecell{Supernode Scale  (\# of NPUs)}} &
    \textbf{\makecell{\# of Nodes}} &
    \textbf{\makecell{\# of Switches}} &
    \textbf{\makecell{Switch Utilization}} \\
    \midrule
    384  & 48 & 56 & 100\% \\
    352  & 44 & 56 & 92\%  \\
    288  & 36 & 42 & 100\% \\
    256  & 32 & 42 & 89\%  \\
    192  & 24 & 28 & 100\% \\
    \bottomrule
  \end{tabular}
\end{table}

\textbf{4) Accommodating Increased Resource Heterogeneity:}
Future AI workloads will require increasingly diverse hardware support within the same execution context. Alongside NPUs and CPUs, next-generation supernodes are likely to incorporate specialized accelerators for tasks such as physics simulation, real-time video processing, lossless data compression, and cryptographic computation. These units are becoming essential components in end-to-end AI pipelines, particularly for multimodal or domain-specific applications. To be efficiently utilized, such heterogeneous resources must share the same high-bandwidth, low-latency interconnect fabric and be accessible as first-class compute peers within the supernode. Supporting this diversity at scale requires both an expanded supernode size and a more flexible interconnect architecture, further reinforcing the trend toward larger, more heterogeneous compute domains that can handle tightly coupled, cross-functional AI workloads.

\subsubsection{Physical Disaggregation and Pooling of CPUs}
\label{sec:physical_cpu_disaggregation_future}

While the current \CMname{}  supernode already achieves a degree of resource flexibility by pooling CPUs and NPUs from its compute nodes (each integrating 4 Kunpeng CPUs and 8 Ascend NPUs), a key future direction for the CloudMatrix architecture involves a more fundamental \textit{physical disaggregation} of CPU and NPU resources, as illustrated in Figure~\ref{fig:cm-conceptual-design}. This envisions a supernode constructed from distinct, specialized node types: NPU-centric nodes densely packed with AI accelerators, and CPU-centric nodes offering substantial general-purpose compute, memory capacity, and I/O capabilities. These heterogeneous node types would be interconnected via the high-bandwidth, low-latency UB network plane, enabling granular, flexible, and scalable resource pooling at the supernode level.

The motivation for physical disaggregation arises from the rigidity of conventional CPU-NPU pairings in fixed node configurations, where static NPU-to-CPU ratios constrain the system's ability to match workload demands. For example, some inference workloads require intensive CPU pre/post-processing or large memory-backed caching, resulting in CPU bottlenecks despite idle NPUs. Conversely, training workloads might saturate NPUs while leaving CPU resources underutilized. In such cases, tightly coupled CPU-NPU configurations lead to suboptimal hardware utilization and inflexible scaling.

Although \CMname{} 's peer-to-peer UB topology already decouples logical resource assignment, enabling flexible CPU-NPU matching across the supernode, physically separating CPU and NPU resources into dedicated resource pools unlocks further advantages:

\textbf{1) Independent and Optimized Scaling:}
Physically separate NPU-centric nodes (e.g., with a minimal local CPU for basic management but maximized NPU density) and CPU-centric nodes (e.g., with many CPU cores, large DRAM capacities, and rich I/O options, serving as the supernode's primary CPU and memory resource pool) could be developed. This allows the NPU compute capacity and the general-purpose CPU/memory capacity of the supernode to be scaled independently and more economically. Datacenter operators could then compose supernodes with highly variable NPU-to-CPU-and-memory ratios, precisely tailored to the dominant workloads (e.g., NPU-rich for training, CPU/memory-rich for data-intensive pre-processing or large-scale EMS caching).

\textbf{2) Enhanced Resource Utilization and Specialization:}
Specialized node designs allow for hardware optimization specific to the primary resource type. NPU nodes could focus on power delivery and cooling for accelerators, while CPU/memory nodes could optimize for memory density, I/O bandwidth, or specific CPU instruction sets. This can lead to better overall efficiency and performance for each resource type compared to a one-size-fits-all hybrid node design.

\subsection{Future Serving System Enhancements}
\label{sec:future_serving_system} 

As the underlying supernode architecture continues to evolve, the LLM serving system must co-evolve to fully leverage these capabilities. A key direction is moving beyond coarse-grained disaggregation (e.g., prefill-decode separation) toward more fine-grained component-level disaggregation and intelligent, adaptive deployment strategies. These approaches aim to improve resource utilization, boost throughput, and support increasingly heterogeneous workloads and hardware configurations.

\subsubsection{Component-Level Disaggregation}
\label{sec:component_disaggregation}

The peer-to-peer serving architecture with prefill-decode-caching disaggregation employed in \CMname{}  has proven effective in separating major phases of LLM inference. However, further improvements are possible by decomposing model execution into even finer-grained components that can be managed, deployed, and scaled independently. We highlight two emerging directions:

\textbf{1) Decode-Attention Disaggregation and Offloading:}
While prefill instances are compute-bound and decode instances are often memory-bound, the Adrenaline system~\cite{liang2025injecting} shows that additional performance gains can be achieved by disaggregating memory-intensive attention computation from the decode path and offloading it to underutilized prefill instances. This approach improves overall memory bandwidth utilization and enables larger batch sizes on decode instances, thereby increasing compute efficiency. It relies on low-latency synchronization, careful co-location of offloaded tasks, and SLO-aware offloading policies. The result is improved throughput without compromising latency, exemplifying how attention disaggregation can unlock latent capacity within existing serving deployments.

\textbf{2) Attention and MoE Disaggregation:} 
Large-scale MoE models present unique challenges due to sparse expert activation and extreme memory demands. MegaScale-Infer~\cite{zhu2025megascale} proposes disaggregating attention and expert components into separate execution services, enabling different parallelism strategies and hardware mappings. Attention layers, which process every token, are deployed using data-parallelism on memory-optimized nodes, while expert FFNs are distributed via expert parallelism across a dedicated resource pool. This disaggregated execution reduces contention, improves throughput, and allows independent scaling of attention and expert resources, which is critical for efficiently serving trillion-parameter MoE models.

Together, these disaggregation techniques represent a shift toward viewing LLMs as collections of loosely coupled microservices, each with distinct performance profiles. This granularity allows better mapping to heterogeneous hardware and improves load balancing and scalability across a supernode.

\subsubsection{Hybrid and Adaptive Deployment}
\label{sec:hybrid_deployment}

Once LLM inference is decomposed into components, which can be considered as fine-grained microservices, such as attention execution, FFN computation, KV cache management, or MoE expert gating, the serving system gains significant flexibility to adopt more sophisticated deployment strategies. These hybrid and adaptive deployment models enable the system to tailor resource allocation to each component’s unique computational and memory requirements, improving overall utilization and scalability.

\textbf{1) Hardware-aware Microservice Placement:}
Each microservice can be mapped to the most suitable hardware type based on its performance profile. For instance,  attention layers, which are typically memory bandwidth-bound, should be prioritized on NPUs with high memory throughput; compute-intensive FFN modules benefit from allocation on NPUs with strong compute capabilities; and lightweight or latency-tolerant operations, such as KV cache indexing, can be offloaded to pooled CPUs or lower-cost general-purpose accelerators. This fine-grained matching enables more efficient use of heterogeneous hardware and reduces cost without compromising performance.

\textbf{2) Hybrid Microservice Co-location:}
Disaggregated microservices can also be dynamically co-located to improve resource utilization across the supernode. For example, memory-bound attention operations from the decode phase can be offloaded to memory-underutilized prefill instances~\cite{liang2025injecting}. Such hybrid co-location strategies help alleviate resource bottlenecks, improve utilization across phases, and increase effective system throughput, especially under variable or bursty workloads.

\textbf{3) Adaptive and Independent Scaling of Microservices:}
A key advantage of microservice disaggregation is the ability to scale each component independently based on real-time workload characteristics. For example, during the processing of long-context inputs, the attention microservice may experience higher load and be scaled accordingly, without necessitating additional FFN or expert resources. This granularity prevents systemic over-provisioning and allows the system to elastically adapt to workload dynamics.

To fully exploit these capabilities, the serving infrastructure must incorporate a sophisticated orchestration layer capable of continuously profiling system load, predicting performance bottlenecks, and making real-time, SLO-aware scheduling and scaling decisions. This orchestrator serves as the control plane for the hybrid deployment model, ensuring that performance guarantees are met even as workloads and resource availability fluctuate.

In summary, hybrid and adaptive deployment strategies, enabled by component-level disaggregation, represent a promising frontier in LLM serving system design. They allow for more precise resource utilization, seamless load balancing across heterogeneous hardware, and the ability to meet future demands posed by increasingly complex and diverse model architectures.

%% file: sections/7-Conclusion.tex
\section{Conclusion} 
\label{sec:conclusion}

In this paper, we introduce Huawei CloudMatrix, a next-generation AI datacenter architecture that embodies Huawei’s vision for advanced AI infrastructure. We specifically highlight Huawei \CMname{}, the first production-grade implementation of this innovative architectural concept. \CMname{} is an AI supernode engineered to efficiently support large-scale AI workloads, featuring a fully peer-to-peer interconnected hardware design. It integrates 384 Ascend 910 NPUs and 192 Kunpeng CPUs interconnected via an ultra-high-bandwidth, low-latency Unified Bus (UB) network. This unique architecture facilitates dynamic resource pooling, streamlined memory management, and exceptional inter-node communication, effectively addressing the scalability and efficiency challenges common in traditional datacenter architecture.

Leveraging \CMname{}, we propose \system{}, a comprehensive serving solution featuring a peer-to-peer serving architecture that disaggregates the inference workflow into distinct prefill, decode, and caching subsystems. This architecture significantly simplifies scheduling, enhances load balancing, and optimizes resource utilization by enabling uniform access to a shared disaggregated memory pool across all NPUs. We further design and implement advanced hardware-aware techniques, including large-scale expert parallelism (LEP), optimized communication and MLA operators, microbatch-based pipelining, and INT8 quantization. These techniques collectively boost MoE and MLA computation throughput, improve caching efficiency, and deliver substantial gains in overall inference performance.

Our extensive evaluations with the DeepSeek-R1 model demonstrate that \system{} achieves remarkable throughput, delivering 6,688 tokens/s per NPU in the prefill stage and 1,943 tokens/s per NPU during decoding, while consistently maintaining a low latency below 50 ms per output token. These results correspond to compute efficiencies of 4.45 tokens/s/TFLOPS for prefill and 1.29 tokens/s/TFLOPS for decode, both of which surpass the published efficiencies of leading frameworks like SGLang on NVIDIA H100 and DeepSeek on H800. Furthermore, \system{} effectively manages the throughput-latency trade-off, capable of sustaining a 538 tokens/s throughput even under a stricter sub-15 ms TPOT constraint. The INT8 quantization strategy further retains accuracy comparable to DeepSeek’s official API across a wide array of benchmarks.

Looking forward, several exciting directions emerge for further enhancing \CMname{}. Future work includes integrating and unifying the VPC and RDMA network planes for even more streamlined interconnectivity, scaling to larger supernode configurations, and pursuing deeper disaggregation and pooling of CPU resources. Additionally, finer-grained component-level disaggregation and adaptive deployment strategies present promising avenues for achieving even greater flexibility, efficiency, and scalability in AI datacenter infrastructures.

Collectively, our findings validate Huawei CloudMatrix as a highly effective, scalable, and performance-optimized platform for deploying large-scale AI workloads, setting a benchmark for future AI datacenter infrastructures.